\documentclass[11pt]{article}

\pdfoutput=1 
\usepackage{jheppub}

\usepackage{amsmath,amssymb,amsfonts}
\usepackage{hyperref}
\hypersetup{colorlinks,citecolor=blue,urlcolor=blue,linkcolor=blue}
\usepackage{graphicx}
\usepackage{xcolor}

\usepackage{caption}
\usepackage{subcaption}

\usepackage{nicefrac}
\usepackage{mathrsfs}
\usepackage{mdframed}
\usepackage{ulem}
\usepackage{dsfont}
\usepackage{units}
\usepackage[english]{babel}
\usepackage{float} 
\usepackage{braket}
\def\beq{\begin{equation}\begin{aligned}}
\def\eeq{\end{aligned}\end{equation}}

\begin{document}


 \title{The Nonperturbative Hilbert Space of Quantum Gravity With One Boundary}

\author[1,2,3]{Vijay Balasubramanian}
\author[4]{Tom Yildirim}
\affiliation[1]{Department of Physics and Astronomy, University of Pennsylvania, Philadelphia, PA 19104, U.S.A.}
\affiliation[2]{Theoretische Natuurkunde, Vrije Universiteit Brussel and International Solvay Institutes, Pleinlaan 2, B-1050 Brussels, Belgium}
\affiliation[3]{Rudolf Peierls Centre for Theoretical Physics, University of Oxford,Oxford OX1 3PU, U.K.}
\affiliation[4]{Department of Physics, Keble Road, University of Oxford, Oxford, OX1 3RH, UK}
\emailAdd{vijay@physics.upenn.edu}
\emailAdd{tom.yildirim@physics.ox.ac.uk}

\abstract{  We discuss a basis for the nonperturbative Hilbert space of quantum gravity with one asymptotic boundary.   We use this basis to show that the Hilbert space for  gravity with two disconnected boundaries factorizes into a product of two copies of the single boundary Hilbert space.}

\date{May 2025}

\maketitle
\newpage

\section{Introduction}
In this paper we address a puzzle about quantum gravity. The  puzzle concerns gravity in universes with two asymptotic boundaries. If the cosmological constant is negative, holographic duality requires that the underlying Hilbert space is the product of two single-boundary gravity Hilbert spaces.
However, the gravity path integral with these boundary conditions includes a sum over wormholes that connect the asymptotic regions. These wormholes seem to threaten factorization of the Hilbert space \cite{Harlow:2015lma,Harlow:2018tqv}.  This is our  puzzle: does the two-boundary Hilbert space of quantum gravity factorize into single-boundary Hilbert spaces?

The puzzle can be resolved if we have access to a  complete basis of gravity states.  Recently, such a basis was constructed by cutting open the Euclidean gravitational path integral for universes with two boundaries \cite{Balasubramanian:2022gmo,Balasubramanian:2022lnw,Climent:2024trz,Balasubramanian:2024yxk,Toolkit}, even though this path integral seems to only be an effective description calculating coarse-grained observables of an underlying fine-grained theory \cite{Saad:2019lba,Saad:2019pqd,Sasieta:2022ksu,Marolf:2020xie,deBoer:2023vsm,deBoer:2024mqg}. Here, using the toolkit constructed in \cite{Toolkit} for deriving equality between quantities in the underlying fine-grained theory with access only to the gravity path integral, we will show that the shell states discussed in \cite{Chandra:2022fwi} span the entire single-boundary Hilbert space. This explicit basis  allows us to show that the complete two-boundary Hilbert space factorizes into a product of two single-boundary Hilbert spaces.  Previous work had shown that traces taken in the Hilbert space associated to two-boundary black holes are products of factors, but did not establish what those factors were or what the mechanism of factorization was (see \cite{Boruch:2024kvv} for 2d JT gravity, and \cite{Balasubramanian:2024yxk} for a leading saddlepoint analysis in higher dimensions).

Four sections follow. Sec.~\ref{sec:QTFTPI} reviews aspects of the gravitational path integral and the toolkit of \cite{Toolkit}.  Sec.~\ref{sec:1s-span}   constructs a basis for the single-boundary Hilbert space. Sec.~\ref{sec:factorisation} uses this basis to demonstrate that the two-boundary Hilbert space is a product of two one-boundary Hilbert spaces.  
We conclude in Sec.~\ref{sec:discussion} with a summary and discussion.

\section{The gravitational path integral} \label{sec:QTFTPI}

\subsection{Review}
\label{sec:review}
\begin{figure}[h]
    \centering
    \includegraphics[width=0.5\linewidth]{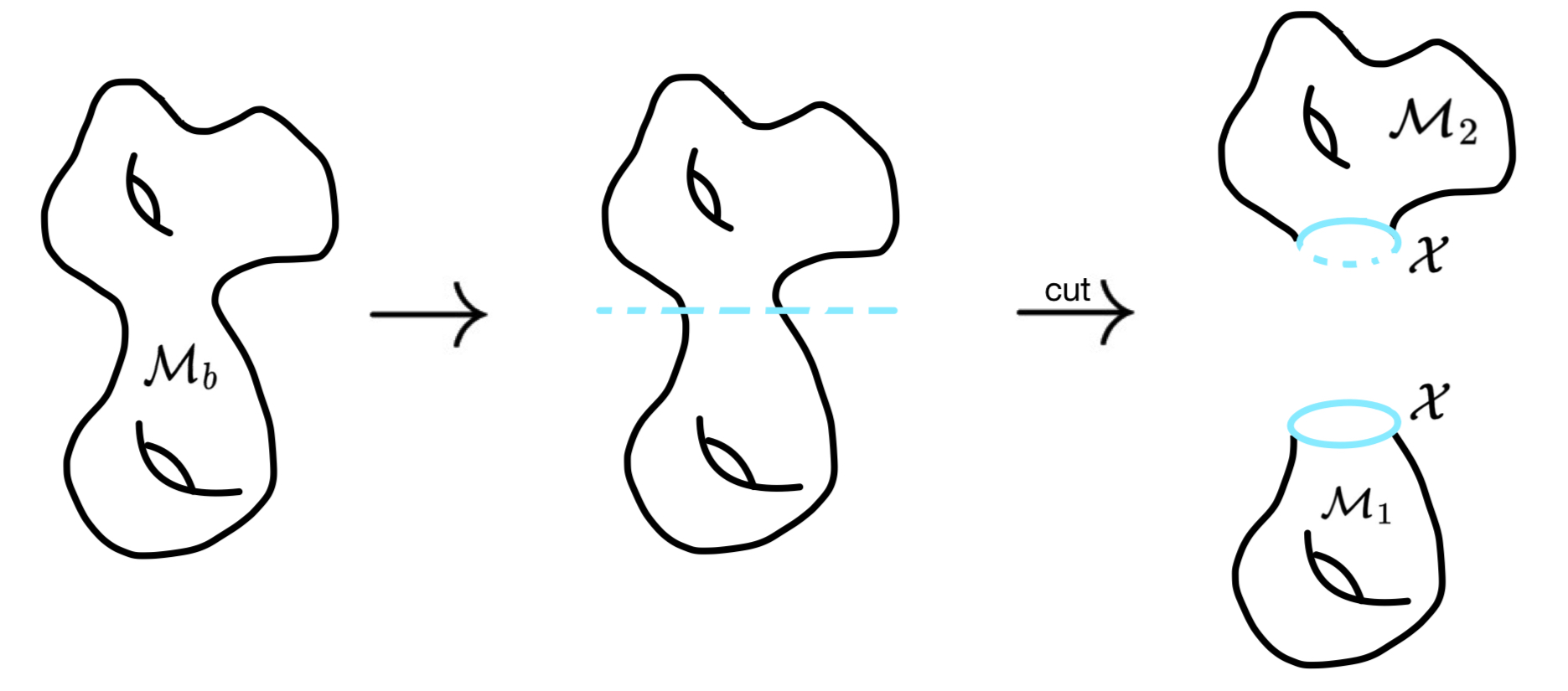}
    \caption{Cartoon of cutting a gravity path integral boundary condition along  $\mathcal{X}$ to define a state.}
    \label{fig:cutpathint}
\end{figure}

The Euclidean gravity path integral is a map from  boundary conditions -- topology, metric and matter sources (collectively $\Sigma_b$) -- to the complex numbers.  We calculate it by summing over bulk topologies, geometries ($g$), and  fields ($\phi$), compatible with the  boundary conditions: 
\begin{align}
\zeta[\Sigma_b] &= \int_{g,\phi \to \Sigma_b} \mathcal{D}g\mathcal \, \mathcal{D}\phi \, e^{-I_{bulk}[g,\phi]} \nonumber
\\
I_{bulk} &= -\frac{1}{16\pi G_{N}}\int_{\mathcal{M}} \sqrt{g}( R -2\Lambda + \mathcal{L}_{matter})  -\frac{1}{8\pi G_{N}}\int_{\partial\mathcal{M}= \mathcal{M}_b} \sqrt{h}K  \, .
\label{eq:gravoverlap} 
\end{align}
Here $I_{bulk}$ is the gravitational action; for example the Einstein-Hilbert action, $\mathcal{M}$ is a bulk manifold filling in the boundary, and a sum over topologies is  understood. This is an effective description below a cutoff for a more complete fundamental theory.  To construct states we formally cut open the boundary conditions put into the Euclidean path integral along some cut $\mathcal{X}$ as in Fig.~\ref{fig:cutpathint}. The two parts produce functionals of  data on the cut, defining states $\ket{M_1}$ and $\bra{M_2}$ such that  $\braket{M_2|M_1} = \zeta[\Sigma_b]$. Note that defining a state by partitioning boundary data is a standard way of producing states in the \textit{bulk} quantum gravitational Hilbert space. For instance, such methods were used in \cite{Iliesiu:2024cnh} and \cite{Marolf:2020xie}  to construct  states in JT gravity and a topological toy model respectively, and in this work we will apply this method in higher dimensional Einstein-Hilbert gravity. Crucially, we do not assume any kind of dual boundary CFT construction.  It is clear our states live in the bulk Hilbert space as we can, in principle, compute the wavefunctionals $\Psi_{\mathcal{M}_{1}}(h,\phi) \equiv\braket{h,\phi|\mathcal{M}_{1}}$ between our states and states defined by some bulk time-slice  data $h_{\mu \nu},\phi$ by performing the path integral subject to these joint boundary conditions. For example, consider the gravity path integral subject to  the periodic asymptotic boundary condition $\mathbb{S}^{1}_{\beta}\times \mathcal{X}$ where $S^{1}_{\beta}$ denotes a boundary circle of length $\beta$.  We will denote this gravitational path integral as ${Z(\beta)}$ and call it the ``Z partition function''.  By cutting this path integral in two it can be interpreted as computing the norm of a state $|\beta\rangle$ defined as ${Z(\beta)}=\braket{\beta|\beta}$.

We can construct states $\ket{i}$ by performing the path integral after inserting  operators  $\mathcal{O}_i$ on the boundary 
(\cite{Marolf:2017kvq,Marolf:2020xie,Colafranceschi:2023moh}).  Inserting $\mathcal{O}_i^\dagger$ instead constructs $\bra{i}$, or equivalently  $\ket{i^*}$.  To calculate the overlap $\braket{j|i}$ we sew the boundary conditions defining $\bra{j}$ and $\ket{i}$ together along $\mathcal{X}$ and integrate over all geometries satisfying this completed boundary condition. To compute the square of this overlap, $\braket{i|j}\braket{j|i}$, we  insert $\mathcal{O}_{i,j}$ on two copies of the boundary manifold and sum over all  geometries and topologies that fill in the boundary conditions, including connected topologies (wormholes) interpolating between the two boundaries.  These wormholes can lead to an apparent paradox,  
\begin{equation} 
\braket{i|j}= \delta_{ij} ~~~~\mathrm{while}~~~
\braket{i|j}\braket{j|i}= \delta^2_{ij} + Z_{WH} \, ,
\label{eq:coarseinner}
\end{equation}
where $Z_{WH}$ is the wormhole contribution.
Following 
\cite{Saad:2019lba,Saad:2019pqd,Sasieta:2022ksu,Marolf:2020xie,deBoer:2023vsm,deBoer:2024mqg,Penington:2019kki,Almheiri:2019qdq,Balasubramanian:2022gmo,Balasubramanian:2022lnw,Climent:2024trz,Balasubramanian:2024yxk,Marolf:2024jze} we interpret this situation as arising because the gravitational path integral is an effective theory  computing coarse-grained averages of quantities in 
an underlying fine-grained theory.  We  therefore write an overbar over quantities computed using the gravitational path integral, e.g.,
\begin{equation}\label{eq:WH2}
\overline{\braket{i_1|j_i}\braket{i_2|j_2} \cdots \braket{i_n|j_n}}  \, .
\end{equation}

Let $\mathcal{H}_{\kappa} \equiv span\{ |i\rangle , i= i,2, ... \kappa \}$ be the span of the states described above. Overlaps between these states are summarized by the Gram matrix $G_{ij} \equiv \braket{i|j}$, whose rank equals the dimension of $\mathcal{H}_{\kappa}$.  To construct an orthonormal basis from the overcomplete $\ket{i}$ basis we write 
\beq \label{Shellbasis}
|v_\gamma\rangle= G^{-1/2}_{ji}U_{i\gamma}|j\rangle \, ,
\eeq
where $U$ is a unitary transformation such that $G=UDU^{\dagger}$ with $D$ diagonal (details in Appendix~A of \cite{Balasubramanian:2024yxk,Toolkit}).   Here and in what follows repeated indices are summed.  Following \cite{Boruch:2024kvv,Balasubramanian:2024yxk}, we define the inverse of the Gram matrix via analytic continuation of $G^{n}_{ij}$ at positive integer $n$ 
\beq \label{eq:shellTr}
 G^{-1}_{ij} \equiv \lim_{n\to -1} G^{n}_{ij}.
\eeq
This procedure inverts nonzero eigenvalues while preserving zero eigenvalues.  The trace over $\mathcal{H}_{\kappa}$ is independent of $U$: 
\beq \label{eq:trace}
Tr_{\mathcal{H}_{\kappa}}(O) = G^{-1}_{ij}\langle j|O|i\rangle \, .
\eeq
Thus, even if $\{\ket{i}\}$ is overcomplete  we can use these states to compute operator traces. The resolution of the identity on $\mathcal{H}_{\kappa}$  becomes 
\beq \label{eq:id}
\mathds{1}_{\mathcal{H}_{\kappa}} = \sum_{\gamma}|v_\gamma\rangle\langle v_\gamma| = G^{-1}_{ij} |i\rangle\langle j|.
\eeq

Now suppose that we want to show that two quantities $\mathcal{A}$ and $\mathcal{B}$ are equal in the fine-grained theory, but  only have access to the gravitational path integral.  Since the latter computes coarse grained averages $\bar{\mathcal{A}}$ and $\bar{\mathcal{B}}$ as discussed above, it is not enough to show that $\bar{\mathcal{A}} - \bar{\mathcal{B}} = 0$.  We have to additionally show that $\overline{(\mathcal{A}-\mathcal{B})^2} = 0$, as this means that $\mathcal{A} - \mathcal{B} = 0$ for all the fine-grained configurations that are being averaged.   Explicit computation of the relevant path integrals is generally intractable, except in the saddlepoint approximation.  But, as pointed out in \cite{Toolkit}, if we only want to show that $\mathcal{A} = \mathcal{B} $ without explicitly calculating either quantity, then it suffices to show that each contribution to $\mathcal{A}$ is matched by an equal contribution to $\mathcal{B}$ and vice versa.   Following \cite{Toolkit} we can achieve this in three steps : ({\bf 1}) Construct an overcomplete basis of states by inserting operators into the Euclidean path integral;  ({\bf 2}) Take a limit in which the basis is infinitely overcomplete, leading to drastic simplifications of the topologies contributing to amplitudes;   ({\bf 3}) Demonstrate an action-preserving bijection between geometries contributing to the path integrals of interest after this topological simplification. This task can be simplified by insertions of the identity (\ref{eq:id}). The authors of \cite{Toolkit} used the two-sided shell states of 
\cite{Sasieta:2022ksu,Balasubramanian:2022gmo,Balasubramanian:2022lnw,Antonini:2023hdh} (details in Appendix~\ref{sec:2Sss} and  \cite{Balasubramanian:2024rek}) to demonstrate this procedure for universes with two boundaries, and showed these states indeed form an (over-) complete basis for this space.  We will extend this approach to universes with a single boundary.
\subsection{The factorisation puzzle}\label{sec:puzzles}

\begin{figure}[h]
  \centering
    \begin{subfigure}{0.6\linewidth}
     \centering
    \includegraphics[width=\linewidth]{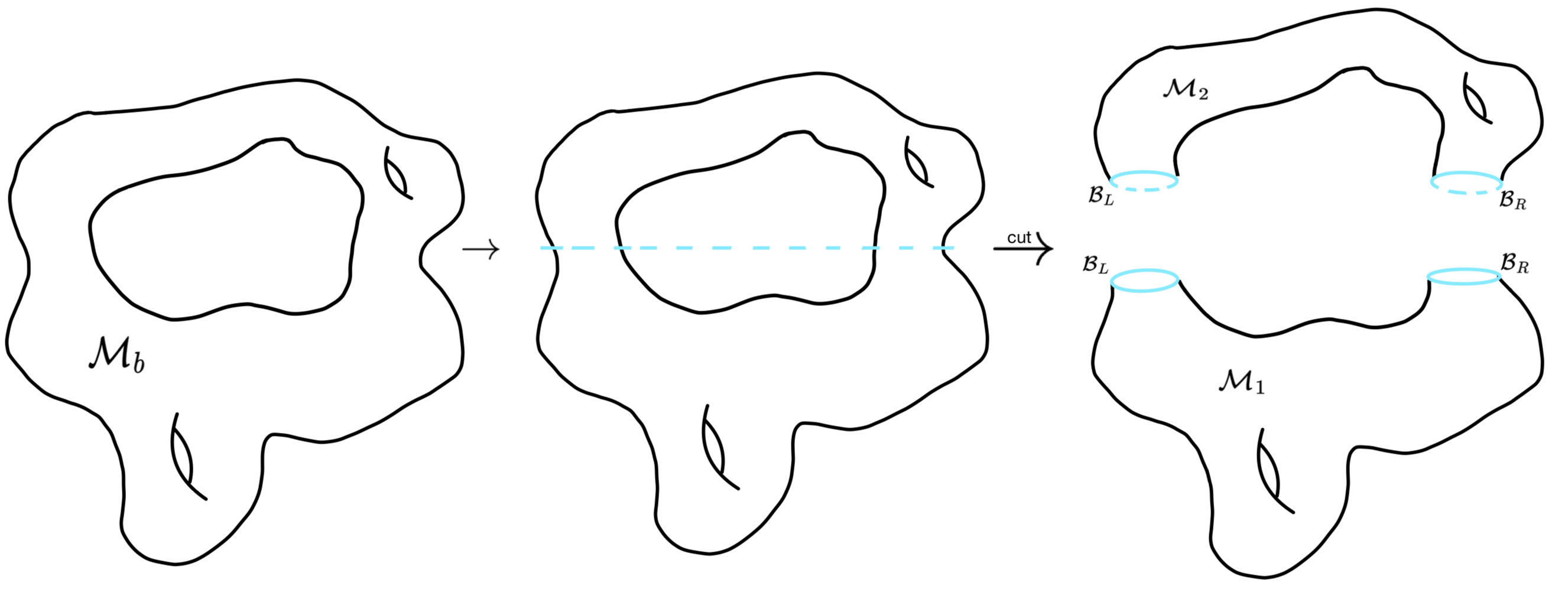}
    \caption{}
    \label{fig:singleBcut}
    \end{subfigure}
   \begin{subfigure}{0.7\linewidth}
   \centering
   \includegraphics[width=\linewidth]{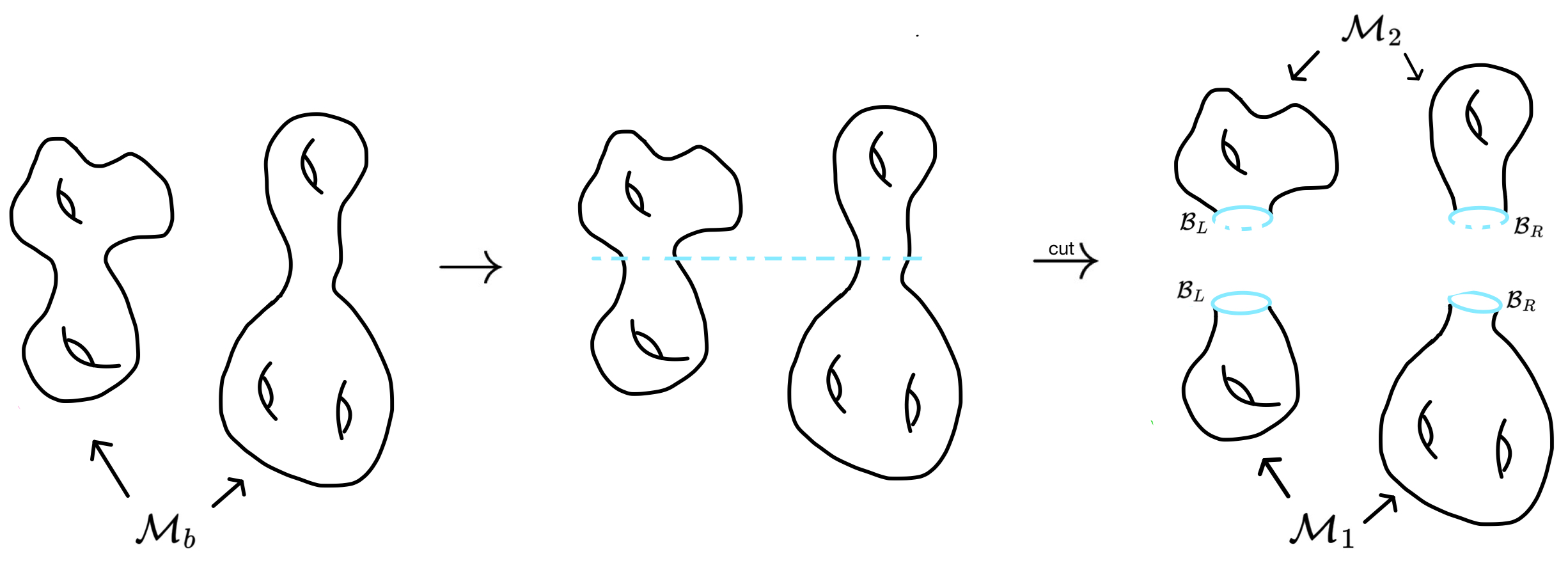}
   \caption{}
  \label{fig:tensorcut}
\end{subfigure}
\caption{Two distinct ways of preparing states in $\mathcal{H}_{L \cup R}$. ({\bf a})  Construction of an example state in $\mathcal{H}_{LR}$.  ({\bf b})  Construction of an example state in $\mathcal{H}_{\mathcal{B}_L}\otimes \mathcal{H}_{\mathcal{B}_R}$.   Note that here we are drawing the {\it boundary conditions} of the Euclidean geometry.  The bulk geometry between the boundary components $\mathcal{B}_{L,R}$ may be either be connected (like a spatial section of the eternal black hole) or disconnected (like a product of two thermal AdS geometries).
}
\end{figure}

We can build states for gravity with two boundaries by cutting open the boundary conditions for Euclidean gravity path integral (perhaps with operators inserted on the boundary) along a cut with two disconnected components $\mathcal{X}= \mathcal{B}_L \cup \mathcal{B}_R$. The latter analytically continue to two disconnected components of the Lorentzian 
 asymptotic boundary.  We will call the full two-sided Hilbert space $\mathcal{H}_{L \cup R}$. States in $\mathcal{H}_{L \cup R}$ come in two categories. First, we could cut open a \textit{single} connected closed boundary manifold, as in Fig.~\ref{fig:singleBcut}. Let us call the span of these states $\mathcal{H}_{LR}$.  Second, we could cut open two disconnected closed boundary manifolds,  as in Fig.~\ref{fig:tensorcut}. Since these disconnected components specify separate path integrals for single-boundary gravity theories, this construction prepares states in a tensor product Hilbert space  $\mathcal{H}_{\mathcal{B}_L}\otimes \mathcal{H}_{\mathcal{B}_R}$. For states in $\mathcal{H}_{LR}$ there can exist bulk geometries contributing to the path integral in which the two components of the cut are connected through the bulk. Moreover, as we are also summing over topologies in the path integral, this can also be the case for geometries contributing to overlaps in $\mathcal{H}_{\mathcal{B}_L}\otimes \mathcal{H}_{\mathcal{B}_R}$.
 
 Generically, we might expect that the full Hilbert space would be the direct sum $\mathcal{H}_{LR} \oplus \mathcal{H}_{\mathcal{B}_L}\otimes \mathcal{H}_{\mathcal{B}_R}$ or perhaps $\mathcal{H}_{L \cup R}=\{u+v\;|\;u\in \mathcal{H}_{LR},\, v\in \mathcal{H}_{\mathcal{B}_L}\otimes \mathcal{H}_{\mathcal{B}_R}\}$. But holography with AdS boundary conditions requires that the full Hilbert space with two disconnected Lorentzian boundaries has a tensor product structure because the dual CFT Hilbert space must. 

\begin{figure}[h]
 \begin{subfigure}[c]{0.32\linewidth}
     \centering
     \includegraphics[width=\linewidth]{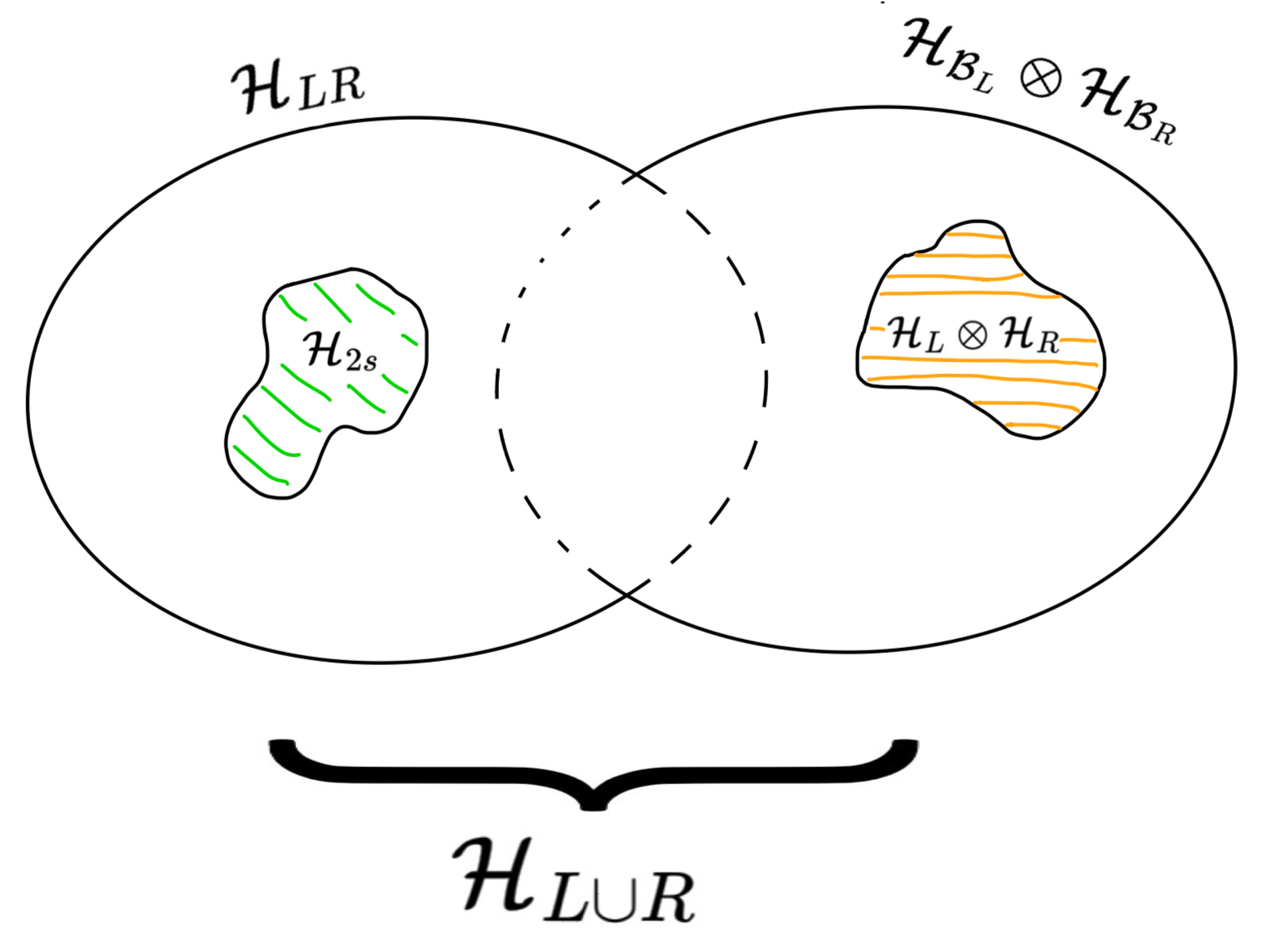}
     \caption{} 
     \label{fig:Hspaces}
 \end{subfigure}
 \hfill
\begin{subfigure}[c]{0.32\linewidth}
    \centering
    \includegraphics[width=\linewidth]{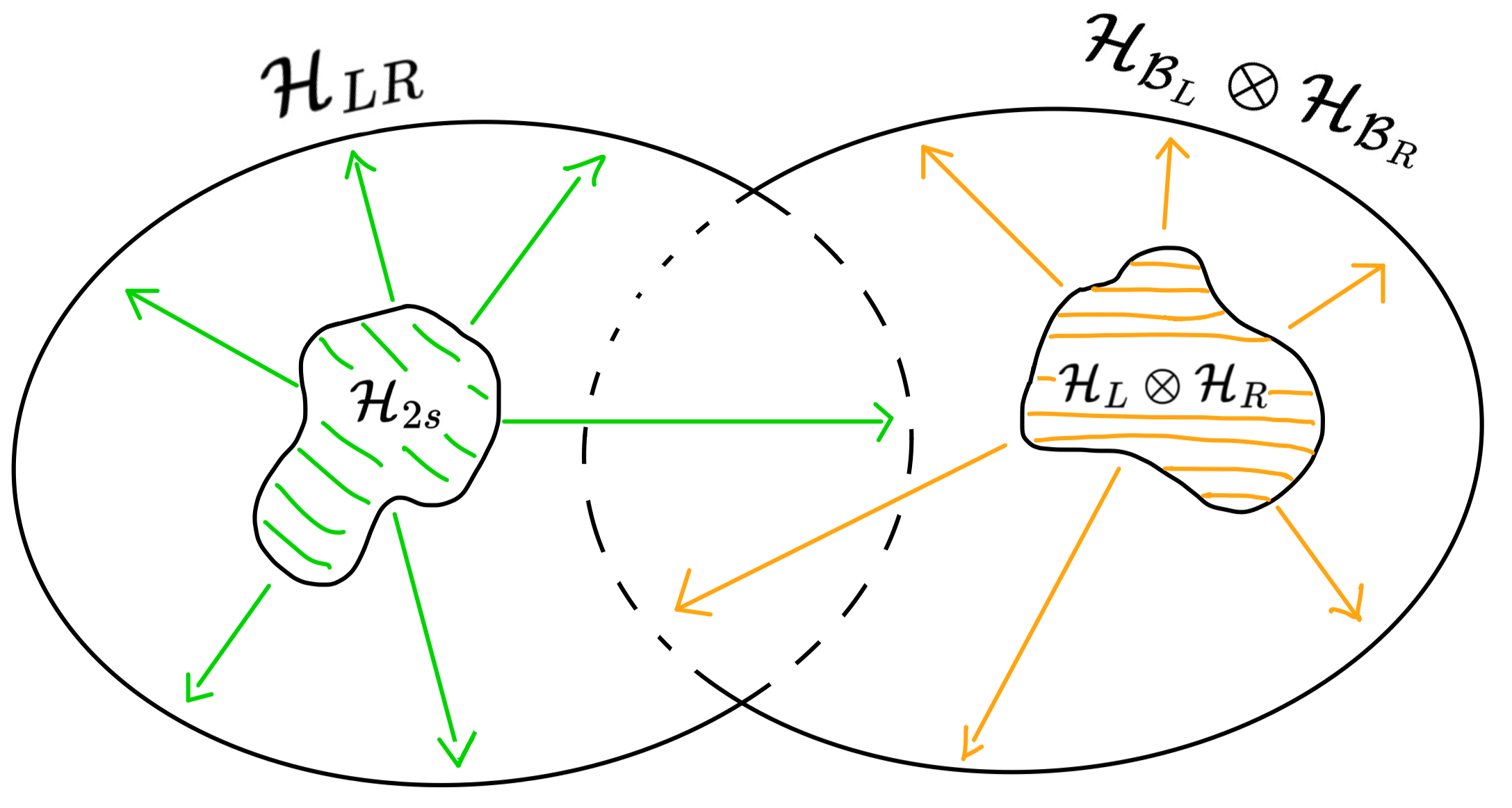}
    \caption{}
    \label{fig:Shellsspan}
    \end{subfigure}
    \hfill
    \begin{subfigure}[c]{0.32\linewidth}
    \centering
    \includegraphics[width=\linewidth]{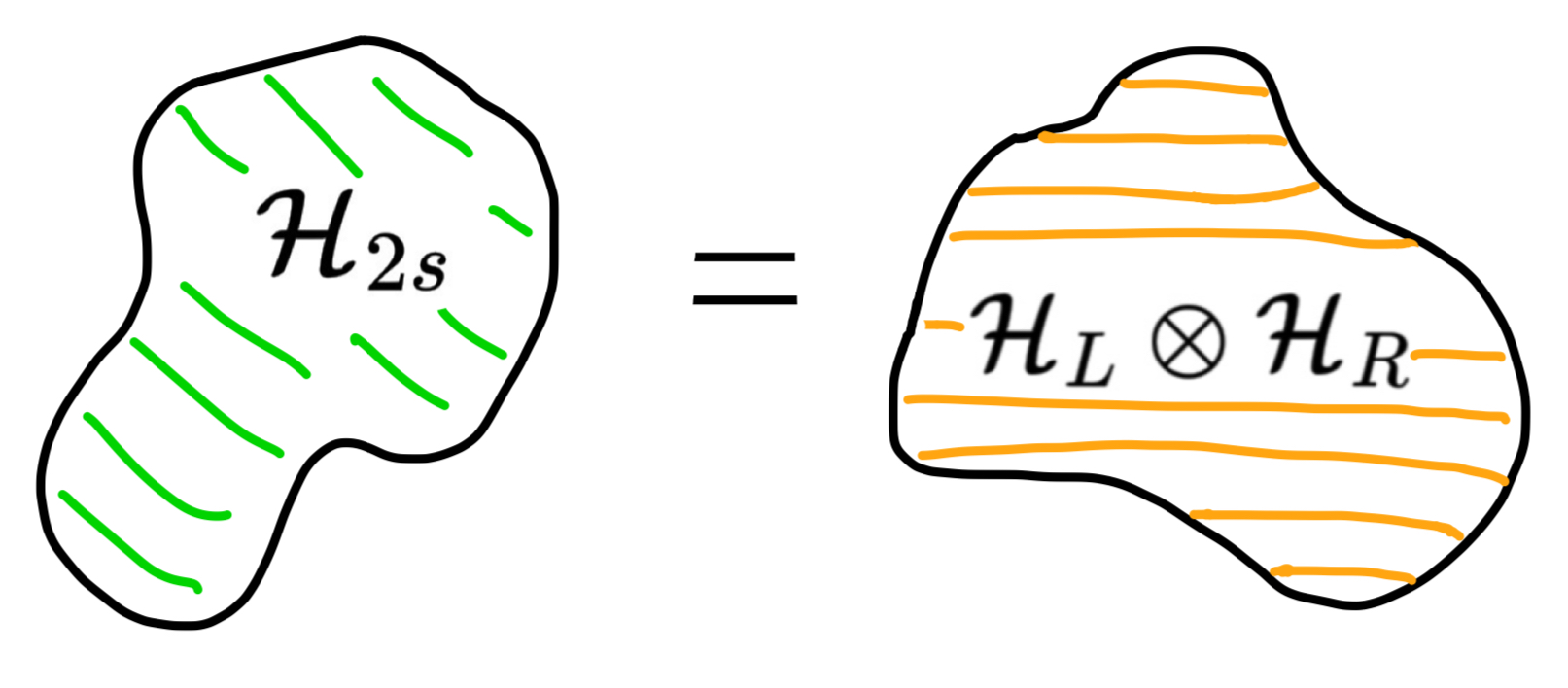}
    \caption{}
    \label{fig:facarg}
\end{subfigure}
\caption{Resolving the factorization problem. ({\bf a}) The two-sided Hilbert space  $\mathcal{H}_{LR}$, the tensor product Hilbert space $\mathcal{H}_{\mathcal{B}{_L}}\otimes\mathcal{H}_{\mathcal{B}{_R}}$, and the total two-sided Hilbert space $\mathcal{H}_{L\cup R}$.  The shaded regions labeled $\mathcal{H}_{2s}$ and $\mathcal{H}_{L} \otimes \mathcal{H}_R$ represent the spans of a set of two-sided ``shell states'' and a set of tensor products of single-sided ``shell states'' respectively, both of which are constructed from the Euclidean path integral.  See \cite{Sasieta:2022ksu,Balasubramanian:2022gmo,Balasubramanian:2022lnw,Antonini:2023hdh} for 2-boundary shell states and Sec.~\ref{sec:1s-span} for 1-boundary shell states.
({\bf b}) We will argue that the two- and single-sided shell states span $\mathcal{H}_{LR}$ and $\mathcal{H}_{\mathcal{B}_{L,R}}$ respectively when we take the number of shell states to infinity. The two-sided shell state span was shown in \cite{Toolkit}. 
({\bf c}) We will then show that the two-sided shell states spanning $\mathcal{H}_{LR}$  also span $\mathcal{H}_{\mathcal{B}_{L}} \otimes \mathcal{H}_{\mathcal{B}{_R}}$ and vice versa. 
\label{fig:factorizationproblem}}
\end{figure}

 This leads to a Hilbert space factorization puzzle: {\it show
the fine-grained, all-order equality in Fig~\ref{fig:factorizationproblem}:
\beq \label{eq:facprob}
\mathcal{H}_{L \cup R}= \mathcal{H}_{\mathcal{B}_L}\otimes \mathcal{H}_{\mathcal{B}_R}=\mathcal{H}_{LR}. 
\eeq
}

The factorization puzzle has sometimes been posed in Lorentzian signature by asking how we can split the effective field theory Hilbert space in, say, the two-boundary AdS black hole background into factors as expected from AdS/CFT for the complete Hilbert space.  However, this formulation is misleading: the code subspace around a single entangled state in a large factorized Hilbert space need not itself factorize. Likewise the effective field theory Hilbert space in the two-sided black hole background need not factorize. Rather, AdS/CFT says that (\ref{eq:facprob}) should be true in the full, non-perturbative theory, and this is what we seek to show.

There have been some earlier steps towards showing (\ref{eq:facprob}). In two dimensional JT gravity, \cite{Boruch:2024kvv} showed that traces in the two-sided Hilbert space are a product of factors to all orders in $e^{-1/ \mathrm{G}_N}$, and \cite{Balasubramanian:2024yxk} extended the argument to any dimension in the leading saddle approximation. Likewise, \cite{Toolkit} argued that the fine-grained thermal partition function of  $\mathcal{H}_{LR}$  factorises exactly:
\beq \label{eq:trfactorisation}
Tr_{\mathcal{H}_{LR}}(e^{-H_{L}\beta_1}e^{-H_{R}\beta_2}) = Z(\beta_1)\times Z(\beta_2) \, 
\eeq
for asymptotically flat or AdS gravity in any dimension. 
As $H_L,H_R$ commute, it follows from (\ref{eq:trfactorisation}) that the two-sided Hilbert space $\mathcal{H}_{LR}$ has a tensor product representation  (see Sec.~3 in \cite{Balasubramanian:2024yxk}).  More generally, \cite{Colafranceschi:2023moh} used
operator algebras on the Hilbert spaces formed by cutting open a generalized notion of a gravitational path integral, and showed that $\mathcal{H}_{LR}$ must be a product of factors.\footnote{More generally a sum over products of factors $\mathcal{H}_{LR}=\bigoplus_{\mu}\mathcal{H}^{\mu,L}_{LR}\otimes\mathcal{H}^{\mu,R}_{LR}$.}
But to resolve the factorization problem we have to additionally show that the two factors are actually the non-perturbative single-boundary Hilbert spaces, as we will do below.  This does not follow simply from algebraic considerations arising from the axioms in \cite{Colafranceschi:2023moh}, and is therefore a non-trivial constraint on the gravity theory.

There is also another sort of factorization puzzle. Suppose we use the  Euclidean gravity path integral to calculate some amplitude or correlation $\mathcal{A}$. Then the path integral for $\mathcal{A}\mathcal{A}$ may not equal $(\mathcal{A})^2$ because of wormhole contributions.  This failure of factorization is usually attributed to  coarse-graining of the ultraviolet theory implicit in doing a path integral just over metrics and geometries.  Despite this coarse graining there are techniques for using the gravitational path integral to show fine-grained equalities \cite{Toolkit,Boruch:2024kvv}. We will use these techniques to show that it is precisely the wormholes that are responsible for  non-factorization of coarse-grained amplitudes that lead reveal the factorization of the fine-grained two-boundary Hilbert space. A similar observation was made in \cite{Boruch:2024kvv}.

\section{Single-sided shell states} \label{sec:1s-span}
To derive the factorisation (\ref{eq:facprob}) we require a complete basis for the single-boundary Hilbert space $\mathcal{H}_{\mathcal{B}}$. We start by arguing that a set of  $\kappa \to \infty$ single-boundary shell states (which we will refer to as $\mathcal{H}_{L},\mathcal{H}_{R}$ for $L,R$ copies of this set) span the full single-sided Hilbert space $\mathcal{H}_{L,R} =\mathcal{H}_{\mathcal{B}_{L,R}} $ and therefore $\mathcal{H}_{\mathcal{B}_L}\otimes \mathcal{H}_{\mathcal{B}_R}=\mathcal{H}_{L}\otimes\mathcal{H}_{R}$, see Fig.~\ref{fig:Shellsspan}. In particular, we will construct a basis for the single-sided Hilbert space  
consisting of states
obtained by a cutting open boundary manifolds with topology $\mathbb{R}\times \mathbb{S}^{d-1}$, so that the cut $\mathcal{B}$ has topology $\mathbb{S}^{d-1}$. In the remainder of this paper we simply denote this Hilbert space as $\mathcal{H}_\mathcal{B}$. If we consider two copies of this space we will label them $\mathcal{H}_{\mathcal{B}_{L}}$ and $\mathcal{H}_{\mathcal{B}_{R}}$.

\subsection{Construction}\label{sec:construction}
\begin{figure}[h]
    \centering
    \begin{subfigure}{0.45\linewidth}
    \centering
    \includegraphics[width=0.3\linewidth]{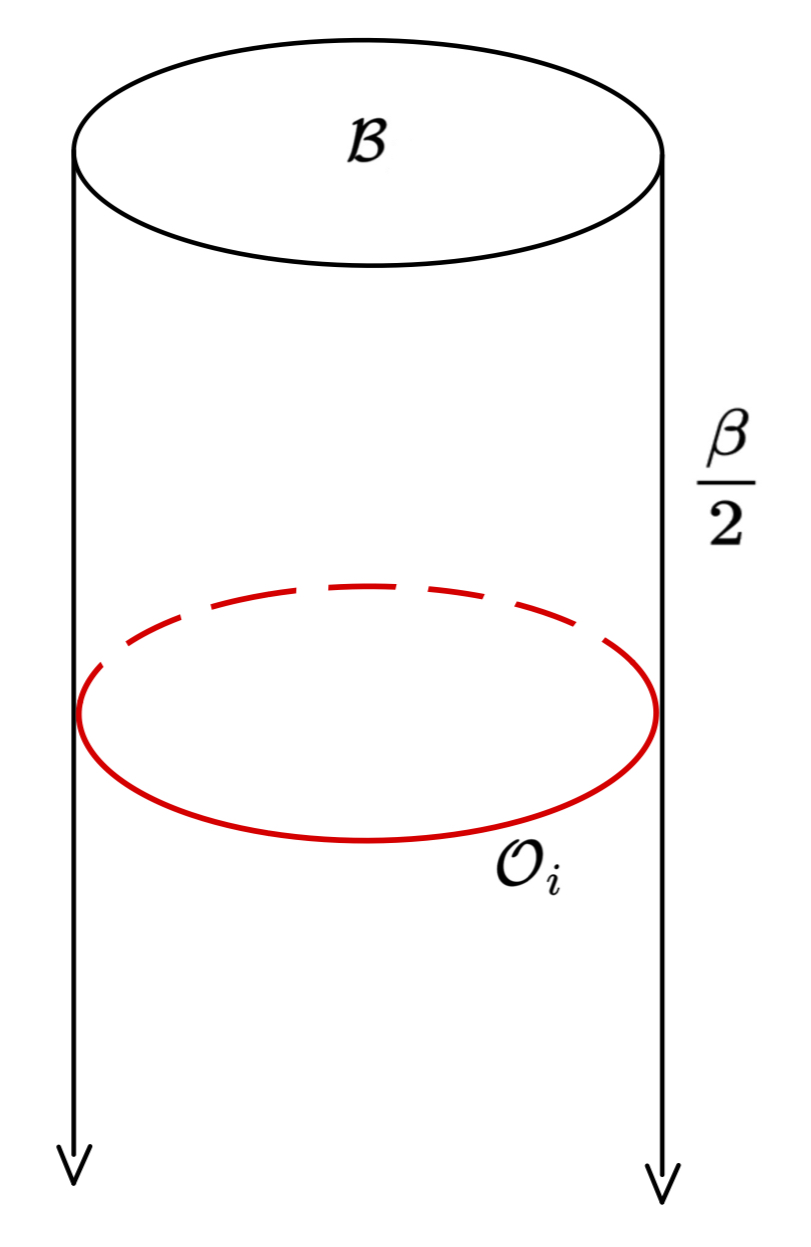}
    \caption{}
    \end{subfigure}
     \centering
    \begin{subfigure}{0.45\linewidth}
    \centering
    \includegraphics[width=0.3\linewidth]{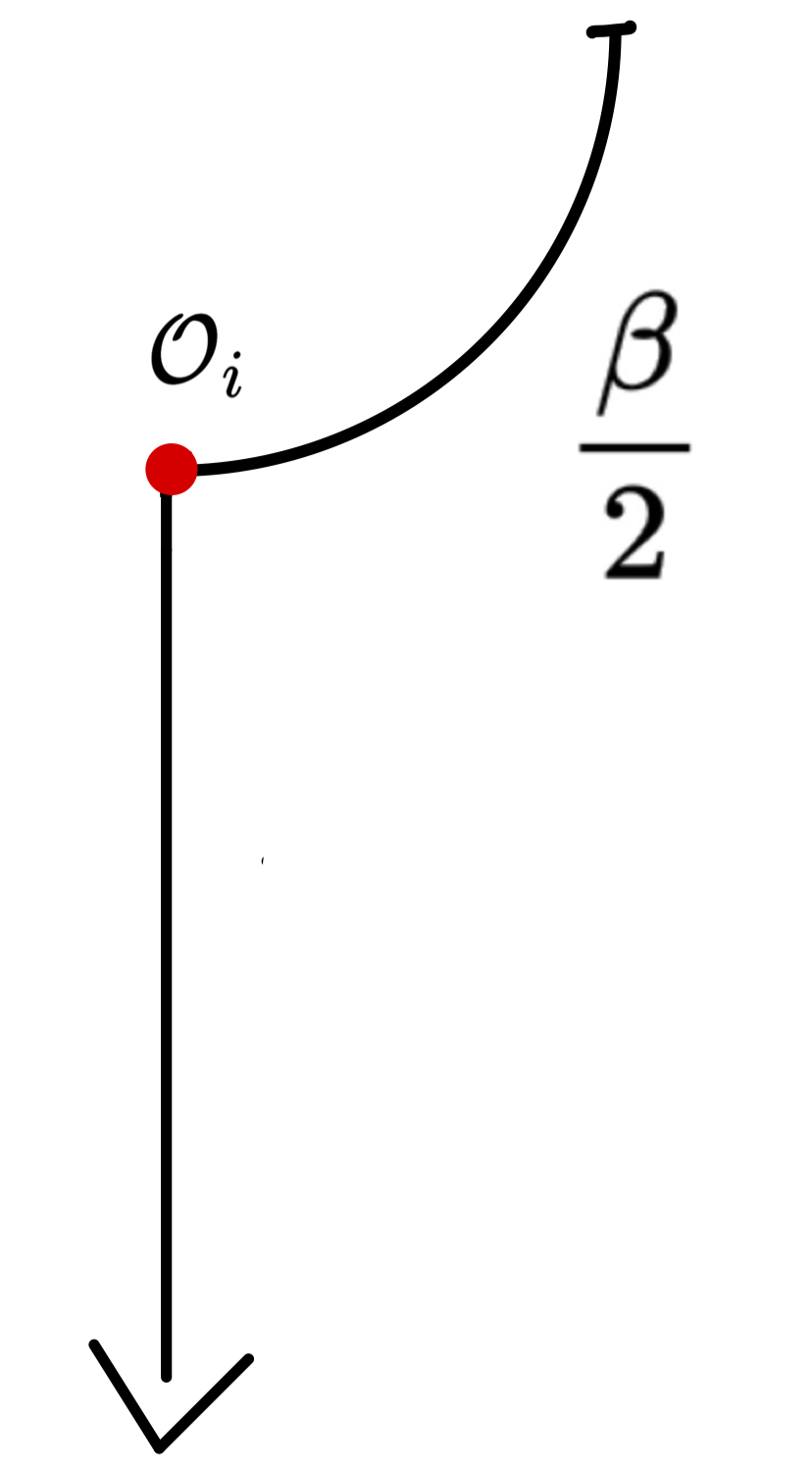}
    \caption{}
    \end{subfigure}

    \caption{Cut-open path integral boundary condition defining the single-sided shell states.  ({\bf a}) Euclidean boundary with topology $\mathbb{R}^{<0}\times\mathbb{S}^{d-1}$ for preparation of the shell states. In AdS/CFT we can also perform the path integral in the boundary CFT with insertion of a $\mathbb{S}^{d-1}$ symmetric operator dual to the shell.  The shell operator $\mathcal{O}_{i}$ is pictured in red and $\beta_{}/2$ is the Euclidean ``preparation time''. ({\bf b}) Euclidean boundary with the $\mathbb{S}^{d-1}$ suppressed. We adopt this convention for the rest of the paper, and sometimes depict this boundary with a curve or a kink to clarify diagrams. }

     \label{fig:sinlge_def}
\end{figure}

Our candidate basis states for $\mathcal{H}_{\mathcal{B}}$ are the single-sided thin matter shell states discussed in \cite{Chandra:2022fwi}.  To define these states we fix a half-boundary with topology $\mathbb{R}^{<0}\times \mathbb{S}^{d-1}$ (where $\mathbb{R}^{<0}$ is the half line) and 
insert an $\mathbb{S}^{d-1}$ symmetric heavy ($m \sim \mathit{O}(1/G_{N})$) dust shell operator  $\mathcal{O}_{S}$ separated by a boundary length $\frac{\beta}{2}$ from the cut $\mathcal{B}$.  The cut-open path integral boundary condition defining such states thus has a  Euclidean boundary with time  $\tau$ extending from $-\infty$ to  $\tau = 0$,  followed by insertion of $\mathcal{O}_S$  at $\tau =0$, and finally another Euclidean time section up to $ \tau =\beta/2$. This boundary condition is depicted in Fig.~\ref{fig:sinlge_def} where the arrows denote a half infinite Euclidean time segment.  We can obtain an infinite family of shell states in this way by varying the mass $m_i$ of the shell operators $\mathcal{O}_{i}$ to produce states  $|i\rangle$. We will see below that the Lorentzian continuation of states prepared in this way will either be a black hole with a shell behind the horizon or a thermal geometry along with an additional closed universe.  In either case the preparation time parameter $\beta$ is related to temperature of the noncompact part of the Lorentzian geometry.

Our construction will apply to both asymptotically flat and AdS universes. But for concreteness we will focus on  the asymptotically AdS case, where there is a dual CFT defined on the boundary with geometry $\mathbb{S}^{d-1} \times \mathbb{R}$. Although our analysis is formulated entirely in the bulk and remains agnostic about any specific boundary dual, it is instructive to briefly review how these states are constructed on the CFT side.   From the CFT  perspective the shell states are defined by preparing the vacuum state, acting with an operator $\mathcal{O}_S$ and subsequently evolving with the CFT Hamiltonian $H_{CFT}$ for a Euclidean time $\frac{\beta}{2}$ :
\beq
|\Psi_S \rangle = e^{-\frac{\beta}{2}H_{CFT}}\mathcal{O}_S |0\rangle.
\eeq
The shell states are created by operators $\mathcal{O}_{S}$ that are holographically dual to heavy spherically-symmetric dust shells inserted at the asymptotic boundary of Eucldiean AdS$_{d+1}$.  Such operators are constructed in the CFT by taking products of $n$ scalar operators $\mathcal{O}_{\Delta}$ of dimension $\Delta \sim \mathit{O}(1)$. To make sufficiently heavy states with $m \sim \mathit{O}(1/G_{N})$ we will need  to insert $n \sim l^{d-1}/G$ operators  at spherically symmetric points on the $\mathbb{S}^{d-1}$, where $l$ is the AdS radius.\footnote{See \cite{Kudler-Flam:2025cki,Liu:2025cml} for recent discussion regarding the existence of the large N limit of these states in the CFT dual.}

Many path integrals of interest in this paper will be subject to boundary conditions given by the Euclidean half line $[-\infty,0]$ followed by a series of operator insertions and Euclidean time evolutions up to some time $T_{E}$ and finally another infinite segment half line $[T_{E},\infty]$. Formally this should be though of the limit of time evolutions $[-\alpha,0]$ and $[T_{E},\alpha + T_{E}]$ where $\alpha \to \infty$. We will refer to the time interval $T_{E}$ as the ``length of the strip'' and denote such an asymptotic  boundary condition as $S(T_{E})$. For example, the difference in length of the asymptotic boundaries $S(T_{E})$ and ${S(0)} $ is given by  $ \lim_{\alpha \to \infty}2\alpha +T_{E} - 2\alpha =T_{E} $. 

\begin{figure}
    \centering
    \includegraphics[width=0.5\linewidth]{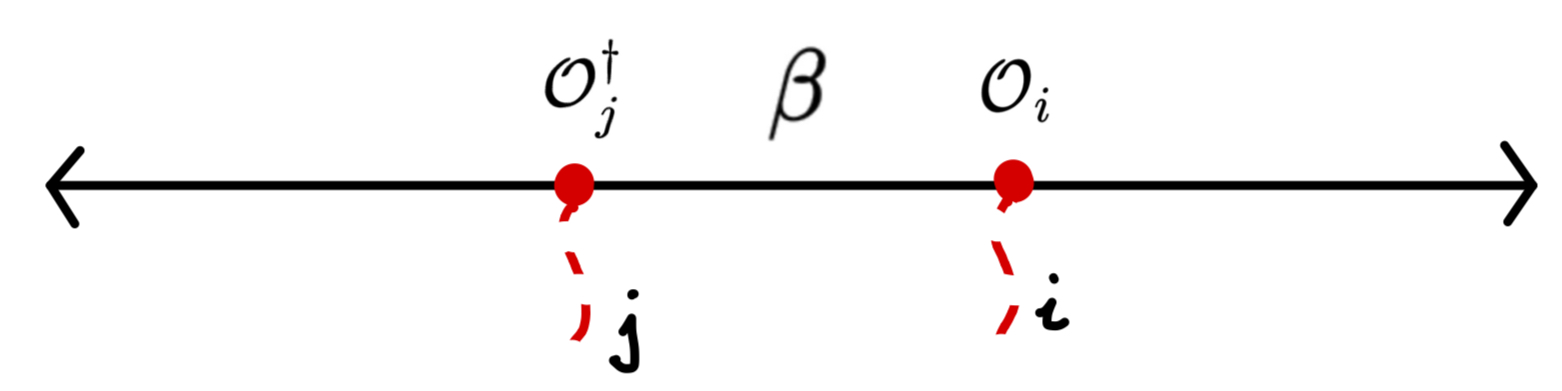}
    \caption{Shell-strip asymptotic boundary condition for the overlap $\braket{j|i}$ consisting of the line $\lim_{\alpha \to \infty} [-\alpha, \beta + \alpha]$ on which  $\mathcal{O}_{i}$ and  $\mathcal{O}^{\dagger}_{j}$ are inserted at $\tau =0 $ and  $\tau=\beta$ respectively.}
    \label{fig:1s_shellbdry}
\end{figure}

We will want to compute the Gram matrix $G_{ij}\equiv \langle i|j\rangle$ of overlaps between shell states.  The ket in this overlap is constructed by the path integral  in Fig.~\ref{fig:sinlge_def}a, while the bra is defined by a similar path integral on the time interval $0<\tau<\infty$. The boundary condition defining $G_{ij}$ consists of a boundary time interval $[-\infty,0]$ followed $\mathcal{O}_i$ insertion, followed by $\beta_R$ boundary time evolution, followed by an $\mathcal{O}^{\dagger}_j$ insertion and another infinite segment half line $[\beta,\infty]$ (Fig.~\ref{fig:1s_shellbdry}). Using the above terminology this boundary condition is therefore a strip of length $\beta$ separating the  $\mathcal{O}_{j}$ and  $\mathcal{O}^{\dagger}_{i}$ operator insertions, and we will refer to such boundary conditions as \textit{shell strip} boundaries.

The coarse-grained overlap $\overline{\langle i|j \rangle}$ is evaluated by the path integral with the shell strip boundary condition (Fig.~\ref{fig:singe_norm}).  Following~\cite{Balasubramanian:2022gmo},  shell states with different mass can be made orthogonal by taking their inertial mass differences to be arbitrarily large.  This is because we expect that it will take  $|m_{i}-m_{j}|$ bulk interactions in Planck units for such shells to annihilate, leading to $\overline{\langle i|j \rangle} =\delta_{ij} Z_{1}$.  Later we will show that  wormholes in the path integral that are stabilized by the shell matter can modify the overlap to $\overline{|\langle i|j \rangle|^2}=\overline{\langle i|j \rangle\langle j|i \rangle} = Z_{2} + \delta_{ij}Z_{1}^2$, where $Z_{2}$ is contribution from  the wormhole saddlepoint  (Fig.~\ref{fig:Single_wormhole}). These non-perturbative corrections are of order $e^{-1/G_N}$. We therefore pick shell labels $\{m_i\}$ with spacings $|m_i-m_j|$ large enough that (i) any perturbative overlap suppression is subleading to $e^{-1/G_N}$, and (ii) each shell’s local energy density $\rho$, curvature invariants $R,\;R_{\mu\nu}R^{\mu\nu}$, and typical quantum wavelengths satisfy $\rho^{1/4},\;R^{1/2},\;\lambda^{-1}\ll M_{\rm Pl}$. With these constraints the semiclassical saddle analysis and the resulting structure of the Gram matrix remain within the validity of the gravitational EFT.

\subsection{Norm and wormhole saddles} \label{sec:1s_WHs}
\begin{figure}
    \centering
    \includegraphics[width=0.8\linewidth]{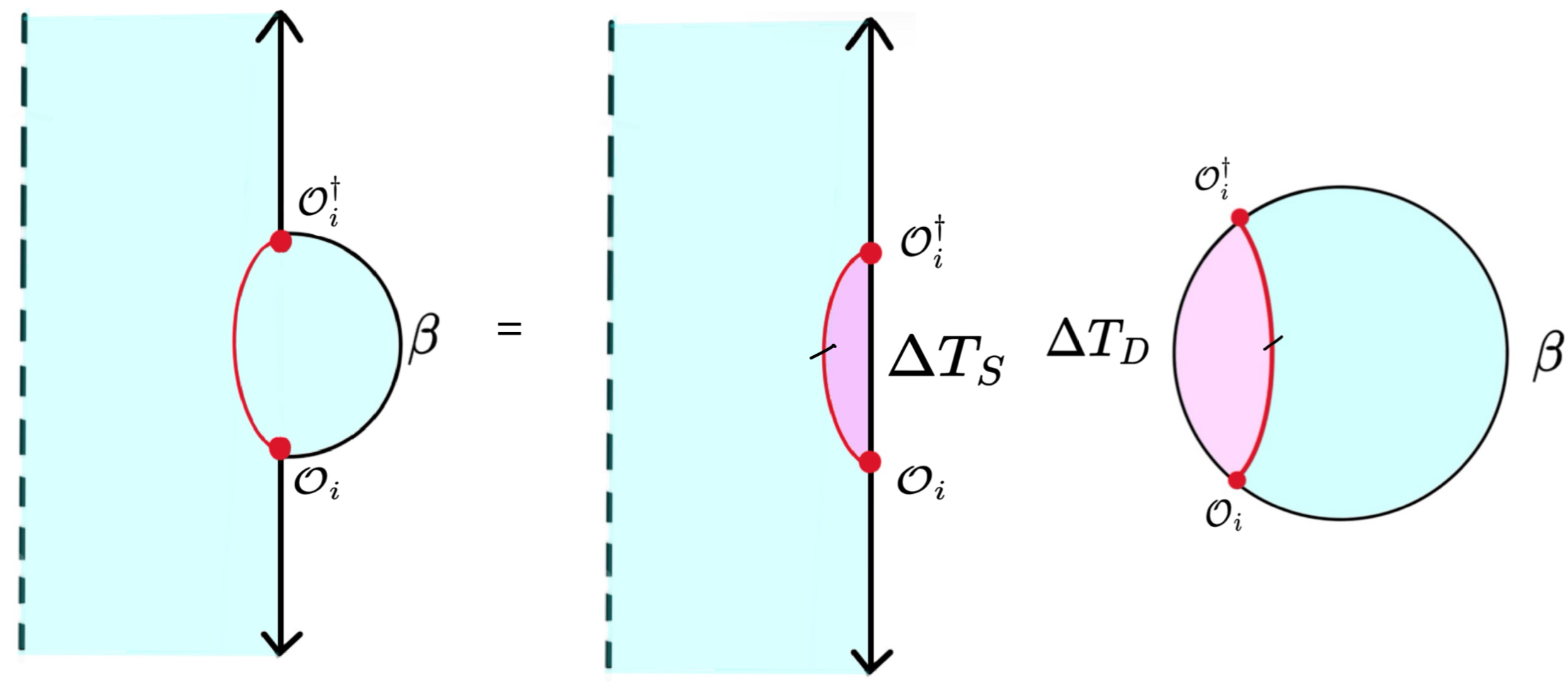}
    \caption{The saddle for the norm $\overline{\braket{i|i}}$ is constructed by considering the shell propagating on a disk and strip separately for some propagation times $\Delta T_{S,D}$ and then gluing them together along the $i$-shell worldvolume by discarding the shell homology region (purple). The junction conditions dynamically determine $\Delta T_{S,D}$ to yield an on shell glued geometry. We have suppressed the angular directions in these diagrams, and represent the radius-time plane with the dashed line representing the origin.}
    \label{fig:singe_norm}
\end{figure}

The path integral for the overlap $\overline{\langle i|j \rangle} =\delta_{ij} Z_{1}$ can be evaluated in the saddlepoint approximation. Saddles satisfying the shell strip boundary condition are obtained by filling in one side of the shell-worldvolume with a saddle geometry of the Z partition function (defined in Sec.~\ref{sec:review}) and the other with the Euclidean vacuum geometry with non-compact Euclidean time (which we refer to as the ``strip") respectively. This procedure can be done for any Z partition saddle --  we refer to any such saddle as a ``disk''. These two geometries are glued together along the trajectory of the shell using the Israel junction conditions \cite{Israel:1966rt}, see Fig.~\ref{fig:singe_norm}.  In particular, we consider a Euclidean AdS strip with a shell operator 
$\mathcal{O}_{i}$ on the boundary at $\tau=0$ and the conjugate operator $\mathcal{O}^{\dagger}_{i}$ inserted at time $\tau=\Delta T_{S}$. Similarly consider a Euclidean AdS disk with boundary insertions $\mathcal{O}_{i}$ and $\mathcal{O}^{\dagger}_{i}$ separated by a time 
$\Delta T_{D}$ on a disk of boundary length $\beta +\Delta T_{D}$. The saddle geometry is then constructed by discarding the shell homology regions (purple regions in Fig.~\ref{fig:singe_norm}) from the strip and disk and gluing together the shell world-volumes across the disk and strip via the junction conditions, which dynamically determine $\Delta T_{S}$ and $\Delta T_{D}$ such that the resulting geometry satisfies the equations of motion. 

\paragraph{Large shell mass limit} 
The junction conditions for the single-sided shell states can be obtained from those for the two-sided shell states \cite{Balasubramanian:2022gmo,Toolkit}  by taking one of the two-sided preparation temperatures, say $\beta_L$, to infinity. The two sided-shell states are reviewed in Appendix~\ref{sec:2Sss} and, as discussed there, in the heavy shell mass ($m_i \to \infty$) limit the propagation times $\Delta T_{1,2}$ on the $L$ and $R$ disks go to zero, and the shell homology regions pinch off so that each shell contributes a factor $Z_{m_i}\sim e^{-2(d-1)log(\mathrm{G_N} m_i)}$ that is a universal function of its mass. In \cite{Balasubramanian:2022gmo,Balasubramanian:2022lnw,Antonini:2023hdh}   this behavior was shown to be independent of which saddle of the Z-partition function is glued along the shell. Furthermore, as this result is independent of $\beta_L, \beta_R$ and the junction conditions are local, by considering $\beta_L \to \infty$ these results  carry over to the single-sided shell state construction.\footnote{The solution to the junction conditions for the single-sided shell states can be found in the appendix of \cite{Chandra:2022fwi}, from which this expected behavior is confirmed.}

Hence, in the $m_i \to \infty$ limit the bulk turning point of the shell approaches the asymptotic boundary and $\Delta 
T_{S},\Delta T_{D} \to 0$. Therefore the shell homology regions pinch off and the remainder grows to cover the entire strip of  length\footnote{This is ``renormalized'' strip length as defined in Sec.~\ref{sec:construction}.} zero and disk of length $\beta$, and the shell contribution a universal factor $Z_{m_i}$. Therefore the
overlap is simply given by:
\beq \label{eq:overlap}
\overline{\langle i|j \rangle} = \delta_{ij} Z_{m_i} \times 
\overline{Z(\beta)}\times \overline{S}(0)
\eeq
where $\overline{Z(\beta)}$ the Z partition function and $\overline{S}(0)$ the zero length strip path integral.\footnote{The regularization of the strip will drop out of all calculations in this paper by normalising the shell states, so we will not specify it.} As explained in detail in \cite{Toolkit} and Appendix~\ref{appendixb}, the form of (\ref{eq:overlap}) is expected to remain valid in the heavy shell mass limit even after perturbative bulk corrections are included. In this regime, corrections to the LHS simply map onto corresponding corrections to the path integrals $\overline{Z(\beta)}$ and $\overline{S}(0) $ appearing on the RHS. This expectation was verified explicitly in \cite{He:2025neu} by embedding the shell construction into a doubly holographic model. The same reasoning applies to all other factorized expressions of the form (\ref{eq:overlap}) that appear in the remainder of this work.

\begin{figure}[h]
\begin{subfigure}[c]{0.48\linewidth}
        \centering
        \includegraphics[width=\linewidth]{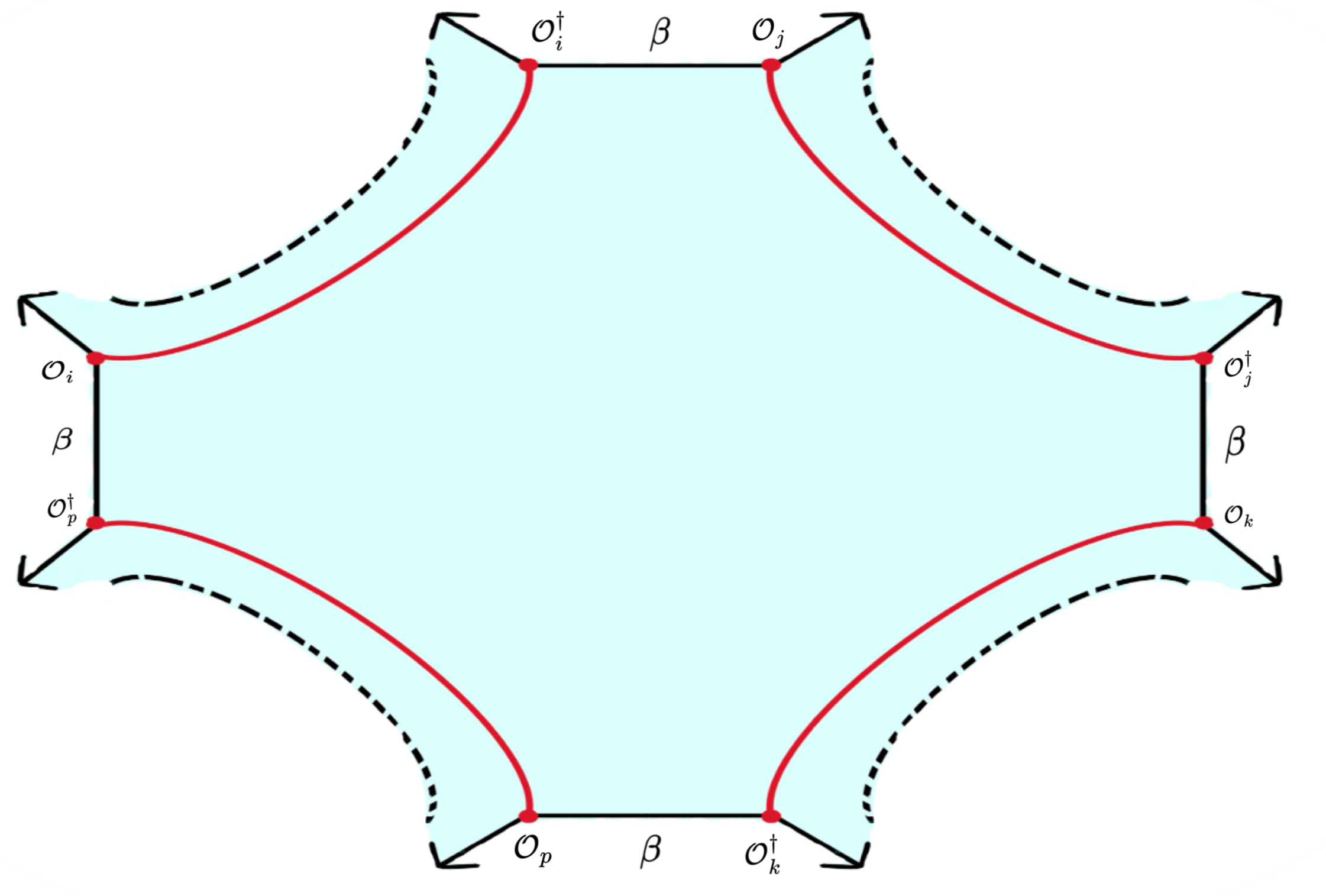}
        \caption{}
        \label{fig:Single_wormhole}
\end{subfigure}
\hfill
\begin{subfigure}[c]{0.48\linewidth}
    \centering
    \includegraphics[width=\linewidth]{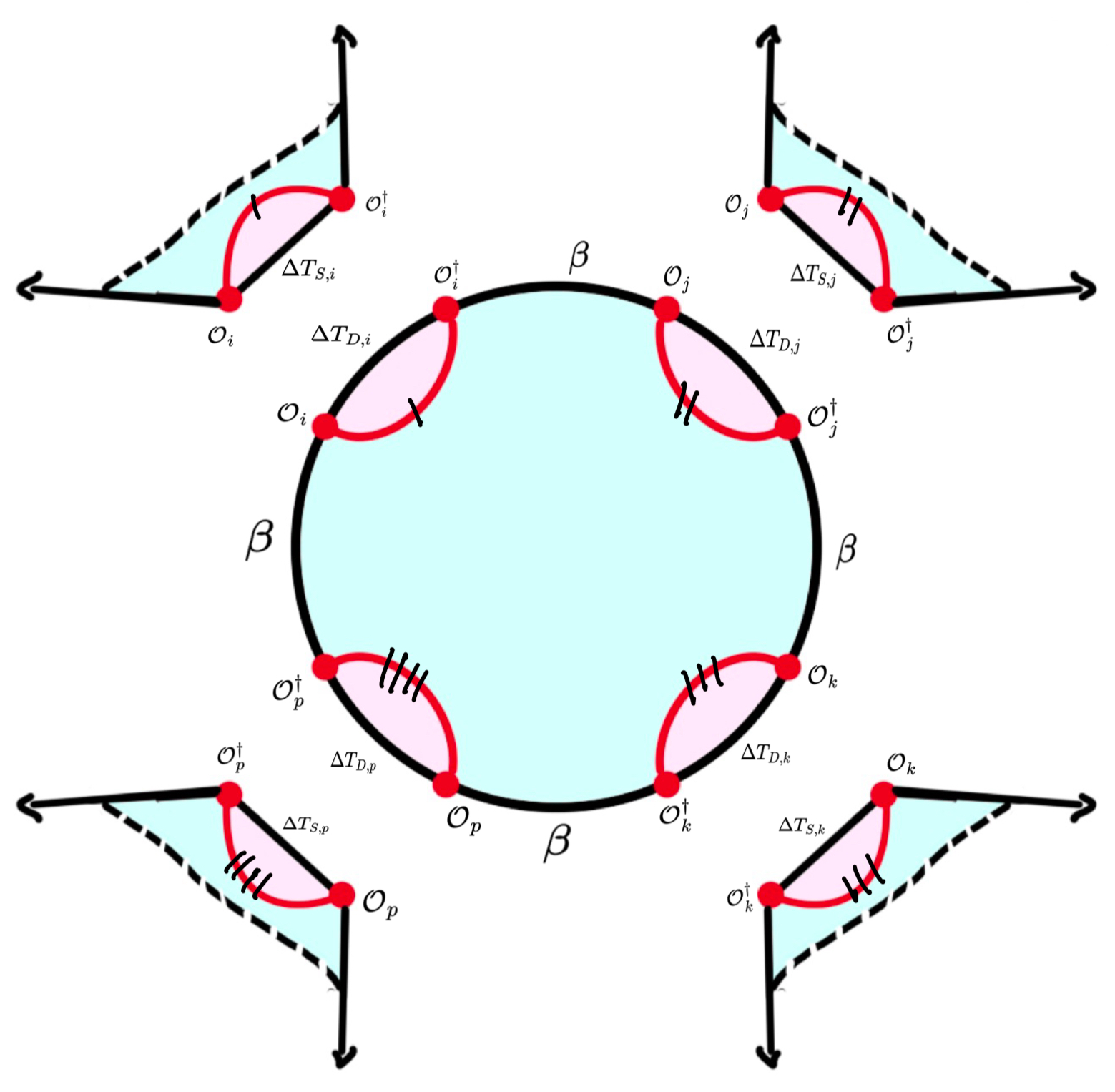}
    \caption{}
    \label{fig:planarWHconstruct}
\end{subfigure}
\caption{The fully connected single-sided shell wormholes saddles for $\overline{(G^n)_{ii}}$ are constructed by gluing $n$ strips into the disk along the shell worldvolumes, depicted here for $n=4$. }
\end{figure}

\paragraph{Wormholes} \label{sec:1sWH} 
Generalizing this construction, we construct wormhole saddlepoints contributing to the amplitude $\overline{(G^n)_{ii}}=\overline{\braket{i|j_1}\braket{j_1|j_2} \cdots \braket{j_n|i}}$  (no sum on $i,j_k$) by connecting $n$ asymptotic boundaries (arising from each factor in $(G^n)_{ii}$) with a single multiboundary wormhole. The $k$-th shell can only propagate through the bulk of the wormhole from $\mathcal{O}_{k}$ to $\mathcal{O}^{\dagger}_{k}$ on different asymptotic boundaries, and hence the wormhole connectivity reflects the index structure of $\overline{(G^n)_{ii}}$ (Fig.~\ref{fig:Single_wormhole}).  The wormhole saddle is constructed analogously to the norm by gluing together $n$ copies of the Euclidean strip into a disk along the shell word-volumes  (Fig.~\ref{fig:planarWHconstruct}). In particular, the saddle geometries satisfying this boundary condition can be constructed for a given $n$ by starting with a disk with $n$ shells inserted on the asymptotic boundary. The shell operator $\mathcal{O}_{S_i}$ is separated from $\mathcal{O}^{\dagger}_{S_i}$ time a boundary time $\Delta T_{D,i}$, and $\mathcal{O}^{\dagger}_{S_i}$ from $\mathcal{O}_{S_{i+1}}$ by a boundary time  $\beta$ in a circular pattern. For each shell, a separate shell-strip is also constructed by placing a shell $\mathcal{O}_{S_i}$ on the strip boundary separated from $\mathcal{O}^{\dagger}_{S_i}$ by $\Delta T_{D,i}$ boundary time. The saddle satisfying the boundary consideration is then constructed by cutting off the shell homology regions on the disk and strips and gluing each strip to the disk by identifying corresponding shell worldvolumes using the junction conditions.

In the large shell-mass limit the shell propagation times in the strips and disks limit to zero and hence the saddle is given by the product of the $n$ strips, the  disk of length $n\beta$ and the  universal   shell contributions $\overline{G^{n}_{ii}}= \overline{Z}(n\beta) \times \left(\overline{S}(0)\right)^n \times \prod_{i=1}^n Z_{m_i}$. After normalising the shell states to have unit norm ($\ket{k}\rightarrow\ket{k}/\sqrt{Z_1}$), the latter two factors cancel out altogether and we are left with
\beq \label{eq:1sshellwh}
\overline{G^{n}_{ii}}= \frac{\overline{Z}(n\beta)}{\overline{Z}(\beta)^n} \, .
\eeq

\subsection{Saddle geometry}
We can associate a geometry to a shell state by considering the leading gravitational saddlepoint contributing to 
the norm $\overline{\langle i|i\rangle}$. In the large shell mass limit resulting in (\ref{eq:overlap}), the leading saddle for the overlap is thus given by the leading saddle for the Z-partition function, $\overline{Z(\beta)}$, which is either thermal Euclidean AdS if $\beta > \beta _{HP}$  or the large Euclidean AdS Black Hole if $\beta < \beta _{HP}$, where $\beta_{HP}$ denotes the inverse temperature for the Hawking-Page transition \cite{Hawking:1982dh}. The time reflection symmetric bulk slice of these saddles can  be analytically continued to Lorentzian signature to associate a Lorentzian geometry to the state. There are thus two classes of single sided shell states:

\begin{enumerate}
\item $\beta < \beta _{HP}$  : The leading saddle for the disk portion will be the Euclidean AdS black hole. In this case the Lorentzian geometry corresponds to an AdS-Schwarschild black hole with the shell propagating in the interior which is capped off by empty AdS behind the shell. We will refer to these states as \textit{type A} single-sided shell states.

\item $\beta >\beta _{HP}$  : The leading saddle for the disk portion will be thermal AdS. In this case the Lorentzian geometry corresponds to thermal AdS at temperature $\beta$ in addition to a disconnected AdS closed ``Big Crunch" universe in which the shell propagates. We will refer to these states as \textit{type B} single-sided shell states.

\begin{figure}[h]
  \begin{subfigure}[c]{0.45\linewidth}
    \centering
    \includegraphics[width=0.7\linewidth]{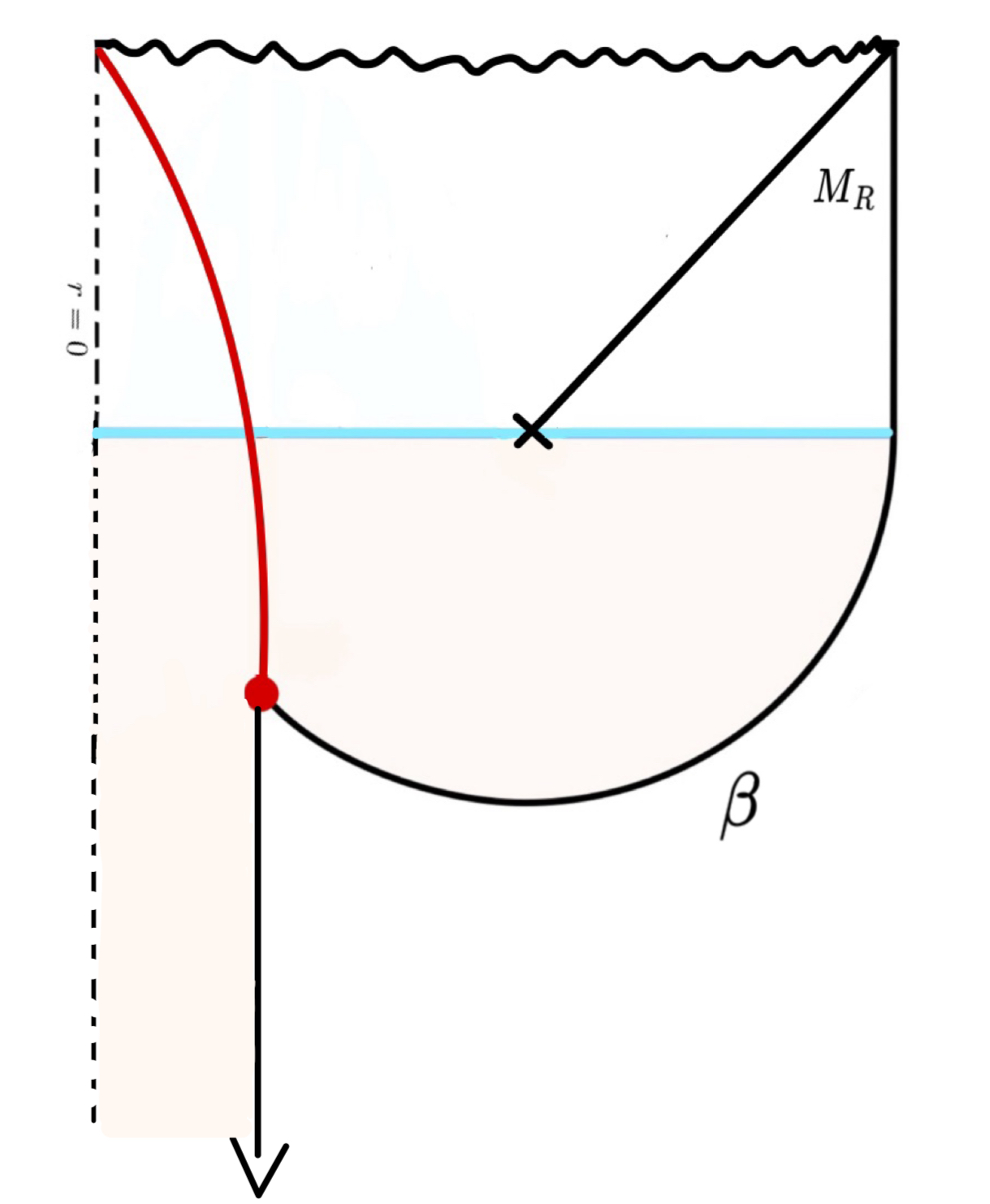}
    \caption{}
    \label{fig:type1}
    \end{subfigure}
    \hfill
 \begin{subfigure}[c]{0.45\linewidth}
    \centering
    \includegraphics[width=0.7\linewidth]{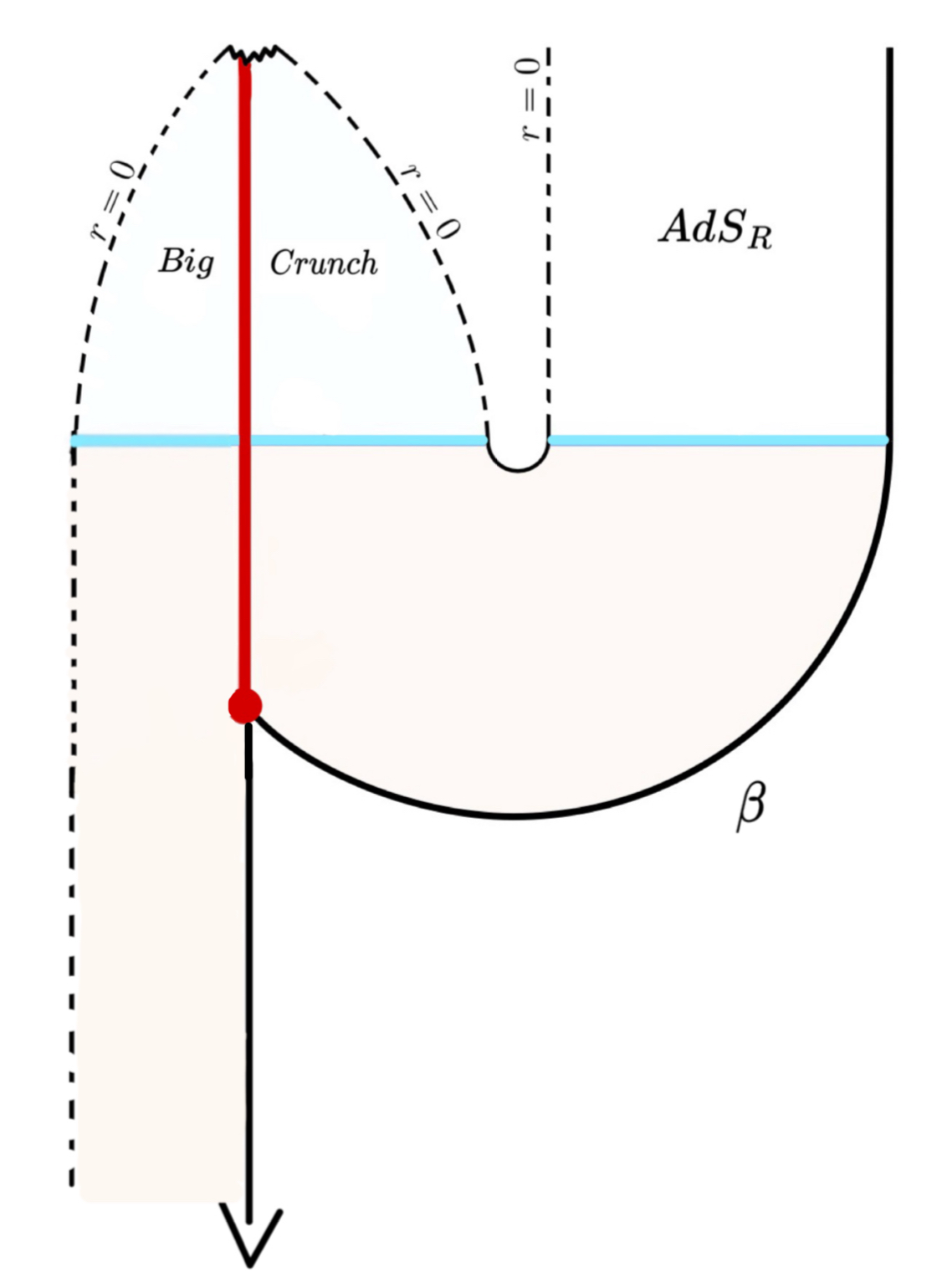}
    \caption{}
    \label{fig:type2}
    \end{subfigure}

\caption{Diagrams showing the analytic continuation of the type A-B single-sided shell states to the Lorentzian signature Penrose diagram.  ({\bf a}) Type A shell state with $\beta < \beta_{HP}$ corresponding to a single-sided black hole. ({\bf b}) Type B shell state with $\beta > \beta_{HP}$ and consisting of thermal AdS with the addition of a disconnected compact Big-Crunch AdS cosmology. }
\end{figure}

\end{enumerate}
In what follows we will denote a single-sided shell state of mass $m_i$ and preparation temperature $\beta_R$ as $|i\rangle_{R}$, and the Hilbert space spanned by $\kappa_R$ such states of varying mass as $\mathcal{H}_R$.

\subsection{A complete basis} \label{sec:complete}
We now we use methods developed in \cite{Toolkit} to establish that the single-sided shell states span the full single-sided Hilbert space $\mathcal{H}_{\mathcal{B}}$ by showing that the projector onto $\mathcal{H}_{R}$ acts as the identity on $\mathcal{H}_{\mathcal{B}}$ at the fine-grained level:
\beq \label{eq:project1}
\mathds{1}_{\mathcal{H}_{\mathcal{B}}} = \prod_{\mathcal{H}_{R}} \equiv G^{-1}_{ij} |i\rangle\langle j|
\eeq
for any shell preparation temperature $\beta_R$. We derive this equality from the gravitational path integral by showing the two path integral equalities
\beq \label{eq:idinsert}
\overline{\langle \Psi|\prod_{{\mathcal{H}_{R}}}|\Psi\rangle}=\overline{\langle \Psi|\Psi\rangle}
\eeq
and
\beq \label{eq:idsqr}
\overline{(\langle \Psi|\prod_{{\mathcal{H}_{R}}}|\Psi\rangle-\langle \Psi|\Psi\rangle)^2}=0.
\eeq
for $\forall |\Psi\rangle \in \mathcal{H}_{\mathcal{B}}$. As we discussed in Sec.~\ref{sec:review}, these equations together imply the fine-grained equality $\langle \Psi|\Psi\rangle=\langle \Psi|\prod_{\mathcal{H}_{R}}|\Psi\rangle$. As the fine-grained inner product must satisfy the Cauchy-Schwartz identity, it then follows that $\mathcal{H}_{R}=\mathcal{H}_{\mathcal{B}}$.

By definition $\mathcal{H}_{\mathcal{B}}$ is the space of states made by cutting open the path integral. Therefore all states $|\Psi\rangle \in \mathcal{H}_{\mathcal{B}}$ can be written as superpositions of states constructed by boundary time evolution over the half infinite strip followed by insertion of some (possibly non-local) operator $\mathcal{O}_{\Psi}$ on the asymptotic boundary followed by $\tilde{\beta}_R$ boundary time evolution. In particular we denote such states as $|\Psi (\tilde{\beta}_R)\rangle= e^{-\frac{\tilde{\beta}_RH_R}{2}}\mathcal{O}_{\Psi} |0 \rangle$ and define 
\begin{equation}
\langle \Psi|\Psi\rangle \equiv S_{\mathcal{O}_{\Psi}}( \tilde{\beta}_R) \, ,
\label{eq:SOnotation}
\end{equation}
where the vacuum state $|0 \rangle$ should be understood as  representing the half line boundary condition\footnote{In holographic theories the bulk-boundary dictionary maps this to the path integral for the CFT vacuum.} and $\mathcal{O}_{\Psi}$ may include boundary time evolution. This does not restrict the states under consideration but rather sets up our notation. 

The path integral (\ref{eq:idinsert}) is $\overline{\langle \Psi|\prod_{\mathcal{H}_{R}}|\Psi\rangle}= \lim_{n \to -1 } \overline{G^{n}_{ij}\langle \Psi|i\rangle\langle j|\Psi\rangle}$ where each shell overlap insertion $\langle i|j\rangle$  corresponds to a shell strip boundary condition (Fig.~\ref{fig:1s_shellbdry}). Similarly, the boundary condition for the $\langle \Psi|i\rangle$ insertion consists of a boundary time interval $[-\infty,0]$, an $\mathcal{O}_i$ insertion at $\tau =0$, followed by an $\mathcal{O}^{\dagger}_{\Psi}$ insertion at $\tau = \frac{\beta_R+\tilde{\beta}_R}{2}$  and finally another infinite segment half line $[\frac{\beta_R+\tilde{\beta}_R}{2},\infty]$.\footnote{As noted above this should be understood as the boundary condition $\lim_{\alpha \to \infty} [\frac{\beta_R+\tilde{\beta}_R}{2},\frac{\beta_R+\tilde{\beta}_R}{2}+\alpha]$.} We refer to these as \textit{operator strip} boundaries. 

As the full single-sided Hilbert space $\mathcal{H}_{\mathcal{B}}$ is expected to be infinite dimensional, we need an infinite number of basis vectors. Hence we take $\mathcal{H}_R$ to be the span of $\kappa_R \to \infty$ number of single-sided shell states obtained by fixing the preparation temperature $\beta_R$ but varying the mass.  As explained in \cite{Toolkit}, in this limit only geometries that do not break any shell index loops contribute to the sum over topologies.

\begin{figure}[h]
    \centering
    \includegraphics[width=0.7\linewidth]{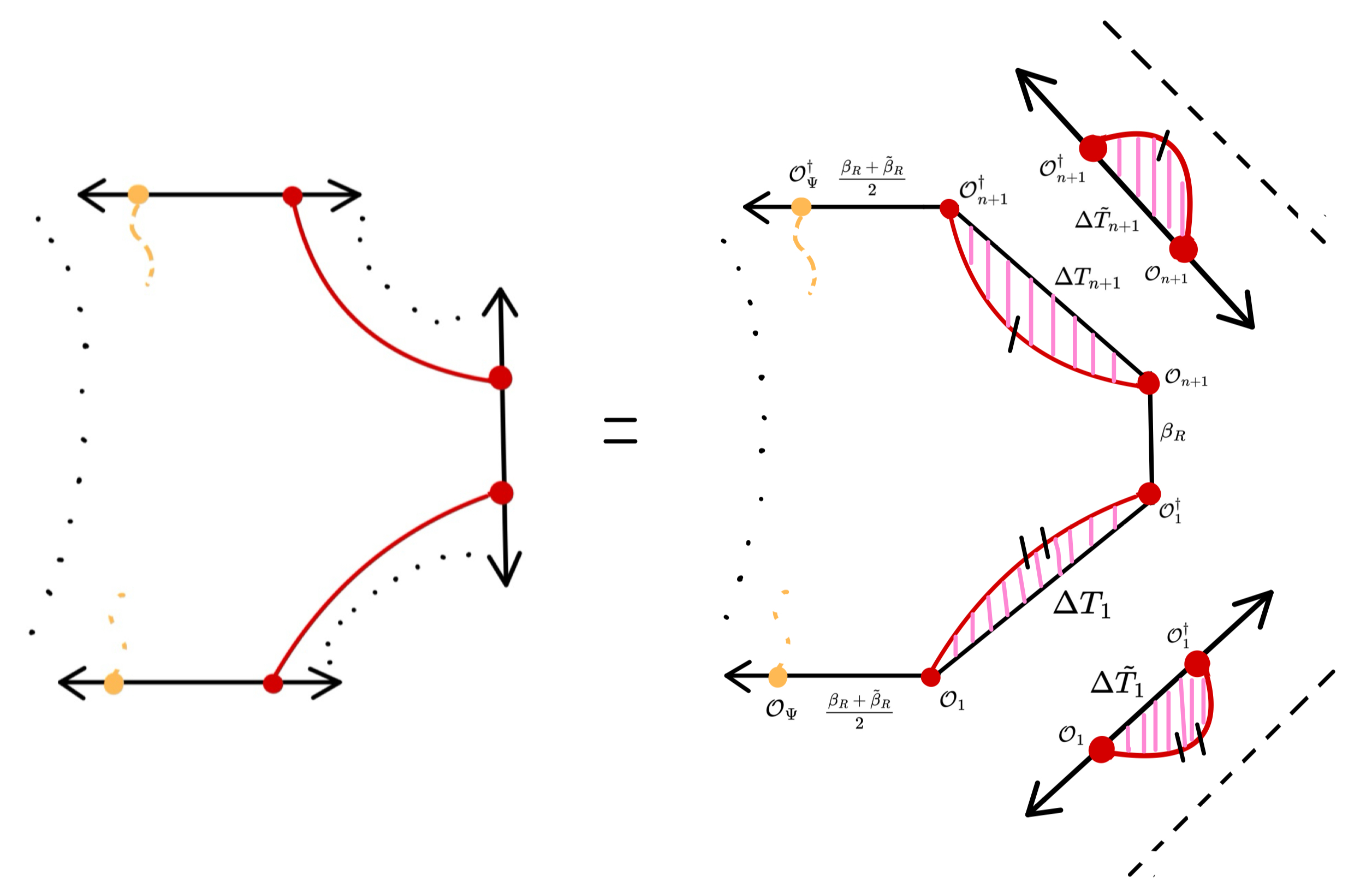}
    \caption{The fully connected wormhole saddle for (\ref{eq:idinnorm1s}) is constructed by gluing $n+1$ shell strips into the central strip, depicted here for $n=1$. }
    \label{fig:StripWH}
\end{figure}

\paragraph{Coarse-grained completeness}
We start by establishing the coarse-grained equality (\ref{eq:idinsert}), which is computed as $\overline{\langle \Psi|\prod_{{\mathcal{H}_{R}}}|\Psi\rangle}=\lim_{n \to -1} \overline{G^{n}_{ij}\langle \Psi|i\rangle\langle j|\Psi\rangle}$,  within the saddle point approximation.  Following \cite{Toolkit}, as we take $\kappa_R \to \infty$ we will use the inverse of the Gram matrix $G^{-1}$ in (\ref{eq:project1}) to account for overlaps between the shell states in the resolution of the identity.  This procedure also takes care of potential overcompleteness in the basis. The  gravity path integral $\overline{G^{n}_{ij}\langle \Psi|i\rangle\langle j|\Psi\rangle}$ contains only a single maximal shell index loop, and hence in the $\kappa_R \to \infty$ limit we only have to consider the maximally connected topology. These geometries connect the $n$ shell boundaries and the two operator boundaries via a single bulk wormhole geometry (Fig.~\ref{fig:StripWH}). We will refer to these wormholes as \textit{strip wormholes}. In principle there can be multiple saddles of this kind depending on the properties of $|\Psi\rangle$, all of which will turn out to be independent of the shell mass index running in the loop. We will show below that such saddles always exist if $\overline{\langle \Psi|\Psi\rangle}$ has a saddle. Denoting the sum over these saddles as $\hat{S}_{n+2}(\Psi)$ we therefore obtain:
\beq \label{eq:idinnorm1s}
\overline{G^{n}_{ij}\langle \Psi|i\rangle\langle j|\Psi\rangle} =\kappa^{n+1} \hat{S}_{n+2}(\Psi)\,.
\eeq

\begin{figure}
    \centering
    \includegraphics[width=\linewidth]{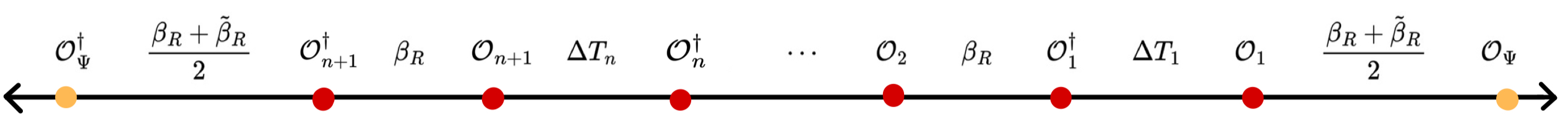}
    \caption{Asymptotic boundary condition for the operator-strip wormhole into which $n+1$ shell strips are glued, forming the connected wormhole contribution to (\ref{eq:idinnorm1s}). }
    \label{fig:stripBC}
\end{figure}

We proceed as follows. Time evolve the state $|\Psi\rangle$ by an amount $\frac{(n+1)\beta_R}{2}$. Consider saddlepoints contributing to the norm of this time evolved state, which in the notation of (\ref{eq:SOnotation}) computes $\overline{S_{\mathcal{O}_{\Psi}}( \tilde{\beta}_R+(n+1)\beta_R)}$. From any such a saddle we can obtain a strip wormhole saddle and vice versa, as we will now show by construction. The strip wormhole saddles are constructed by gluing $n
+1$ shell strips into one central strip along the shell worldvolumes (see the right side of Fig.~\ref{fig:StripWH}).  This is in contrast with the wormholes for $\overline{G^n}_{ii}$ discussed in Sec.~\ref{sec:1sWH} which involved gluing $n$ shell strips into a central disk. The to-be-glued shell strip boundary condition for the $i$-th shell consists of the half line $[-\infty,0]$, followed by $\mathcal{O}_{S_i},\mathcal{O}^{\dagger}_{S_i}$ insertions separated by $\tilde{T}_i$ and another half line. The central strip geometry has an asymptotic boundary condition consisting of the $[-\infty,0]$ half line, a $\mathcal{O}_{\Psi}$ operator insertion at $\tau =0$,  followed by the first shell insertion $\mathcal{O}_{S_1}$ at $\tau= \frac{\tilde{\beta}_R+{\beta}_R}{2}$, and insertion of $\mathcal{O}^{\dagger}_{S_1}$ at time $T_1$ later.   A repeating pattern then follows where every $\mathcal{O}^{\dagger}_{S_{j}}$ insertion is followed $\beta_R$ time evolution, insertion of $\mathcal{O}_{S_{j+1}}$ and $T_{j+1}$ time evolution, followed by $\mathcal{O}^{\dagger}_{S_{j+1}}$ insertion until $\mathcal{O}^{\dagger}_{S_{n+1}}$.  This is further followed by a  $\frac{\tilde{\beta}_R+{\beta}_R}{2}$ time evolution, $\mathcal{O}^{\dagger}_{\Psi}$ and a half line time evolution, completing the strip boundary, see Fig.~\ref{fig:stripBC}. We will denote the sum over saddlepoints for the path integral subject to this central strip boundary condition  as $S_{Central}(\tilde{\beta}_R +\sum_{i}T_i + (n+1)\beta_R)$. 

On both the shell- and central-strips the shells propagate into the bulk before being re-absorbed a time $\tilde{T}_i,T_i$ later respectively. The strip wormhole saddle is obtained by cutting off the shell homology regions on all the strips and gluing the shell strips into the central strip along the corresponding shell worldvolumes via the junction conditions, which dynamically determine $\{T_i,\tilde{T}_i\}$ to yield an on-shell geometry (Fig.~\ref{fig:StripWH}). In the large shell mass limit the propagation times $\{T_i,\tilde{T}_i\}$ go to zero, the shell homology regions on the operator and shell strips pinch off, and the effect of each shell is only to contribute a universal factor $Z_{m_i}$. Hence in this limit the remaining part of the saddle action is given by action of the $n+1$ shell strips resulting in a factor $\overline{S}^{n+1}(0)$ and the  action of the central strip. Note that neither of these contain a contribution from the shells, as the shell action is already accounted for. As the central strip now has no shell insertions and $T_i\to 0$  what remains is simply the action of the $\overline{S_{\mathcal{O}_{\Psi}}}((n+1) \beta_R + \tilde{\beta}_R)$ saddles\footnote{We stress that again $\overline{S}_{\mathcal{O}_{\Psi}}((n+1) \beta_R + \tilde{\beta}_R)$ collectively denotes the sum over saddles to $\overline{\langle \Psi | \Psi\rangle}((n+1) \beta_R + \tilde{\beta}_R) $ and for each of these saddles individually the above construction follows through.} defined above. We therefore obtain $\hat{S}_{n+2}(\Psi)= \prod_{i=1}^{n+1} Z_{m_i}\times\overline{S}^{n+1}(0)\times\overline{S}_{\mathcal{O}_{\Psi}}((n+1) \beta_R + \tilde{\beta}_R)$. After normalising the shell states to unity  we get:
\beq
\hat{S}_{n+2}(\Psi)= \frac{\overline{S}_{\mathcal{O}_{\Psi}}((n+1) \beta_R + \tilde{\beta}_R)}{\overline{Z}(\beta_R)^{n+1}}.
\eeq

The upshot is that given a saddle for the norm of a time-evolved version of the state $\ket{\Psi}$ given by $\overline{S}_{\mathcal{O}_{\Psi}}((n+1) \beta_R + \tilde{\beta}_R)$ the saddles for (\ref{eq:idinnorm1s}) can automatically be constructed in the large shell mass limit and are independent of the shell index running in the loop, because of the universal pinching-off behavior of the shell insertions in this limit. Taking  $n \to -1$, we find
\beq \label{eq:span}
\overline{\langle \Psi|\prod_{\mathcal{H}_{R}}|\Psi\rangle}= \lim_{n \to -1 } \kappa^{n+1} \frac{\overline{S}_{\mathcal{O}_{\Psi}}((n+1) \beta_R + \tilde{\beta}_R)}{\overline{Z}(\beta_R)^{n+1}} = \overline{S}_{\mathcal{O}_{\Psi}}(\tilde{\beta}_R) \equiv \overline{\langle \Psi|\Psi\rangle} \, ,
\eeq
showing (\ref{eq:idinsert}) within the saddlepoint approximation  to the path integral.

\paragraph{Fine-grained completeness} 
Next, we need to also show (\ref{eq:idsqr}) to obtain the fine-grained equality (\ref{eq:project1}). If $\overline{\langle \Psi|\Psi\rangle^2}\neq\left(\overline{\langle \Psi|\Psi\rangle}\right)^2$ due to connected wormhole contribution to the LHS supported by $\mathcal{O}_\Psi$ matter insertions there will be multiple saddles of different topology for each term in (\ref{eq:idsqr}).  We consider the family of wormhole saddles $\overline{\langle \Psi|\Psi\rangle \langle \tilde\Psi| \tilde \Psi \rangle}|_{conn}$ where $|\tilde\Psi\rangle$ is a state obtained by boundary time evolution of $|\Psi\rangle$ by an amount $\Delta_R$; we denote this as $\overline{\langle \Psi|\Psi\rangle \langle\tilde\Psi |\tilde\Psi\rangle}|_{conn}\equiv \overline{S_{\Psi \tilde\Psi}(\Delta_R)}$. Similar to the discussion above, in the large shell mass limit any of the $\mathcal{O}_\Psi$ matter supported wormhole saddles relevant for (\ref{eq:idsqr}) can be obtained by gluing shell strips into the $S_{\Psi \tilde\Psi}(\Delta_R)$ wormhole for different values of $\Delta_R$. 

\begin{figure}[h]
    \centering
    \includegraphics[width=\linewidth]{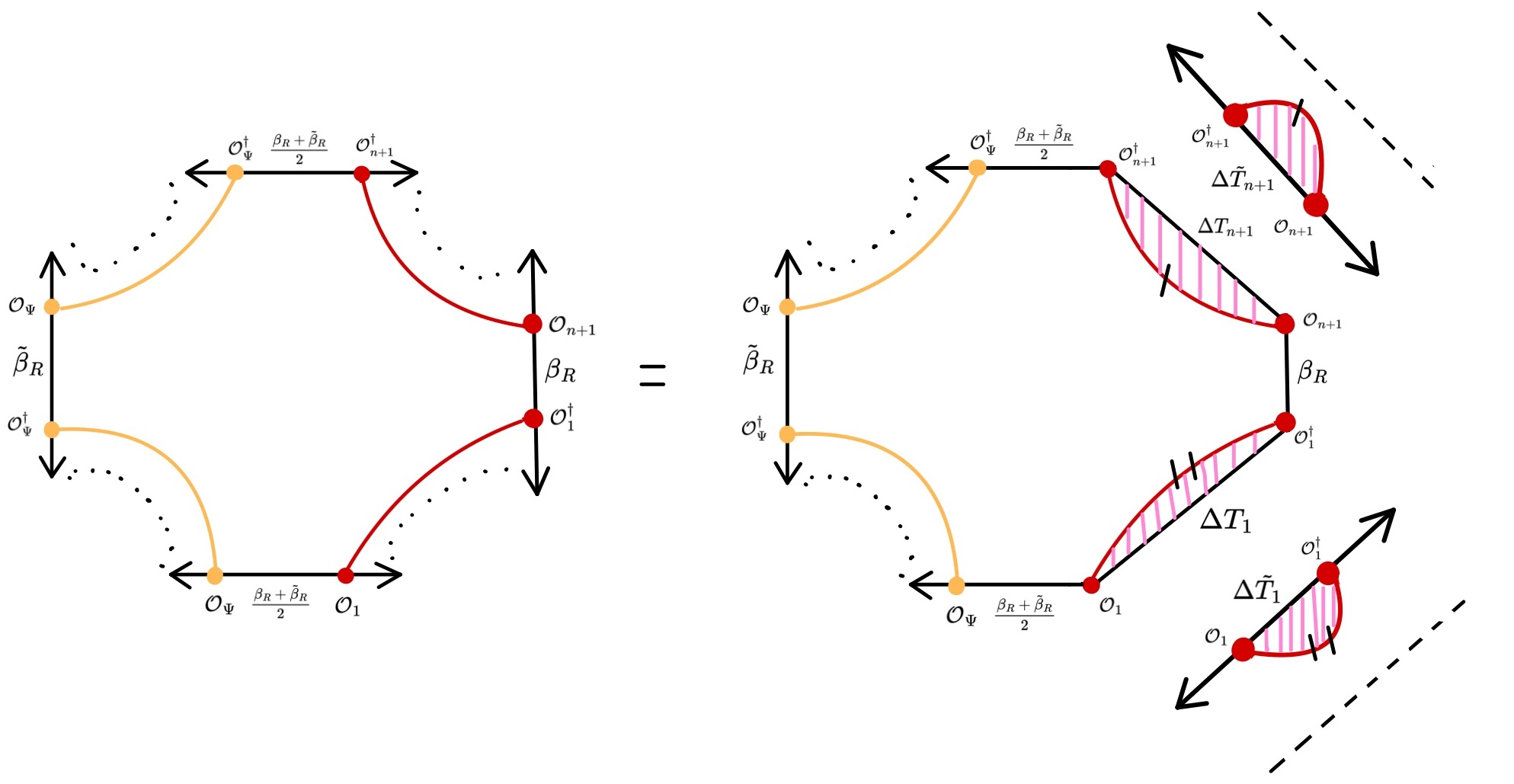}
    \caption{The additional fully connected wormhole contributions to (\ref{eq:idsqr}) can be constructed analogously to those for (\ref{eq:idinnorm1s}) by gluing $n+1$ shell strips into the central bulk.}
    \label{fig:sqrproof}

\end{figure}

We are considering the $\kappa_R \to \infty$ limit in which the shell index loop cannot be broken. Taking, for example, the cross term in (\ref{eq:idsqr})
\beq \label{spancross}
\overline{\langle \Psi|\prod_{\mathcal{H}_{R}}|\Psi\rangle\langle \Psi|\Psi\rangle}= \lim_{n \to -1}\overline{G^{n}_{ij}\langle \Psi|i\rangle \langle j|\Psi\rangle\langle \Psi|\Psi\rangle} \, 
\eeq 
there can then be two types of saddle: saddles connecting the various $\mathcal{O}_{\Psi}$ insertions across $\langle \Psi|\prod_{\mathcal{H}_{R}}|\Psi\rangle$ and $\langle \Psi|\Psi\rangle$, and ones that do not. For the latter we simply have $\overline{\langle \Psi|\prod_{\mathcal{H}_{R}}|\Psi\rangle\langle \Psi|\Psi\rangle}|_{disconn}=\overline{\langle \Psi|\prod_{\mathcal{H}_{R}}|\Psi\rangle}\times \overline{\langle \Psi|\Psi\rangle}$ which by (\ref{eq:span}) equals $\left(\overline{\langle \Psi|\Psi\rangle}\right)^2$. The fully connected contributions can be constructed by starting with the wormhole connecting
the operator boundary $\braket{\Psi|\Psi}$ to boundary of the central strip (Fig.~\ref{fig:stripBC}), and again gluing in $n+1$ shell strips along the shell worldvolumes (Fig.~\ref{fig:sqrproof}) into the central strip as above.  In the large shell mass limit the shell homology regions pinch off and the shells contribute universally to the action. The action is therefore given by that of $n+1$ strips of length zero in addition to the action of the wormhole without any shell insertions, which is simply the action of the $\overline{S_{\Psi \tilde\Psi}\left((n+1)\beta_R\right)}$ saddles defined above. Hence, after normalising the shell states we obtain 
\beq
 \lim_{n \to -1}\overline{G^{n}_{ij}\langle \Psi|i\rangle \langle j|\Psi\rangle\langle \Psi|\Psi\rangle}|_{conn}=   \lim_{n \to -1} \kappa^{n+1}\frac{\overline{S_{\Psi \tilde\Psi}\left((n+1)\beta_R\right)}}{\overline{Z}(\beta_R)^{n+1}} = \overline{S_{\Psi \tilde\Psi}(0) }\equiv \overline{\langle \Psi|\Psi\rangle^2}|_{conn}\, .
\eeq

Hence the connected and disconnected saddles to (\ref{spancross}) account for the connected and disconnected saddles to $\overline{\langle \Psi|\Psi\rangle\langle \Psi|\Psi\rangle}$,
and we have $\overline{\langle \Psi|\prod_{\mathcal{H}_{R}}|\Psi\rangle\langle \Psi|\Psi\rangle}=\overline{\langle \Psi|\Psi\rangle\langle \Psi|\Psi\rangle}$. A similar argument applies to the term $\overline{\left(\langle \Psi|\prod_{\mathcal{H}_{R}}|\Psi\rangle\right)^2}=\overline{\langle \Psi|\Psi\rangle\langle \Psi|\Psi\rangle}$, where now the fully-connected contributions are constructed by gluing $n+1$ shells strips into each of two central strip boundaries connected by the $\mathcal{O}_{\Psi}$-supported wormhole. Therefore in the $n \to -1$ limit each term in (\ref{eq:idsqr}) limits to $\overline{\langle \Psi|\Psi\rangle^2}$, proving (\ref{eq:idsqr}) .

\paragraph{Conclusion} Since we have shown both (\ref{eq:idinsert}) and (\ref{eq:idsqr}) we obtain the fine grained equality
\beq \label{Hshellspan}
\mathcal{H}_{\mathcal{B}}=\mathcal{H}_{R}
\eeq
in the saddlepoint approximation to the path integral.  The relation holds if we take $\mathcal{H}_{R}$ to be the span of $\kappa_R \to \infty$ shell states of fixed preparation temperature but varying mass.
Interestingly, (\ref{Hshellspan})  holds  for \textit{any} choice of preparation temperature, since the derivations of (\ref{eq:idinsert}) and (\ref{eq:idsqr}) were both independent of the temperature used to prepare the shell states.\footnote{Moreover the above argument extends trivially to $\langle \Psi|\Phi\rangle =\langle \Psi|\prod_{\mathcal{H}_R}|\Phi\rangle + \mathcal{O}(e^{G^{0}_N})$ for any $|\Psi \rangle, |\Phi \rangle \in \mathcal{H}_{\mathcal{B}}$.} We could therefore take this basis to consist of type A shell states, which contain a horizon, or type B shell states which do not but instead have a compact universe. As a $\kappa_R \to \infty$ set of either of these states spans, {\it a type A black hole geometry can be written as a superposition of horizonless type B geometries and vice versa.} In \cite{Toolkit} we reached the same conclusion for the two-sided shell states and proceeded to argue that the notion of a fixed geometry associated to a state made by cutting the gravity path integral must be emergent in the semiclassical limit. Our results here show that the same observations apply to the single-boundary gravity theory. Finally, we can use the ``surgery methods'' of the toolkit proposed in \cite{Toolkit} to extend the argument that {\it the shell states span $\mathcal{H}_{\mathcal{B}}$ to all orders in the fine-grained theory (see appendix~\ref{sec:allOspan}).}

\section{The two-sided Hilbert space is a product of single-sided Hilbert spaces}\label{sec:factorisation}

We showed in \cite{Toolkit} that all states of quantum gravity with two boundaries  constructed by the Euclidean gravity path integral with a single connected boundary $\mathcal{H}_{LR}$ (Fig.~\ref{fig:singleBcut}) are in the span $\mathcal{H}_{2s}$ of the two-sided shell states reviewed in Appendix~\ref{sec:2Sss}. Likewise, it follows from  Sec.~\ref{sec:1s-span} that the states of two-boundary quantum gravity constructed by cutting the Euclidean path integral with a boundary consisting of two closed disconnected components
$\mathcal{H}_{\mathcal{B}_L} \otimes \mathcal{H}_{\mathcal{B}_R}$ (Fig.~\ref{fig:tensorcut}) are in the span of $\mathcal{H}_L \otimes \mathcal{H}_R$ of products of the single-sided shell states.  In this section we will show that $\mathcal{H}_{2s} = \mathcal{H}_L \otimes \mathcal{H}_R$, which in turn shows the factorisation of the full Hilbert spaces $\mathcal{H}_{LR}=\mathcal{H}_{\mathcal{B}_L} \otimes \mathcal{H}_{\mathcal{B}_R} $.

\begin{figure}
    \centering
    \begin{subfigure}[c]{0.3\linewidth}
    \includegraphics[width=\linewidth]{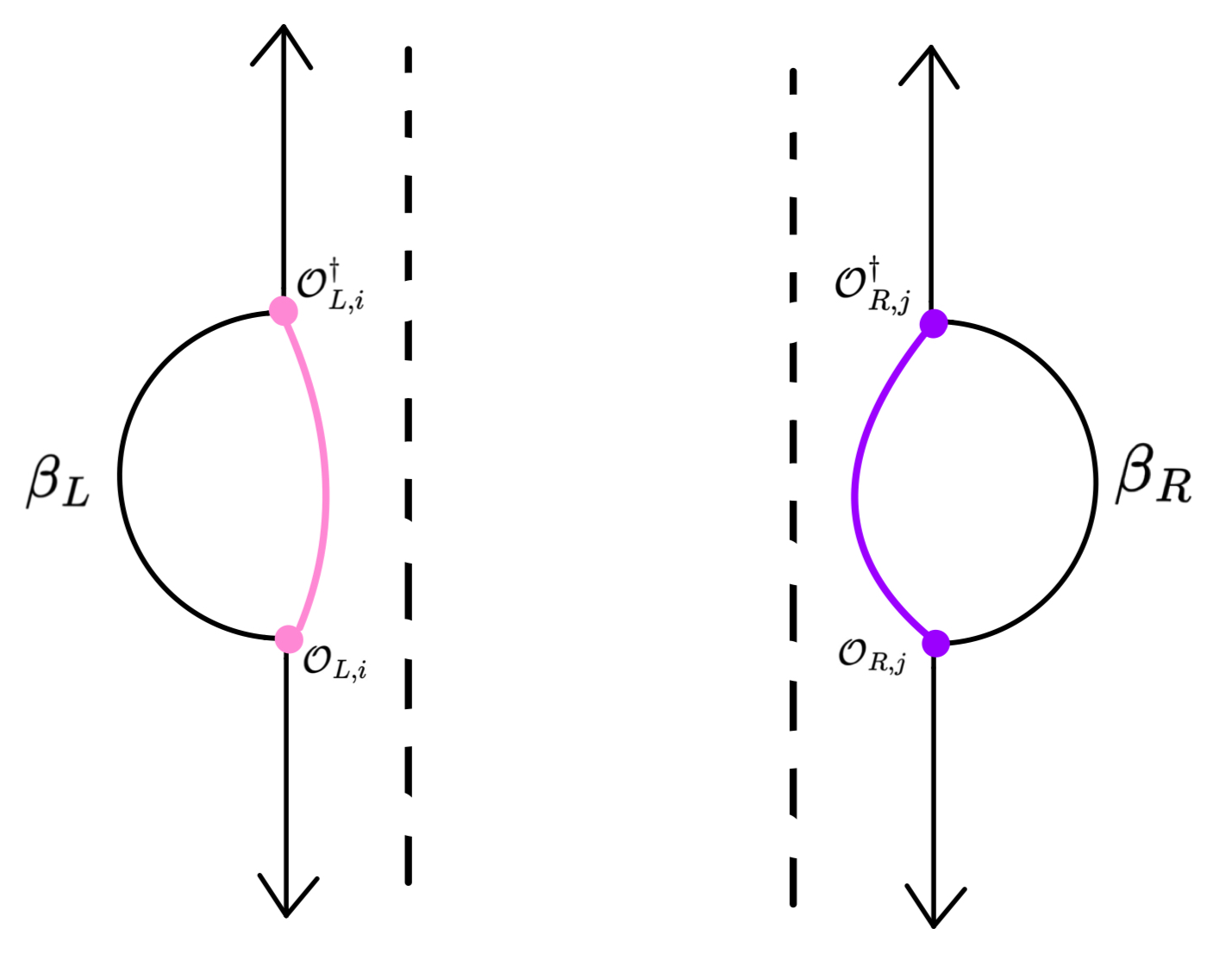}
    \caption{}
     \end{subfigure}
     \begin{subfigure}[c]{0.3\linewidth}
    \includegraphics[width=\linewidth]{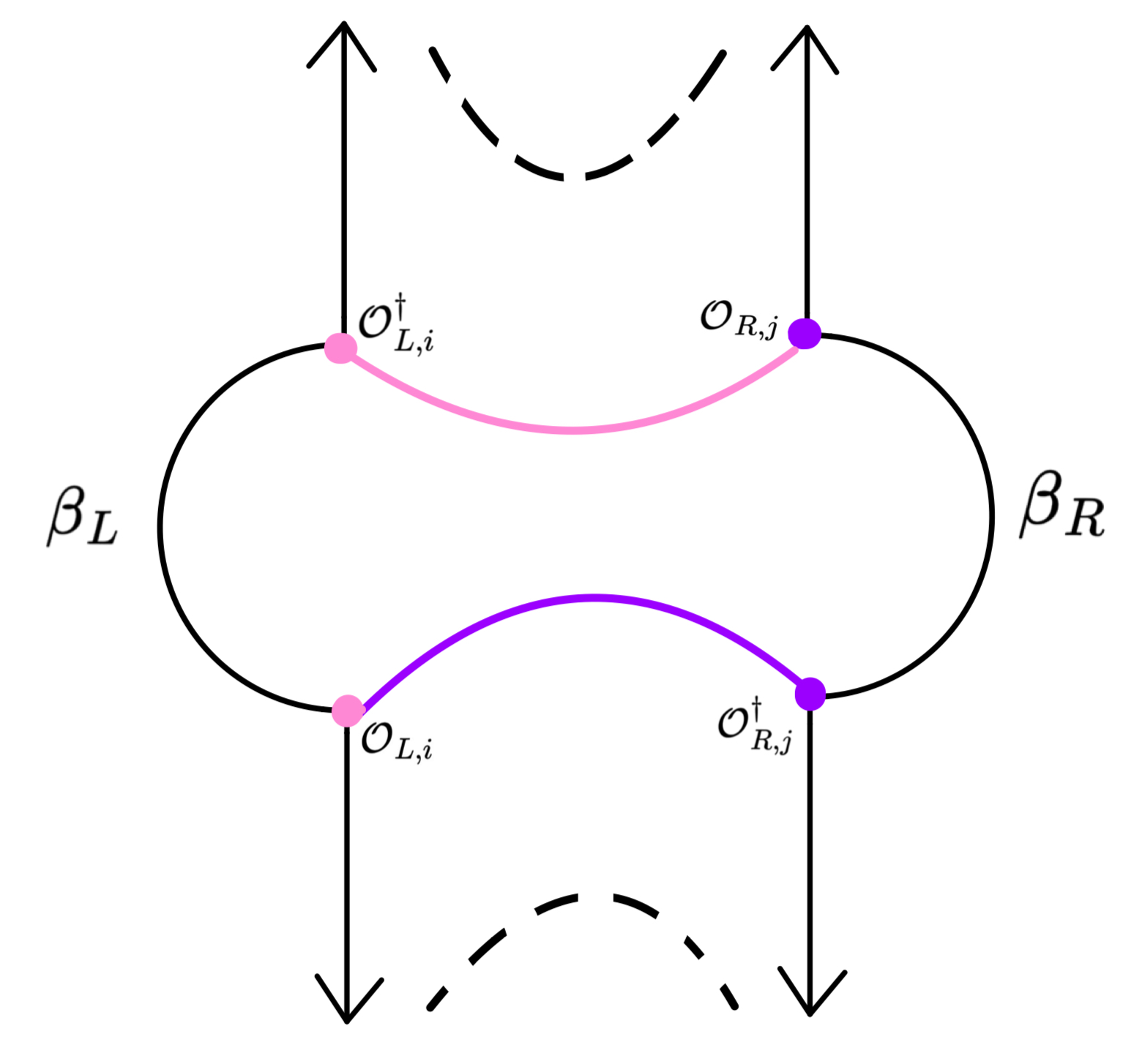}
        \caption{}
     \end{subfigure}
     \begin{subfigure}[c]{0.35\linewidth}
    \includegraphics[width=\linewidth]{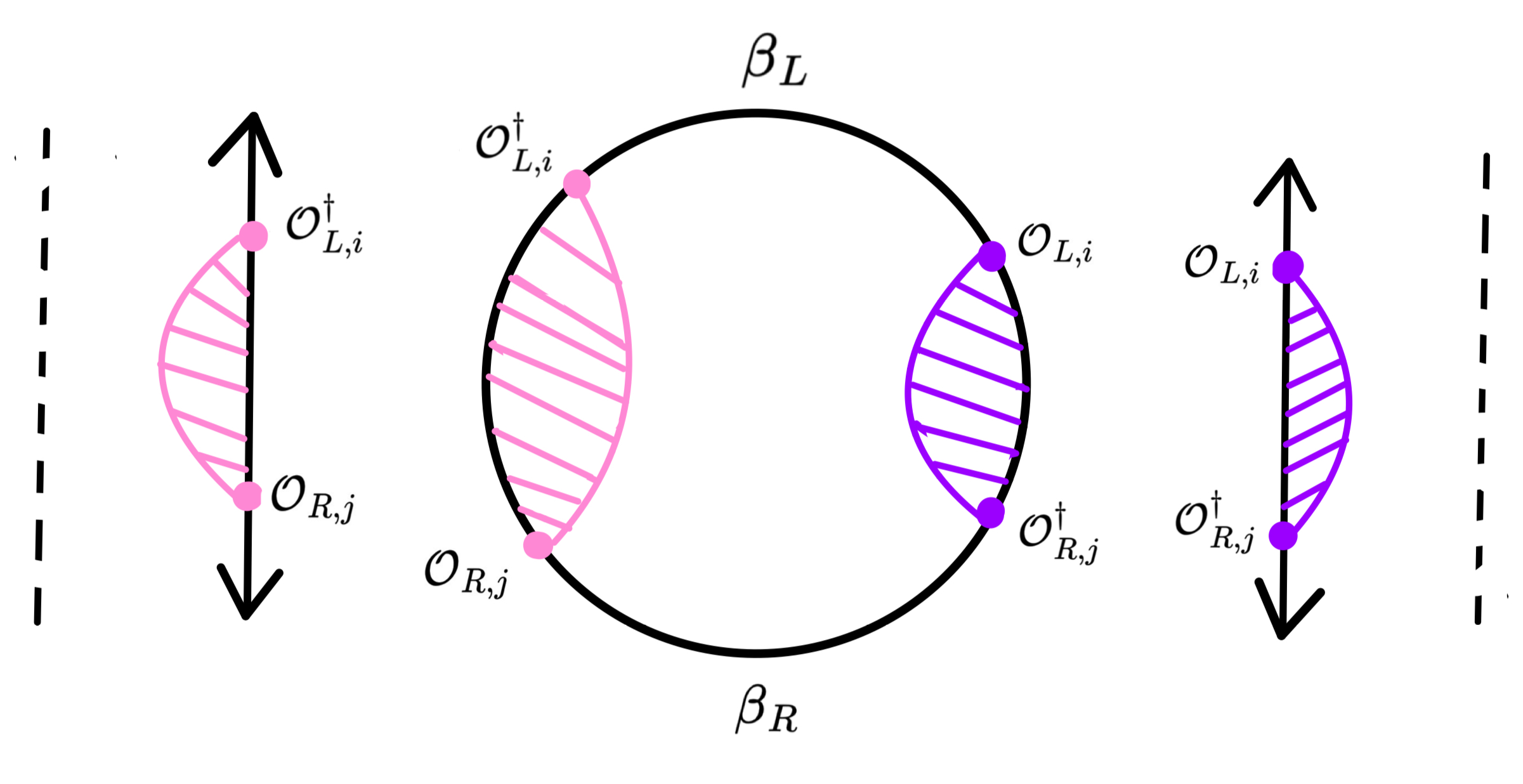}
        \caption{}
     \end{subfigure}

    \caption{Saddle geometries for $\overline{\langle i|i \rangle_L\langle j|j \rangle_R}$: ({\bf a}) Disconnected contribution corresponding to two copies the saddle (\ref{eq:overlap}). ({\bf b}) Connected wormhole contribution present when $i=j$. ({\bf c}) This wormhole contribution is constructed by gluing $L,R$ strips into a disk. }
    \label{fig:tensornorm}
\end{figure}

First define the $L,R$ single-sided shells states with preparation temperatures $\beta_{L,R}$ respectively and the two-sided shell states with temperatures $\tilde{\beta}_{L,R}$. We denote a two-sided  state defined by the shell operator insertion $\mathcal{O}_i$ in boldface as $\ket{\bf i}_{2s}$, and $L,R$ single-sided  states defined by that same operator as $\ket{ i}_{L,R}$.  First we want to determine the normalization of the product state $\ket{i}_L \otimes \ket{j}_R$.   There are two saddlepoint contributions -- the product $\overline{\langle i|i\rangle_L} \times \overline{\langle j|j \rangle_R}$ of the norms of the $L$ and $R$ states, and the wormhole shown in  Fig.~\ref{fig:tensornorm}.  The wormhole only contributes if the shell creating the $L$ state can annihilate with the shell creating the $R$ state, which is possible only if $m_i = m_j$ in view of the large inertial mass gaps between shells (see Sec.~\ref{sec:review}). Thus we have:
\beq \label{eq:tensor_norm}
\overline{\langle i|i \rangle_L\langle j|j \rangle_R} = \overline{Z}(\beta_L)\overline{Z}(\beta_R)\overline{S(0)}^2 + \delta_{ij} \overline{Z}(\beta_L+\beta_R)\overline{S(0)}^2 \, ,
\eeq
where $\bar{Z}$ denotes the Z-partition function (see Sec.~\ref{sec:review}) and $\bar{S}$ denotes the path integral with the strip boundary condition (see Sec.~\ref{sec:1s-span}). 
 For simplicity will choose the bases for  $\mathcal{H}_{L,R}$ to contain shells of different mass so that the wormhole contribution is absent.

\begin{figure}[h]
    \centering
    \includegraphics[width=0.7\linewidth]{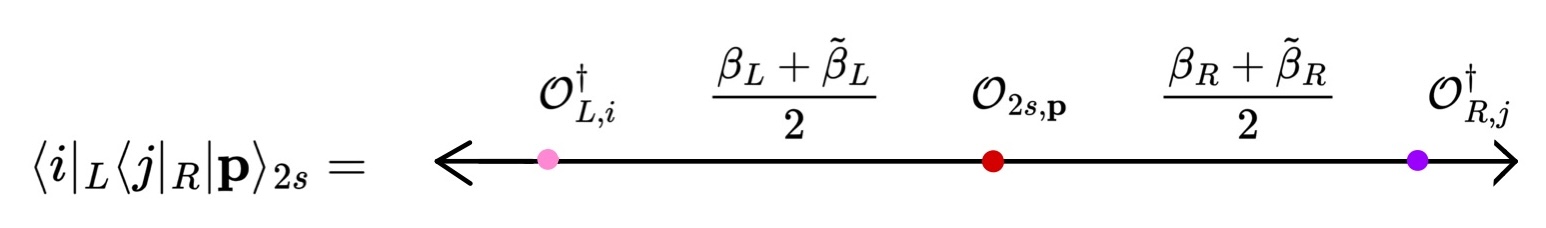}
    \caption{Boundary condition for the overlap between a two-sided shell state and the tensor product of single-sided shell states: $\langle i|_L \langle j|_R|\mathbf{p}\rangle_{2s}$.}
    \label{fig:tensorspanbc}
\end{figure}

\begin{figure}
    \centering
    \begin{subfigure}[c]{0.48\linewidth}
      \includegraphics[width=\linewidth]{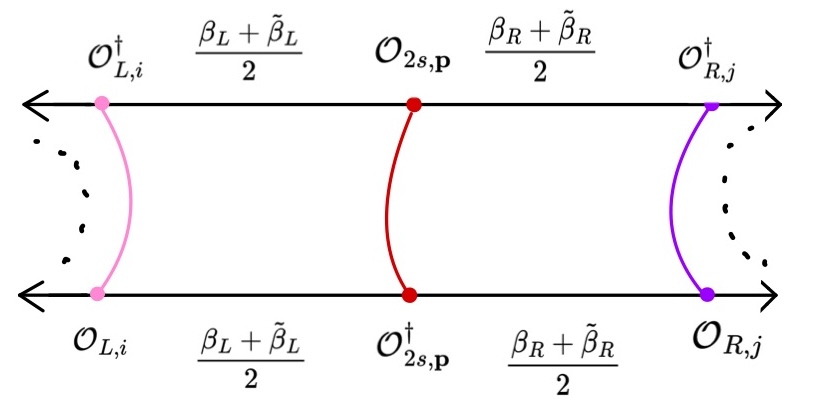}
    \caption {}
      \end{subfigure}
   \hfill
    \begin{subfigure}[c]{0.48\linewidth}
    \includegraphics[width=\linewidth]{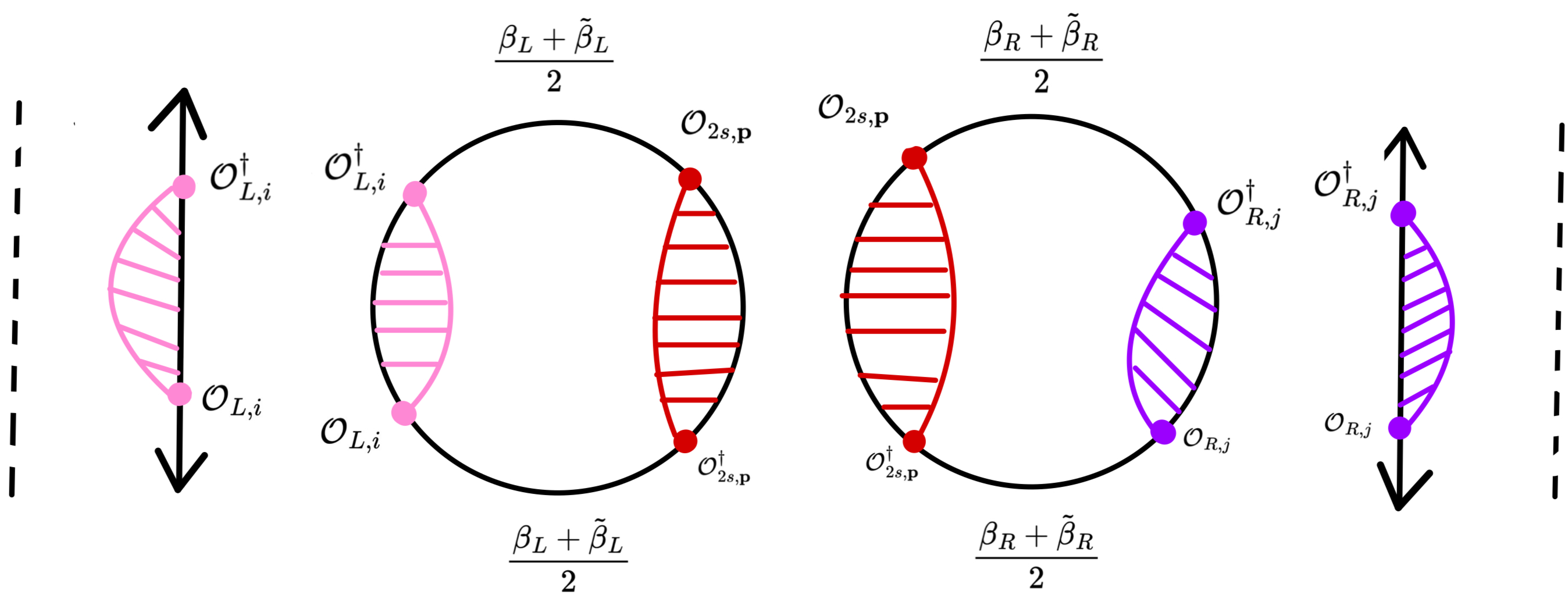}
    \caption{}
    \end{subfigure}
    \caption{ ({\textbf a}) Universal wormhole contribution to $\overline{|\langle i|_L \langle j|_R|\mathbf{p}\rangle_{2s}|^2}$.  ({\textbf b}) This wormhole in constructed by gluing a $L,R$ strip into the $L,R$ disk and then gluing the disk together along the shell worldvolumes.  }
     \label{fig:direct_overlap}
\end{figure}

Next we want to calculate the overlap between a product of one-sided states and a two-sided one: $\langle i|_L \langle j|_R|\mathbf{p}\rangle_{2s}$. The boundary condition for the path integral calculating this overlap consists of a half infinite line, followed by an $\mathcal{O}^{\dagger}_{L,i}$ insertion and time evolution by $\frac{\beta_L}{2}$ from the state $\langle i|_L$, followed by 
$\frac{\tilde{\beta}_L}{2}$  insertion of $\mathcal{O}_{2s,\mathbf{p}}$, and time evolution by $\frac{ \tilde{\beta}_R}{2}$ from $|\mathbf{p}\rangle_{2s}$, and finally $\frac{ {\beta}_R}{2}$ evolution and insertion of $\mathcal{O}^{\dagger}_{R,j}$  half infinite line coming from $\langle j|_R$ (see Fig.~\ref{fig:tensorspanbc}). Strictly speaking, constructing the saddle points for this boundary condition requires a detailed analysis of shell splitting and joining dynamics. However, if we choose the $L,R$ and two-sided shell bases such that all the shell operators in any one of them have  large mass differences, then these splitting interactions will be highly suppressed, as it takes $|m_{\mathbf{p}}-(m_{i}+m_{j})|$ bulk interactions in Planck units to match the shells, resulting in very small overlap. Nevertheless, if we instead consider the square of the overlap $\overline{|\langle i|_L \langle j|_R|\mathbf{p}\rangle_{2s}|^2}$ we find a universal wormhole contribution constructed by gluing a $L,R$ disk together along the $\mathcal{O}_{2s,\mathbf{p}}$ insertion and gluing a strip into each of disks along the insertions  $\mathcal{O}^{\dagger}_{L,j}$, $\mathcal{O}^{\dagger}_{R,i}$ shells respectively as in Fig.~\ref{fig:direct_overlap}. In the large shell mass limit, the shell homology regions again pinch off, and we are left with two disks and two strips. After normalising the states,   this wormhole saddle is therefore given by:
\beq
\overline{|\langle i|_L \langle j|_R|\mathbf{p}\rangle_{2s}|^2}=\frac{\overline{Z}(\beta_L +\tilde{\beta}_L)\overline{Z}(\beta_R +\tilde{\beta}_R)} { \overline{Z}(\beta_L )\overline{Z}(\tilde{\beta}_L)\overline{Z}(\beta_R )\overline{Z}(\tilde{\beta}_R)} \,
\label{eq:onetwooverlap}
\eeq
where the strip contributions have canceled out with  normalizations. 

The non-zero value of (\ref{eq:onetwooverlap}) shows that $\mathcal{H}_{L}\otimes \mathcal{H}_{R}$ and $\mathcal{H}_{2s}$ are not orthogonal.  To show that they are equal we will establish below that $\mathcal{H}_L\otimes \mathcal{H}_R \subseteq \mathcal{H}_{2s}$ and $\mathcal{H}_{2s} \subseteq  \mathcal{H}_L\otimes \mathcal{H}_R$ in the fine-grained theory by demonstrating that: (a) For any $|\Psi_{\otimes}\rangle \in\mathcal{H}_L\otimes \mathcal{H}_R $
\beq \label{eq:tensorinclusion}
\langle \Psi_{\otimes}|\prod_{\mathcal{H}_{2s}}|\Psi_{\otimes}\rangle=\langle \Psi_{\otimes}|\Psi_{\otimes}\rangle ,
\eeq
and (b) For any $|\Phi_{2s}\rangle \in\mathcal{H}_{2s}$ 
\beq \label{eq:2sinclusion}
\langle \Phi_{2s}|\prod_{\mathcal{H}_L}\prod_{\mathcal{H}_R}|\Phi_{2s}\rangle=\langle \Phi_{2s}|\Phi_{2s}\rangle .
\eeq

\subsection{Two-sided shell states the span tensor product of one-sided shell states}
\label{sec:2s_span_1s}

We start with the case where $|\Psi_{\otimes}\rangle$ is a tensor product. To show (\ref{eq:tensorinclusion}) it suffices  that
\beq \label{eq:2s_span_tensor}
\langle i|_{L} \langle j|_{R} \prod_{\mathcal{H}_{2s}}|i\rangle_{L}|j\rangle_{R}=\langle i|i\rangle_{L} \langle j| j\rangle_{R}.
\eeq
where $\prod_{\mathcal{H}_{2s}}$ is the projector onto the two-sided shell Hilbert space, because the product states $\ket{i}_L\ket{j}_R$ form a basis of $\mathcal{H}_{\mathcal{B}_L}\otimes \mathcal{H}_{\mathcal{B}_R}$. Following \cite{Toolkit} and Sec.~\ref{sec:review}, the projector $\prod_{\mathcal{H}_{2s}}$ can be resolved in terms of the shell states as
\beq \label{eq:2s_span_tensor_sc}
\overline{\langle i|_{L} \langle j|_{R} \prod_{\mathcal{H}_{2s}} |i\rangle_{L}|j\rangle_{R}} = \lim_{n \to -1 } \overline{G^n_{2s,\mathbf{kl}} \langle i|_{L} \langle j|_{R} |\mathbf{k}\rangle_{2s}\langle _{2s} \mathbf{l}|i\rangle_{L}|j\rangle_{R}}.
\eeq
where the Gram matrix $G$ accounts for overcompleteness and non-orthogonality of the shell basis. The boundary condition for the path integral computing  $G^n_{2s}$ corresponds to the product of $n$ two-sided shell state overlaps $\braket{\mathbf{j}|{\mathbf{i}}}$. Each overlap involves a periodic boundary with 
$\mathcal{O}_{2s,\mathbf{i}}$ and $\mathcal{O}^{\dagger}_{2s,\mathbf{j}}$ insertions separated by $\tilde{\beta}_L,\tilde{\beta}_R$ boundary time (see Fig.~\ref{fig:shell_bdry} and Appendix~\ref{sec:2Sss})  while the  $\langle j|_{L}\langle l|_{R} |\mathbf{k}\rangle_{2s}$ insertions corresponds to the mixed-shell boundary condition in Fig.~\ref{fig:tensorspanbc}.

\begin{figure}
    \centering
    \begin{subfigure}[c]{\linewidth}
    \centering
    \includegraphics[width=0.6\linewidth]{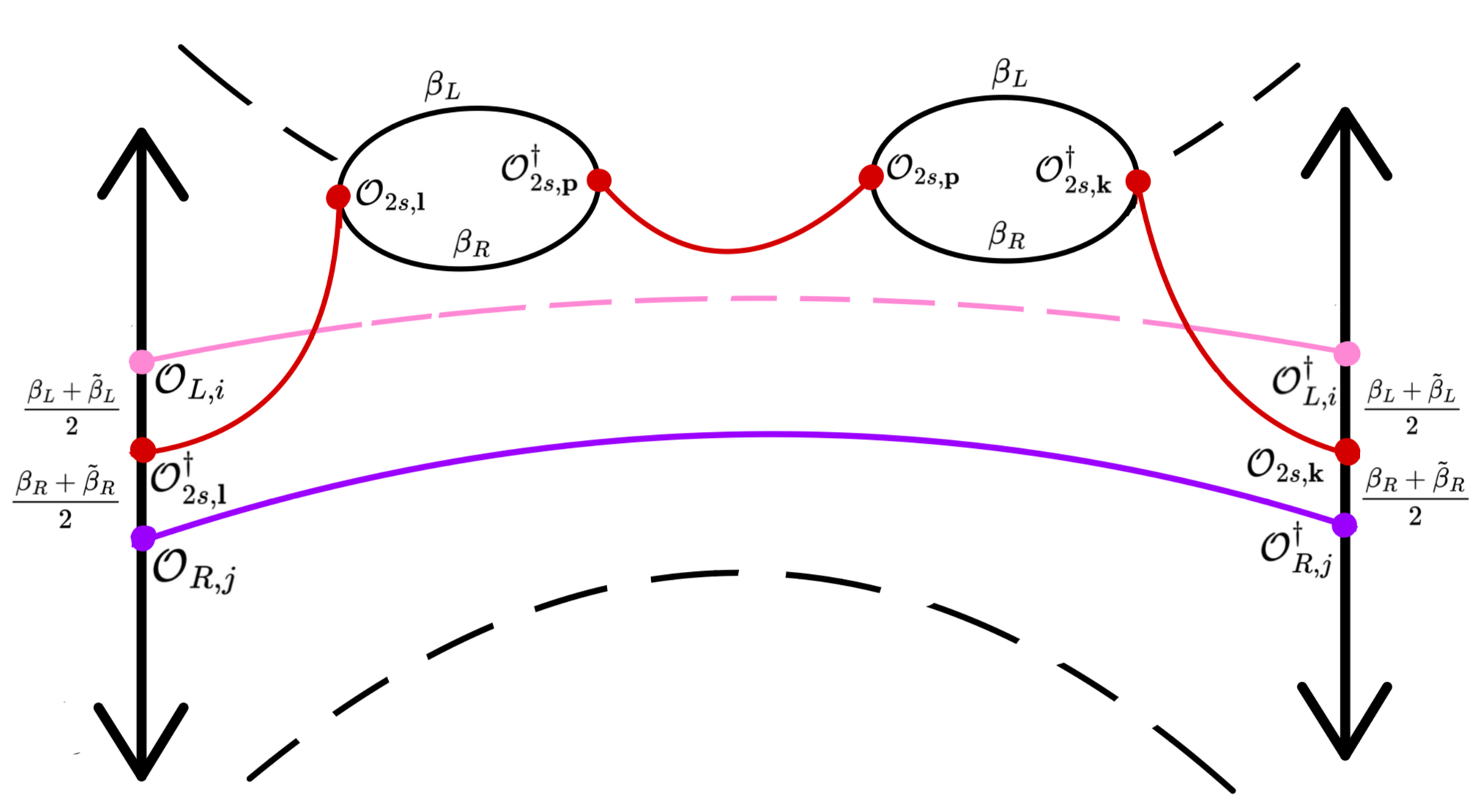}
    \caption{}
    \label{fig:2s_span_tensor_sc}
    \end{subfigure}
    \hfill
    \begin{subfigure}[c] {\linewidth}
    \centering
        \includegraphics[width=0.7\linewidth]{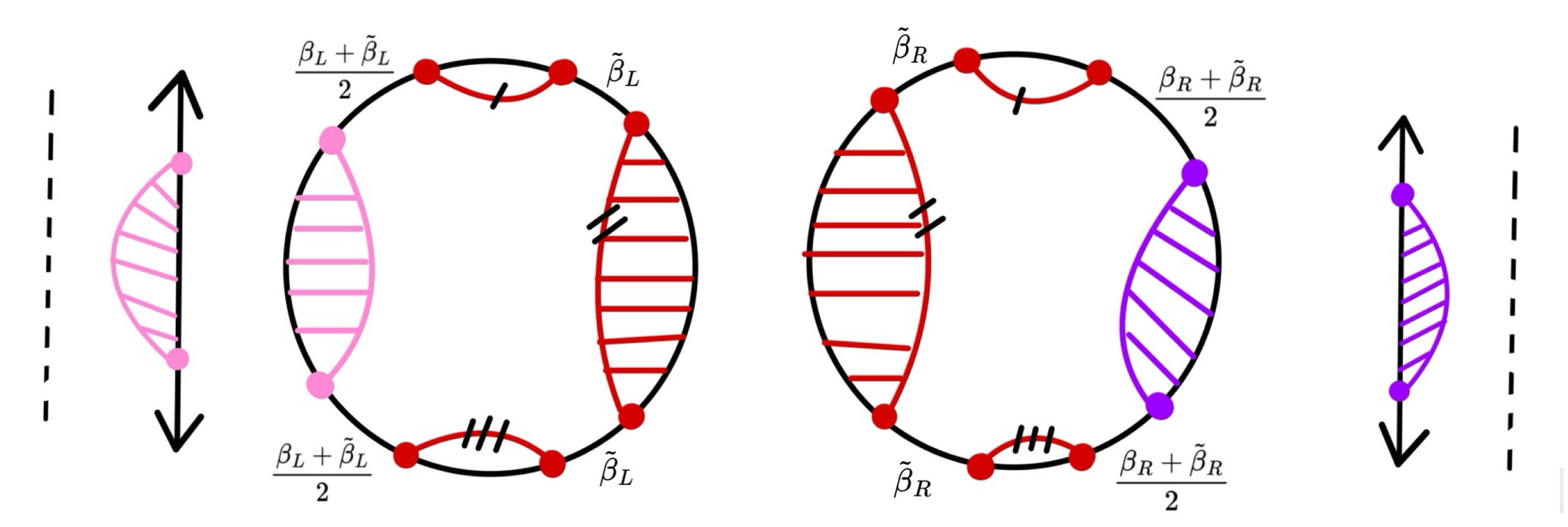}
        \caption{}
        \label{fig:2s_span_tensor_sc_const}
    \end{subfigure}
    \caption{ ({\textbf a}) The fully connected contribution to (\ref{eq:2s_span_tensor_sc}) depicted for $n=2$. ({\textbf b}) These saddles are constructed by gluing an $L,R$ disk together along the worldvolume of $n+1$ two-sided shell wordvolumes and in addition gluing an $L,R$ strip into the respective disks along the single-sided shell insertions. }
\end{figure}

To evaluate (\ref{eq:2s_span_tensor_sc}) we take a limit where the number of 2-sided shells included in the basis becomes infinite ($\kappa_{2s} \to \infty$). In this limit 
only geometries with a maximal number of boundary index loops contribute  in path integral computations \cite{Toolkit}.  For (\ref{eq:2s_span_tensor_sc}) this means that the $n$ two-shell boundaries and the two mixed-shell boundaries must be connected through a single bulk (Fig.~\ref{fig:2s_span_tensor_sc}).  Saddle points satisfying this boundary condition can be constructed by starting with  $L,R$ disks, each containing $n+1$ two-sided shell insertions and one $L,R$ single-sided shell insertion respectively.  An $L,R$ shell strip is glued into the $L,R$ disk along the single-sided shell insertions, and the two disks are then glued together along the remaining $n+1$ two-shell worldvolumes (Fig.~\ref{fig:2s_span_tensor_sc_const}). In the large single- and two-sided shell mass limit the shell homology regions pinch off, so that the shells contribute universally to the action. The wormhole saddle is then:
$ \overline{S}_L(0) \times \overline{S}_R(0) \times \overline{Z}(\beta_L+(n+1)\tilde{\beta}_L)\times\overline{Z}(\beta_R+(n+1)\tilde{\beta}_R)\times Z_{m_i,L}\times Z_{m_j,R}\times\prod_{k=1}^{n+1}Z_{m_k,2s}$. 
After normalizing the two-sided shell states, this wormhole saddle is independent of the shell indices running in the loop and we therefore obtain $\overline{G^n_{2s,\mathbf{kl}} \langle i|_{L} \langle j|_{R} |\mathbf{k}\rangle_{2s}\langle _{2s} \mathbf{l}|i\rangle_{L}|j\rangle_{R}}=$
\beq \label{eq:2sspantensorwh}
\kappa^{n+1}\frac{\overline{S}_L(0) \times \overline{S}_R(0) \times \overline{Z}(\beta_L+(n+1)\tilde{\beta}_L)\times\overline{Z}(\beta_R+(n+1)\tilde{\beta}_R)\times Z_{m_i,L}\times Z_{m_j,R}}{\overline{Z}(\tilde{\beta}_L)^{n+1}\overline{Z}(\tilde{\beta}_R)^{n+1}}
\eeq
Finally, we take $n \to -1$ and find
\beq \label{eq:abc}
\overline{\langle i|_{L} \langle j|_{R} \prod_{\mathcal{H}_{2s}} |i\rangle_{L}|j\rangle_{R}}= \overline{S}_L(0) \overline{Z}(\beta_L) Z_{m_i,L}\overline{S}_R(0)\overline{Z}(\beta_R)Z_{m_j,R}=\overline{\langle i|i\rangle_{L} \langle j| j\rangle_{R}}
\eeq
showing that $\overline{\langle \Psi_{\otimes}|\prod_{\mathcal{H}_{2s}}|\Psi_{\otimes}\rangle}=\overline{\langle \Psi_{\otimes}|\Psi_{\otimes}\rangle}$ within the saddle approximation.

\paragraph{Fine grained} 
To extend this coarse-grained equality to a fine-grained one (\ref{eq:tensorinclusion}) we have to also show that \beq \label{eq:2sintensorsqr}
\overline{\left(\langle i|_{L} \langle j|_{R} \prod_{\mathcal{H}_{2s}}|i\rangle_{L}|j\rangle_{R}- \langle i|i\rangle_{L} \langle j| j\rangle_{R}\right)^2}=0. 
\eeq 
The projector onto two-sided shell states $\prod_{\mathcal{H}_{2s}}$ is defined via an $n \to -1$ limit as above, and we will show that each term in (\ref{eq:2sintensorsqr}) limits to $\overline{\left(\langle i|i\rangle_{L} \langle j| j\rangle_{R}\right)^2}$. Note first that $\overline{\left(\langle i|i\rangle_{L} \langle j| j\rangle_{R}\right)^2}= \overline{\langle i|i\rangle_{L} \langle j| j\rangle_{R}\langle i|i\rangle_{L} \langle j| j\rangle_{R}}=\overline{\langle i|i\rangle_{L} \langle i|i\rangle_{L}} \times \overline{\langle j| j\rangle_{R}\langle j| j\rangle_{R}} $ where the final equality follows from the discussion around (\ref{eq:tensor_norm}): $L,R$ shell states are chosen to all have different mass, hence the wormholes connecting $L$ to $R$ shells are disallowed. Even so, each factor in $\overline{\langle i|i\rangle_{L} \langle i|i\rangle_{L}} \times \overline{\langle j| j\rangle_{R}\langle j| j\rangle_{R}}$ has connected and disconnected saddles as constructed in Sec.~\ref{sec:1sWH}. For example, for unnormalised shell states, we have 
\beq \label{eq:sqrtsingleside}
\overline{\langle j| j\rangle_{R}\langle j| j\rangle_{R}}= Z_{m_j,R}^2 \overline{S}(0)^2(\overline{Z}(2\beta_R) + \overline{Z}(\beta_R)^2) \eeq where the first term comes from the connected saddles and the second from the disconnected ones, and similarly for $L$.

Now consider, for example, the cross term in (\ref{eq:2sintensorsqr}). By using shell states as above to resolve the projector we obtain
\beq \label{eq:2s_span_tensor_sqr}
\overline{\langle i|_{L} \langle j|_{R} \prod_{\mathcal{H}_{2s}}|i\rangle_{L}|j\rangle_{R}\times\langle i|i\rangle_{L} \langle j| j\rangle_{R}}= \lim_{n \to -1 } \overline{G^n_{2s,\mathbf{kl}} \langle i|_{L} \langle j|_{R} |\mathbf{k}\rangle_{2s}\langle _{2s} \mathbf{l}|i\rangle_{L}|j\rangle_{R}\langle i|i\rangle_{L} \langle j| j\rangle_{R}}.
\eeq
We must consider the $\kappa_{2s} \to \infty$ limit in which the two-sided shell states span, and hence the two-sided shell index loop in $G^n_{2s,\mathbf{kl}}|\mathbf{k}\rangle_{2s}\langle _{2s}\mathbf{l}|$ cannot be broken, as explained in \cite{Toolkit}. As (\ref{eq:2s_span_tensor_sqr}) contains additional  $\ket{i}_L,\ket{j}_R$ shell insertions compared to (\ref{eq:2s_span_tensor_sc}), there are then two types of saddlepoint for (\ref{eq:2s_span_tensor_sqr}), depending on wether the single sided shell insertions are connected or not.  Firstly there are saddles in which the  
$\langle i|i\rangle_{L} \langle j| j\rangle_{R}$ and $\langle 
i|_{L} \langle j|_{R} 
\prod_{\mathcal{H}_{2s}}|i\rangle_{L}|j\rangle_{R}$ boundary insertions are not connected, so that the path integral gives a product of the contributions from each term separately. It was shown in (\ref{eq:abc}) that $\overline{\langle i|_{L} \langle j|_{R} 
\prod_{\mathcal{H}_{2s}}|i\rangle_{L}|j\rangle_{R}}= \overline{\langle i|i\rangle_{L} \langle j| j\rangle_{R}}$. Hence these disconnected contributions give 
contributions of $\overline{\langle i|i\rangle_{L} \langle j| j\rangle_{R}} \times \overline{\langle i|i\rangle_{L} \langle j| j\rangle_{R}}$, which are exactly the disconnected contributions to $\overline{\left(\langle i|i\rangle_{L} \langle j| j\rangle_{R}\right)^2}$. 

\begin{figure}[h]
\begin{subfigure}[c]{\linewidth}
    \centering
    \includegraphics[width=0.6\linewidth]{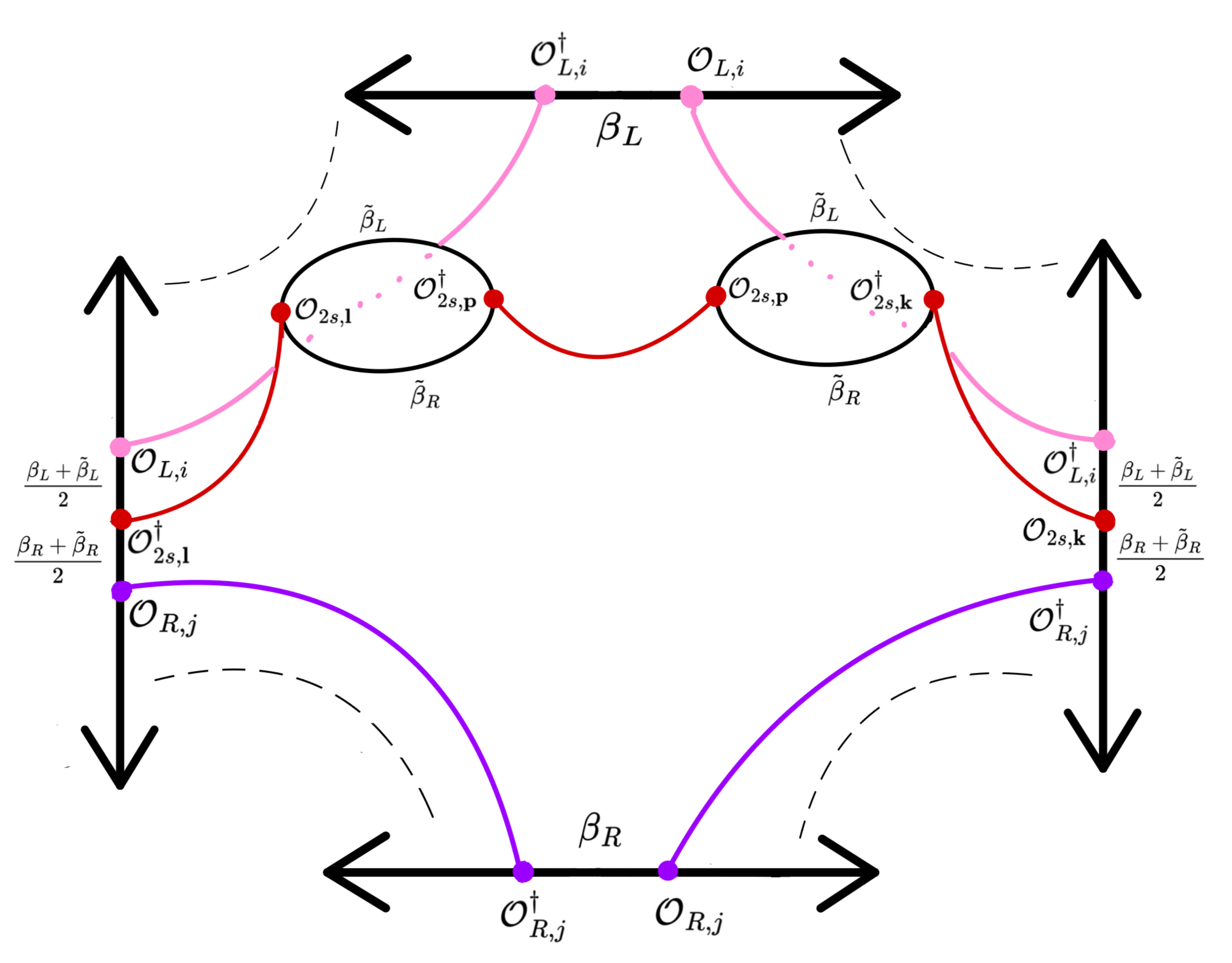}
    \caption{}
    \label{fig:sqrWh}
    \end{subfigure}
    \begin{subfigure}[c]{\linewidth}
    \centering
       \includegraphics[width=0.8\linewidth]{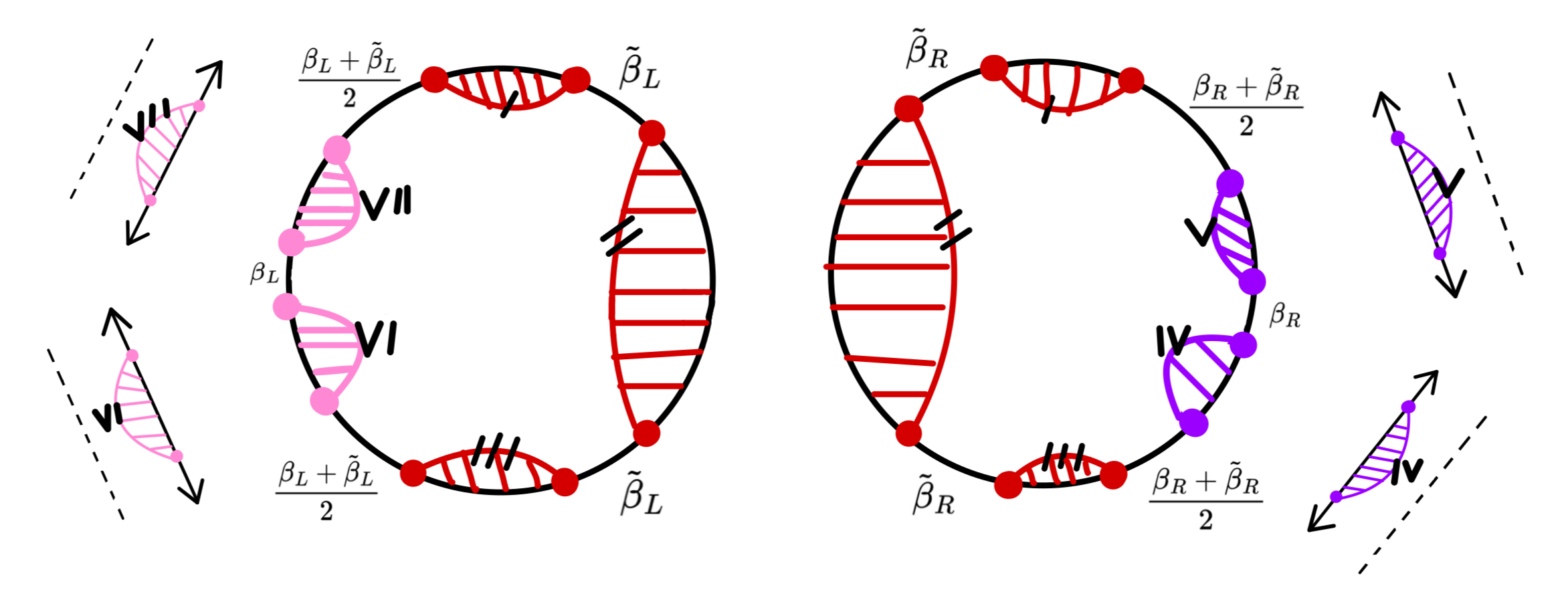}
    \caption{}
    \label{fig:sqrWh2}
    \end{subfigure}
    \caption{({\textbf a}) The fully connected contribution to $ \overline{G^n_{2s,\mathbf{kl}} \langle i|_{L} \langle j|_{R} |\mathbf{k}\rangle_{2s}\langle _{2s} \mathbf{l}|i\rangle_{L}|j\rangle_{R}\langle i|i\rangle_{L} \langle j| j\rangle_{R}}$ depicted for $n=2$. ({\textbf b}) These saddles are constructed by gluing an $L,R$ disk together along the worldvolume of $n+1$ two-sided shell wordvolumes and in addition gluing two $L,R$ strip seperates by $\beta_{L,R}$ into the respective $L,R$ disks along the single-sided shell insertions.}
\end{figure}

 Second, there are saddles connecting the different  $\ket{i}_L,\ket{j}_R$ shell insertions (Fig.~\ref{fig:sqrWh}). These wormhole saddles are constructed analogously to those for (\ref{eq:2s_span_tensor_sc}) by including an additional $i$- shell on the $L$ disk separated by boundary time $\beta_L$ from the original $i$- shell then  gluing in another shell strip along the added $i$ shell, and similarly for the $R$ disk (Fig.~\ref{fig:sqrWh2}). Thus, in the large shell mass limit, after normalizing the two-sided  states, the wormhole saddle contributes:
\beq \label{eq:LRin2sWH}
\frac{\overline{S}_L(0)^2 \times Z_{m_i,L}^2 \times \overline{Z}(2\beta_L+(n+1)\tilde{\beta}_L) \times\overline{S}_R(0)^2 \times Z_{m_j,R}^2 \times \overline{Z}(2\beta_R+(n+1)\tilde{\beta}_R)}{\overline{Z}(\tilde{\beta}_L)^{n+1}\overline{Z}(\tilde{\beta}_R)^{n+1}} \, .
\eeq
In the $n \to -1$ limit this is the same as the wormhole contribution  to $\overline{\left(\langle i|i\rangle_{L} \langle j| j\rangle_{R}\right)^2}|_{conn}=\overline{\langle i|i\rangle_{L} \langle i|i\rangle_{L}}|_{conn} \times \overline{\langle j| j\rangle_{R}\langle j| j\rangle_{R}}|_{conn}$ where as above $\overline{\langle j| j\rangle_{R}\langle j| j\rangle_{R}}|_{conn}= Z_{m_j,R}^2 \overline{S}(0)^2 \overline{Z}(2\beta_R)$.   Putting everything together we get $\overline{\langle i|_{L} \langle j|_{R} \prod_{\mathcal{H}_{2s}}|i\rangle_{L}|j\rangle_{R}\langle i|_{L} \langle j|_{R} |i\rangle_{L}|j\rangle_{R}}=\overline{\left(\langle i|i\rangle_{L} \langle j| j\rangle_{R}\right)^2}$ as all saddles are accounted for on both sides of the equality. Similarly, 
after accounting for the analogous connected and disconnected saddles,  we find  
\beq
\overline{\langle i|_{L} \langle j|_{R} \prod_{\mathcal{H}_{2s}}|i\rangle_{L}|j\rangle_{R}\langle i|_{L} \langle j|_{R} \prod_{\mathcal{H}_{2s}} |i\rangle_{L}|j\rangle_{R}} = \overline{\left(\langle i|i\rangle_{L} \langle j| j\rangle_{R}\right)^2} \, .
\eeq 
Hence (\ref{eq:2sintensorsqr})  holds, at least within the sum over  saddle points.

The above shows $\overline{\langle \Psi_{\otimes}|\prod_{\mathcal{H}_{2s}}|\Psi_{\otimes}\rangle}=\overline{\langle \Psi_{\otimes}|\Psi_{\otimes}\rangle}$ and $\overline{(\langle \Psi_{\otimes}|\prod_{\mathcal{H}_{2s}}|\Psi_{\otimes}\rangle-\langle \Psi_{\otimes}|\Psi_{\otimes}\rangle)^2}=0$, within the saddlepoint approximation to the path integral.
As the single- and two-sided shell states form a basis for $\mathcal{H}_{\mathcal{B}_L} \otimes \mathcal{H}_{\mathcal{B}_R}$ and $\mathcal{H}_{LR}$ respectively, this establishes the fine-grained equality (\ref{eq:tensorinclusion}) and therefore $\mathcal{H}_{\mathcal{B}_L} \otimes \mathcal{H}_{\mathcal{B}_R} \subseteq \mathcal{H}_{LR}$. We extend this argument to hold as full path integral equalities by using the surgery tool of the toolkit in \cite{Toolkit} in Appendix~\ref{allO2sspan1s}.

\subsection{Tensor product shell states span two-sided shell states  } \label{sec:1s_span_2s}
Similarly we will establish $\mathcal{H}_{2s} \subseteq  \mathcal{H}_L\otimes \mathcal{H}_R$ by showing (\ref{eq:2sinclusion}) for arbitrary $|\Phi_{2s}\rangle$ two-sided shell states, say $|\mathbf{p}\rangle_{2s}$.
This shows (\ref{eq:2sinclusion}) for any state in $\mathcal{H}_{LR}$ because the 2-sided shell states form a basis (as shown in \cite{Toolkit}). Thus we want to use the path integral to show that
\begin{eqnarray}
\label{eq:overlineqtensorspan}
\overline{\langle \mathbf{p}| \prod_{\mathcal{H}_L\otimes \mathcal{H}_R}| \mathbf{p}\rangle_{2s}}- \overline{\langle \mathbf{p}| \mathbf{p}\rangle}_{2s}
&=& 0 
\\
 \label{eq:sqroverlineqtensorspan}
\overline{\left(\langle \mathbf{p}| \prod_{\mathcal{H}_L\otimes \mathcal{H}_R}| \mathbf{p}\rangle_{2s}- \langle \mathbf{p}| \mathbf{p}\rangle_{2s}\right)^2}&=&0,
\end{eqnarray}
for any  $|\mathbf{p}\rangle_{2s}$.

\begin{figure}
    \centering
    \begin{subfigure}{\linewidth}
        \centering
 \includegraphics[width=0.75\linewidth]{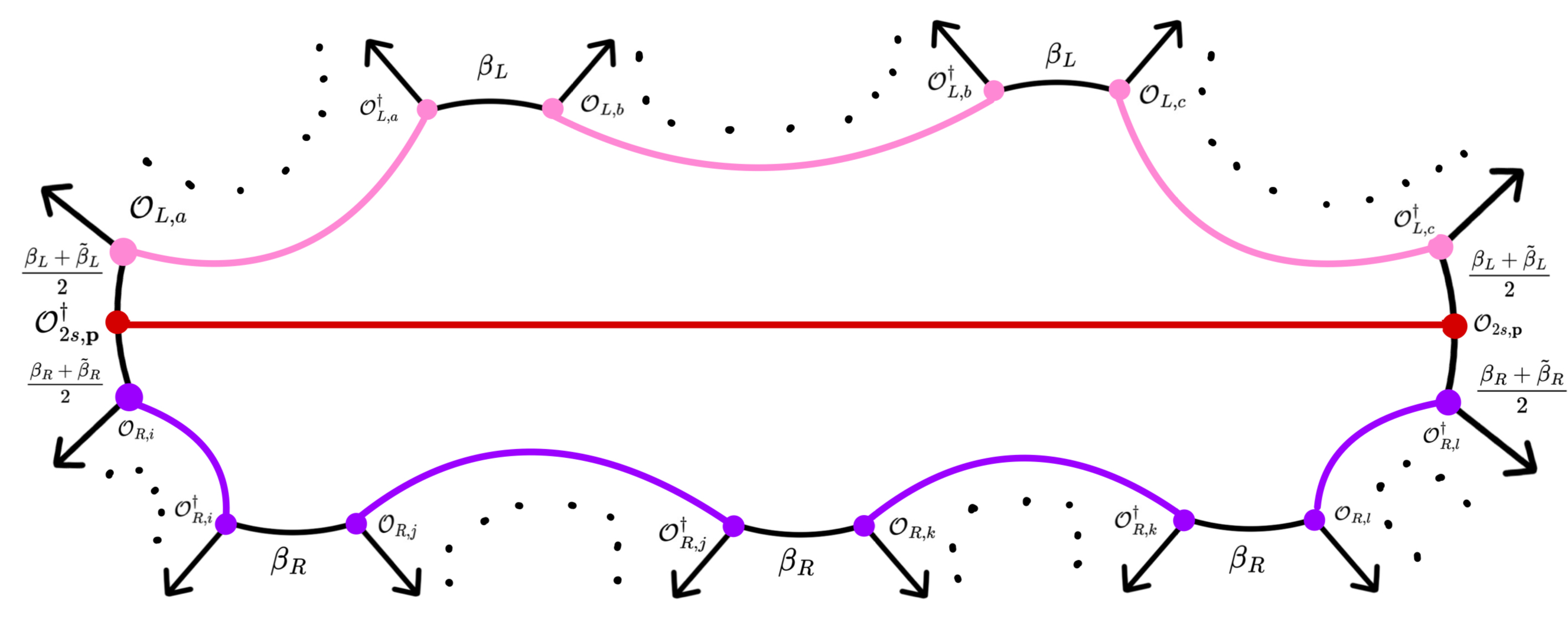}
    \caption{}
    \label{fig:tensorspanbcloop}
    \end{subfigure}
        \hfill
      \begin{subfigure}{\linewidth}
      \centering
    \includegraphics[width=0.7\linewidth]{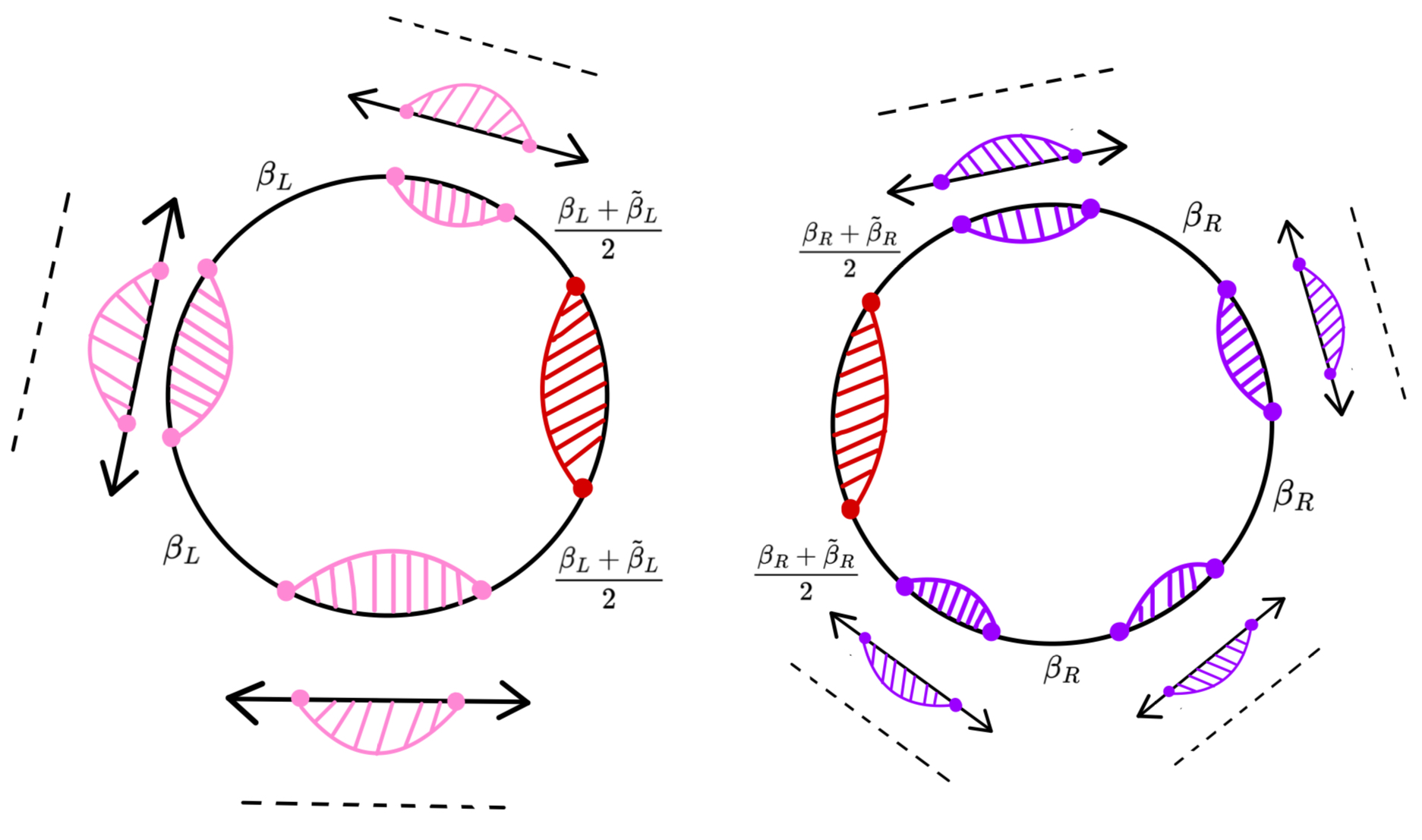}
    \caption{}
    \label{fig:tensor_span_2s_sc}
\end{subfigure}
\caption{({\textbf a}) The fully connected contribution to  (\ref{eq:1s_span_2s}) depicted for $n=2,m=3$. ({\textbf b}) These saddles are constructed by gluing an $L,R$ disk together along the worldvolume of two-sided $\mathbf{p}$-shell wordvolume and in addition gluing $n,m$ strips into the respective $L,R$ disks along the single-sided shell insertions.}
\end{figure}

As the Gram matrix of overlaps for  $\mathcal{H}_L\otimes \mathcal{H}_R$ is the tensor product $G_{L\otimes R} = G_{L} \otimes G_{R}$ the projector onto this space is simply given by 
 $\prod_{\mathcal{H}_L}\prod_{\mathcal{H}_R}$:
\beq \label{eq:1s_span_2s}
\overline{\langle \mathbf{p}| \prod_{\mathcal{H}_L\otimes \mathcal{H}_R}| \mathbf{p}\rangle_{2s} }=\lim_{n,m \to -1}\overline{ G^{n}_{L,ac}G^{m}_{R,il} \langle \mathbf{p}|_{2s} a\rangle_{L} |i\rangle_{R}   \langle c|_{L}\langle l|_{R} |\mathbf{p}\rangle_{2s}}.
\eeq
From Sec.~\ref{sec:1s-span}, path integrals computing $G^n_{L}$ require $n$ shell strip boundaries of length $\beta_L$ (similarly for $G^m_{R}$), and   $\langle j|_{L}\langle l|_{R} |\mathbf{p}\rangle_{2s}$ requires the mixed-shell boundary defined above (Fig.~\ref{fig:tensorspanbc}). 
Also, as we discussed, when we take $\kappa_L, \kappa_R \to \infty$, the single-sided shell states span $\mathcal{H}_{\mathcal{B}_L}\otimes \mathcal{H}_{\mathcal{B}_R}$, and only geometries with a maximal number of shell index loops contribute to the path integral. These geometries must connect the $n+m$ asymptotic strip boundaries with the two mixed-shell boundary in a single bulk (Fig.~\ref{fig:tensorspanbcloop}). Such wormholes are constructed by gluing an $L,R$ disk together along the $\mathcal{O}_{2s,p}$ shell and gluing $n,m$ strips into the $L,R$ disk long the $L,R$ single-sided shell insertions respectively (Fig.~\ref{fig:tensor_span_2s_sc}). Thus, in the large shell mass limit the wormhole saddle contribution to (\ref{eq:1s_span_2s}) is:
\beq
 \overline{Z}(\tilde{\beta}_L+(n+1)\beta_L)\overline{Z}(\tilde{\beta}_R +(m+1)\beta_R)\overline{S}_{L}(0)^{n+1}\overline{S}_{R}(0)^{m+1} Z_{m_\mathbf{p},2s}\prod_{k=1}^{n+1}Z_{m_k,L} \prod_{k=1}^{m+1}Z_{m_k,R} \, .
\eeq
After normalizing the $L,R$ single-sided shell states this  becomes independent of the shell indices running in the loop and we have :
\beq \label{eq:45}
 \overline{ G^{n}_{L,ac}G^{m}_{R,il} \langle \mathbf{p}|_{2s} a\rangle_{L} |i\rangle_{R}   \langle c|_{L}\langle l|_{R} |\mathbf{p}\rangle_{2s}}= \kappa^{n+1}\frac{\overline{Z}(\tilde{\beta}_L+(n+1)\beta_L)\overline{Z}(\tilde{\beta}_R +(m+1)\beta_R) Z_{m_\mathbf{p},2s}}{\overline{Z}(\beta_L)^{n+1}\overline{Z}(\beta_R)^{m+1}} \, .
\eeq
Taking $m,n \to -1$ we recover the norm (\ref{eq:2s_shell_WH}) of the two-sided shell state $\ket{\mathbf{p}}_{2s}$:
\beq
\overline{Z}(\tilde{\beta}_L)\overline{Z}(\tilde{\beta}_R)Z_{m_\mathbf{p},2s}=\overline{\langle \mathbf{p}| \mathbf{p}\rangle}_{2s},
\eeq 
thereby deriving (\ref{eq:overlineqtensorspan}).

\paragraph{Fine grained}To arrive at the fine-grained statement (\ref{eq:2sinclusion}) we have to also show (\ref{eq:sqroverlineqtensorspan}). The logic here closely resembles that of the fine-grained result in Sec.~\ref{sec:2s_span_1s} above. Once again we show that each term in (\ref{eq:sqroverlineqtensorspan}) limits to $\overline{\braket{\mathbf{p}|\mathbf{p}}_{2s}^2} = Z_{m_{\mathbf{p}}}^2 \left( \overline{Z}(\tilde{\beta}_L)^2\overline{Z}(\tilde{\beta}_R)^2 +\overline{Z}(2\tilde{\beta}_L)\overline{Z}(2\tilde{\beta}_R )\right)$ where the first term corresponds to the disconnected contribution $\overline{\braket{\mathbf{p}|\mathbf{p}}_{2s}}\times\overline{\braket{\mathbf{p}|\mathbf{p}}_{2s}} $ and the latter the wormhole contribution, see (\ref{eq:2s_WH_nonormal}) and Appendix.~\ref{sec:2Sss}. 

\begin{figure}[h]
\centering
\begin{subfigure}{0.45\linewidth}
    \centering
    \includegraphics[width=0.8\linewidth]{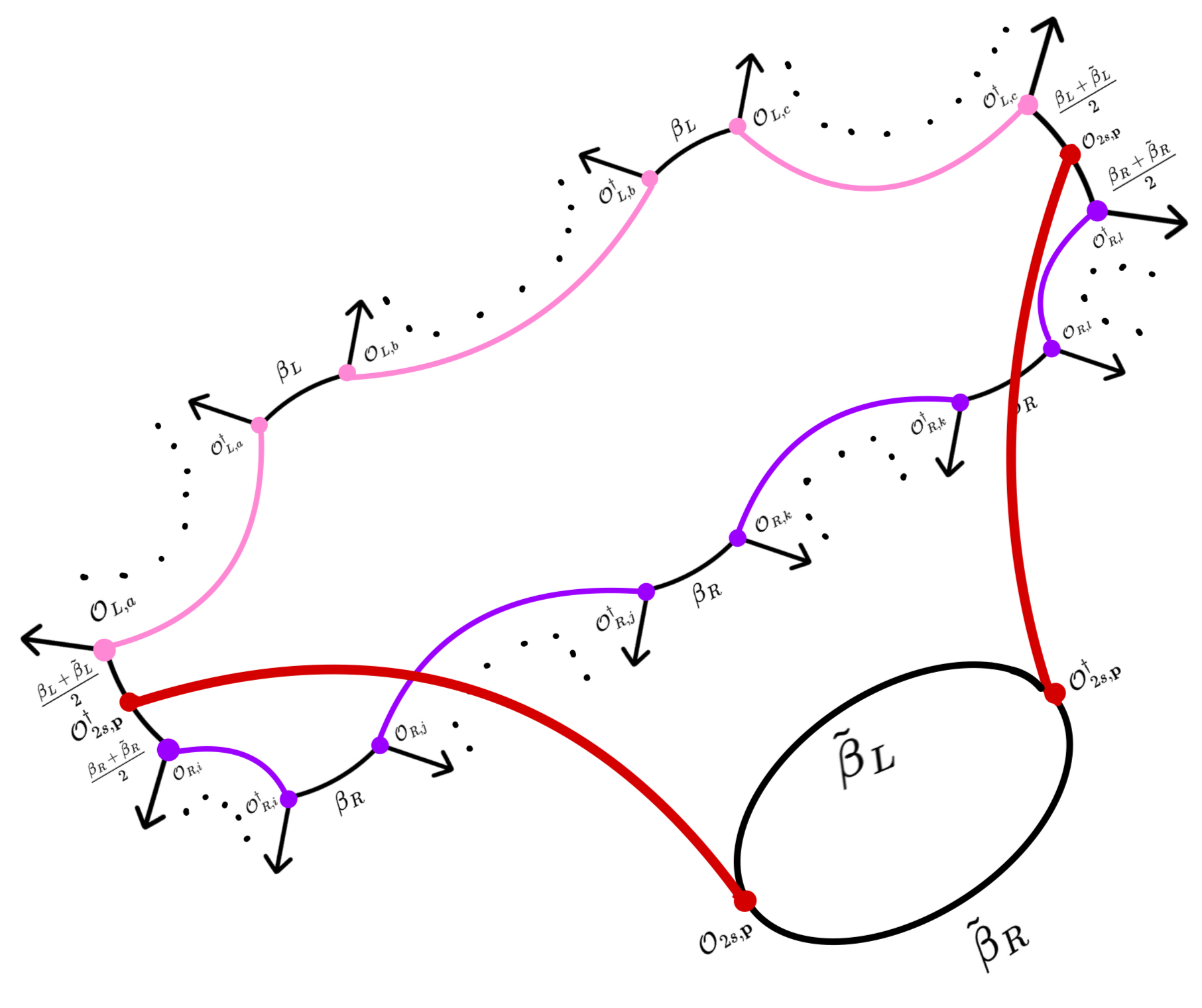}
    \caption{}
    \label{fig:sqrsaddles}
\end{subfigure}
\hfill
\begin{subfigure}{0.45\linewidth}
    \centering
    \includegraphics[width=1.2\linewidth]{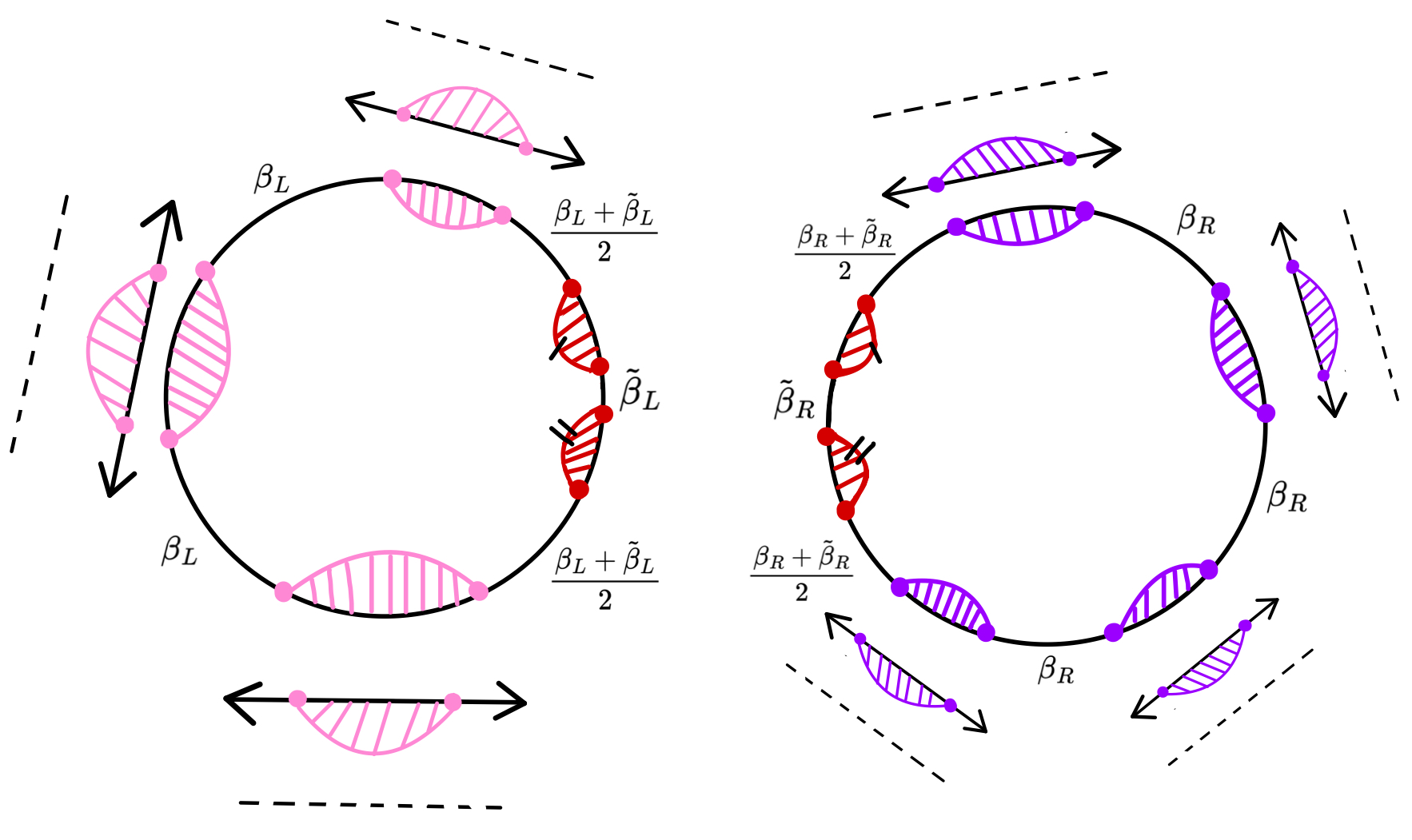}
    \caption{}
    \label{fig:sqrtfullconsaddle}
\end{subfigure}
\caption{({\bf a}) Fully connected 
saddle contributing to (\ref{eq:tensorspansqr}). ({\bf b}) These saddles are constructed by gluing an $L,R$ disk together along the worldvolumes of two copies of the two-sided $\mathbf{p}$-shell wordvolume and in addition gluing $n,m$ strips into the respective $L,R$ disks along the single-sided shell insertions.}
\end{figure}

 Consider for example, the cross term:
\beq \label{eq:crossterm}
\overline{\langle \mathbf{p}| \prod_{\mathcal{H}_L\otimes \mathcal{H}_R}| \mathbf{p}\rangle_{2s}\langle \mathbf{p}| \mathbf{p}\rangle_{2s}}=\lim_{n,m \to -1}\overline{ G^{n}_{L,ac}G^{m}_{R,il} \langle \mathbf{p}|_{2s} a\rangle_{L} |i\rangle_{R}   \langle c|_{L}\langle l|_{R} |\mathbf{p}\rangle_{2s} \langle\mathbf{p}| \mathbf{p}\rangle_{2s}} \, .
\eeq 
Recall that we require the $\kappa_{L,R} \to \infty$ limit in order for the single-sided shell states to span. Hence the single-sided shell index loops cannot be broken \cite{Toolkit}. Thus there are again two classes of saddle; ones in which the multiple $\mathbf{p}$ insertions are connected and ones in which they are not. The disconnected saddles correspond to $\overline{\langle \mathbf{p}| \prod_{\mathcal{H}_L\otimes \mathcal{H}_R}| \mathbf{p}\rangle_{2s}}\times \overline{\langle \mathbf{p}| \mathbf{p}\rangle_{2s}}$, which we have shown around (\ref{eq:45}) to equal $\overline{\langle \mathbf{p}| \mathbf{p}\rangle_{2s}}\times \overline{\langle \mathbf{p}| \mathbf{p}\rangle_{2s}}$. The fully-connected saddles for (\ref{eq:crossterm}) (see Fig.~\ref{fig:sqrsaddles}) account for the connected contribution to $\overline{\braket{\mathbf{p}|\mathbf{p}}_{2s}^2}$. 
The construction of these fully connected wormhole saddles is analogous to that of the saddles for (\ref{eq:1s_span_2s}) apart from the addition of an extra identified $\mathbf{p}$-shell worldvolume on the $L,R$ disks separated by $\tilde{\beta}_{L,R}$ from the other $\mathbf{p}$-shell insertion (see Fig.~\ref{fig:sqrtfullconsaddle}). Upon gluing the $L,R$ disks together, this extra identification results in the additional $\langle \mathbf{p}| \mathbf{p}\rangle_{2s}$ asymptotic boundary. 

 Hence in the large shell mass limit, after normalising the shell states, these connected saddles contribute 
\beq \label{eq:tensorspansqr}
\overline{ G^{n}_{L,ac}G^{m}_{R,il} \langle \mathbf{p}|_{2s} a\rangle_{L} |i\rangle_{R}   \langle c|_{L}\langle l|_{R} |\mathbf{p}\rangle_{2s} \langle\mathbf{p}| \mathbf{p}\rangle_{2s}}|_{conn}=\frac{\overline{Z}(2\tilde{\beta}_L+(n+1)\beta_L)\overline{Z}(2\tilde{\beta}_R +(m+1)\beta_R)Z_{m_\mathbf{p},2s}^2}{\overline{Z}(\beta_L)^{n+1}\overline{Z}(\beta_R)^{m+1}} \, .
\eeq
In the $m,n \to -1$ limit this recovers the connected contribution to the squared norm of the un-normalised $\ket{\mathbf{p}}$ state
$\overline{Z}(2\tilde{\beta}_L)\overline{Z}(2\tilde{\beta}_R )Z_{m_\mathbf{p},2s}^2 = \overline{(\langle \mathbf{p}| \mathbf{p}\rangle)^2}_{2s}|_{conn}$ (see Appendix.~\ref{sec:2Sss}).  Again, $\overline{\langle \mathbf{p}| \prod_{\mathcal{H}_L\otimes \mathcal{H}_R}| \mathbf{p}\rangle_{2s}\langle \mathbf{p}| \prod_{\mathcal{H}_L\otimes \mathcal{H}_R}| \mathbf{p}\rangle_{2s}}=\overline{(\langle \mathbf{p}| \mathbf{p}\rangle)^2}_{2s}$ follows by similar logic. Putting everything together, we have derived the fine grained result (\ref{eq:2sinclusion}), and therefore $\mathcal{H}_{LR} \subseteq  \mathcal{H}_{\mathcal{B}_L}\otimes \mathcal{H}_{\mathcal{B}_R}$, within the saddlepoint approximation.  Using the surgery approach of the toolkit in \cite{Toolkit} the argument  can again be extended to hold for the full path integral (see Appendix~\ref{allOtensorspan2s}).

\paragraph{Conclusion}
Above we have shown that $\mathcal{H}_L \otimes \mathcal{H}_R =\mathcal{H}_{2s}$.   These Hilbert spaces coincide with 
$\mathcal{H}_{\mathcal{B}_L}\otimes \mathcal{H}_{\mathcal{B}_R}$ and $\mathcal{H}_{LR}$ respectively, so $\mathcal{H}_{LR} =\mathcal{H}_{\mathcal{B}_L}\otimes \mathcal{H}_{\mathcal{B}_R}$. This resolves the factorization problem (\ref{eq:facprob}) because  the total two-sided Hilbert space is $\mathcal{H}_{L\cup R}=\mathcal{H}_{\mathcal{B}_L}\otimes \mathcal{H}_{\mathcal{B}_R}  \, \cup \, \mathcal{H}_{LR}$ and these two sets in the union are the same. 

Interestingly, wormhole geometries connecting mixed  single- and two- sided shell state boundaries were crucial in deriving this result. 
The perturbative gravity Hilbert space obtained in the ${G_\mathrm{N}}\to 0$ limit does not factorise because such wormholes  are not included. This is consistent with the observation that the $L,R$ von Neumann algebras of observables, $\mathcal{A}_{L,R}$, describing low-energy back reaction around the two-sided black hole saddle are of type III \cite{Leutheusser:2021qhd,Leutheusser:2022bgi}. Such algebras do not have well-defined traces, states or density matrices. Interestingly, after imposing the gravitational constraints the resulting crossed product $L,R$ algebras are of type II \cite{Chandrasekaran:2022eqq,Chandrasekaran:2022cip,Witten:2021unn}. While type II algebras contain density matrices and traces up to constants, they do not contain any pure states and do not have a representation on a tensor product Hilbert space $\mathcal{H}_{\mathcal{B}_L}\otimes \mathcal{H}_{\mathcal{B}_R} $. The tensor product factorisation (\ref{eq:facprob}) of the nonpertubative, fine-grained two-sided Hilbert space implies that the full nonperturbative $L,R$ algebras must be of type I, for which there exist a unique trace, pure states and tensor product representation. Furthermore, as the two-sided Hilbert space factorises into copies of the single-sided theory, these nonpertubative $L,R$ algebras must be unitarily equivalent to the nonperturbative single-sided operator algebra. 
This conclusion is exactly what is expected if the single-sided theory is described by a holographic dual theory on the asymptotic boundary, as these statements then follow trivially from the dual quantum mechanics. Note however that all the steps leading to the factorisation (\ref{eq:facprob}), are applicable in Euclidean path integral formulation of asymptotically flat quantum gravity too.

\subsection{Disentangling the shell states} \label{sec:twoshell}
The discussion above resolves the generalized version of the Hilbert space factorisation problem introduced in Sec.~\ref{sec:puzzles}. However, initial discussions of the factorisation problem were formulated in terms of an Einstein-Rosen bridge connecting  two asymptotic boundaries of a two-sided black hole \cite{Harlow:2015lma,Harlow:2018tqv,Boruch:2024kvv,Balasubramanian:2024yxk}. Recall that the geometry associated to a state created by cutting open the gravity path integral is determined by the leading saddle computing the norm. So we can only pose the  version of the factorization problem based on the presence of an Einstein-Rosen bridge in a  perturbative semiclassical limit where we keep the leading saddle and ignore the others. By contrast, above we have shown the factorisation (\ref{fig:factorizationproblem}) requires inclusion of non-perturbative saddle contributions that are omitted in this approximation. Indeed, as explained in Sec.~\ref{sec:puzzles}, the fact that some geometries computing the norm of a gravity state may analytically continue to Lorentzian wormholes is not in tension with factorization. Still, one might worry about the fact the geometries of some states in the intrinsically two-sided Hilbert spaces $\mathcal{H}_{LR}$ do not ``look'' like they arise from an underlying tensor product. 

Consider for example the two-sided shell basis for $\mathcal{H}_{LR}$ used above and reviewed in Appendix~\ref{sec:2Sss}. As discussed in \cite{Toolkit}, the semiclassical geometry associated to these states come in three classes, determined by wether the left and right preparation temperatures $\tilde{\beta}_L,\tilde{\beta}_R$ are above/below the Hawking-Page transition scale $\beta_{HP}$. The resulting geometry is either a two-sided black hole with a long wormhole interior, a single-sided black hole with disconnected copy of thermal AdS or two disconnected copies of thermal AdS entangled with a compact universe. The authors of \cite{Toolkit} showed that, non-perturbatively, each of these three sets provides a basis.
The latter two bases consist of disconnected geometries, and  may initially appear factorized because because they are not geometrically connected, but are not actually separable tunable.  For example, none of these states is a product of one-sded black holes.

In holographic settings the two-sided single-shell states map to partially entangled states (PETS) in the dual CFT \cite{Goel:2018ubv}. In general, it is hard to deduce the underlying factorizing structure of a tensor product Hilbert space with access only to such a basis of highly entangled states. Since we have derived that the two-sided Hilbert space must factorize, reducing the entanglement between the $L,R$ Hilbert space factors should produce a basis corresponding to explicitly factorized Lorentzian geometries. We will show that this is indeed the case. To do so we consider an alternative basis for the two-sided Hilbert space consisting of states with an additional shell operator inserted. The two-shell state $|\mathbf{i,j}\rangle_{2s}$ consists of an $O_i$ and $O_j$ shell insertions separated by a Euclidean time $\beta_m$:
\beq
 |\mathbf{i,j}\rangle_{2s}= |e^{-\frac{\tilde{\beta}_LH_L}{2}}\mathcal{O}_{\mathbf{i}}e^{\frac{-\beta_m H_R}{2}}\mathcal{O}_{\mathbf{j}}e^{\frac{-\tilde{\beta}_R H_R}{2}}\rangle.
\eeq
We have written a time evolution by $H_{R}$, but could equally have used $H_{L}$.  \begin{figure}
    \centering
    \includegraphics[width=\linewidth]{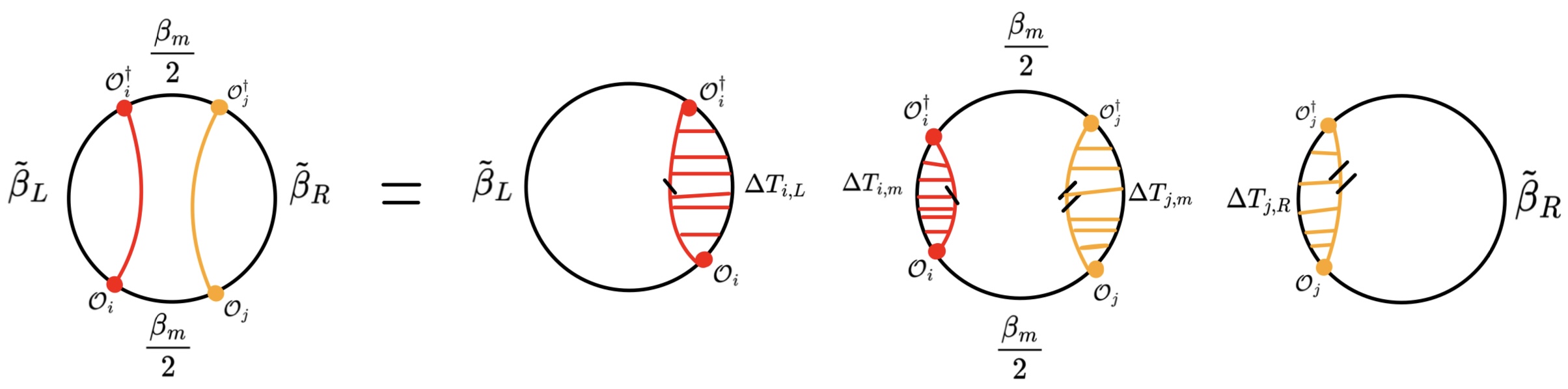}
    \caption{Construction of the saddle for $\overline{\langle \mathbf{i,j}| \mathbf{i,j} \rangle_{2s}}$: the $L$ and $R$ disks are glued together into the central disk by identifying the $\mathbf{i}$ and $\mathbf{j}$ shell worldvolumes. }
    \label{fig:s2-s2-saddle}
\end{figure}
If we have AdS boundary conditions, from the dual CFT perspective, varying $\beta_m$ amounts to tuning the entanglement between the $L,R$ Hilbert space factors. 
A simple adaptation of the argument that the single-shell states span $\mathcal{H}_{LR}$ given in \cite{Toolkit} shows that a $\kappa \to \infty$ set of the two shell states also form a basis of the full two-sided Hilbert space $\mathcal{H}_{LR}$. In particular, whenever  a geometry is glued together along the shell world-volume in the single-shell case the analogous operation for the two shell case is to glue together two distinct shell worldvolumes via an additional disk containing each shell separated by a boundary time $\beta_m/2$ on either side.  Compare for example Fig.~\ref{fig:s2-s2-saddle} with Fig.~\ref{fig:shell_norm} in Appendix~\ref{sec:2Sss}. In the large shell mass limit the only difference is the presence of additional factors of $\overline{Z}(\beta_m)$. In the completeness argument this will induce a factor $\lim_{n \to -1} \overline{Z}(\beta_m)^{n+1}$, leaving the conclusion unchanged.  Hence the two-shell states span $\mathcal{H}_{LR}$ regardless of the value of $\beta_m$.

\begin{figure}
    \centering
     \begin{subfigure}[c]{0.45\linewidth}
    \includegraphics[width=\linewidth]{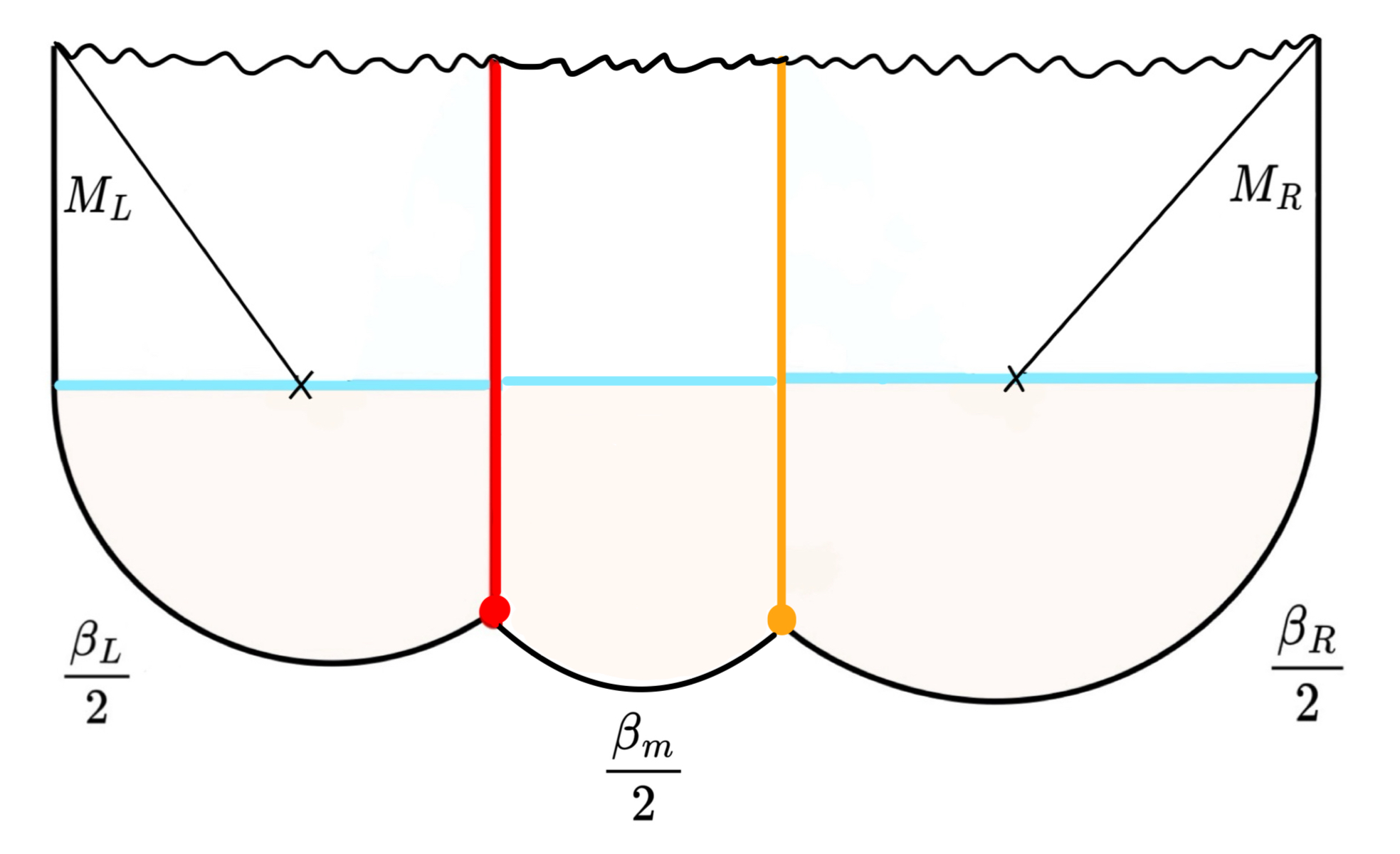}
    \caption{}
    \label{fig:long_WH_2s_2s_BH}
\end{subfigure}
\begin{subfigure}[c]{0.45\linewidth}
    \centering
    \includegraphics[width=\linewidth]{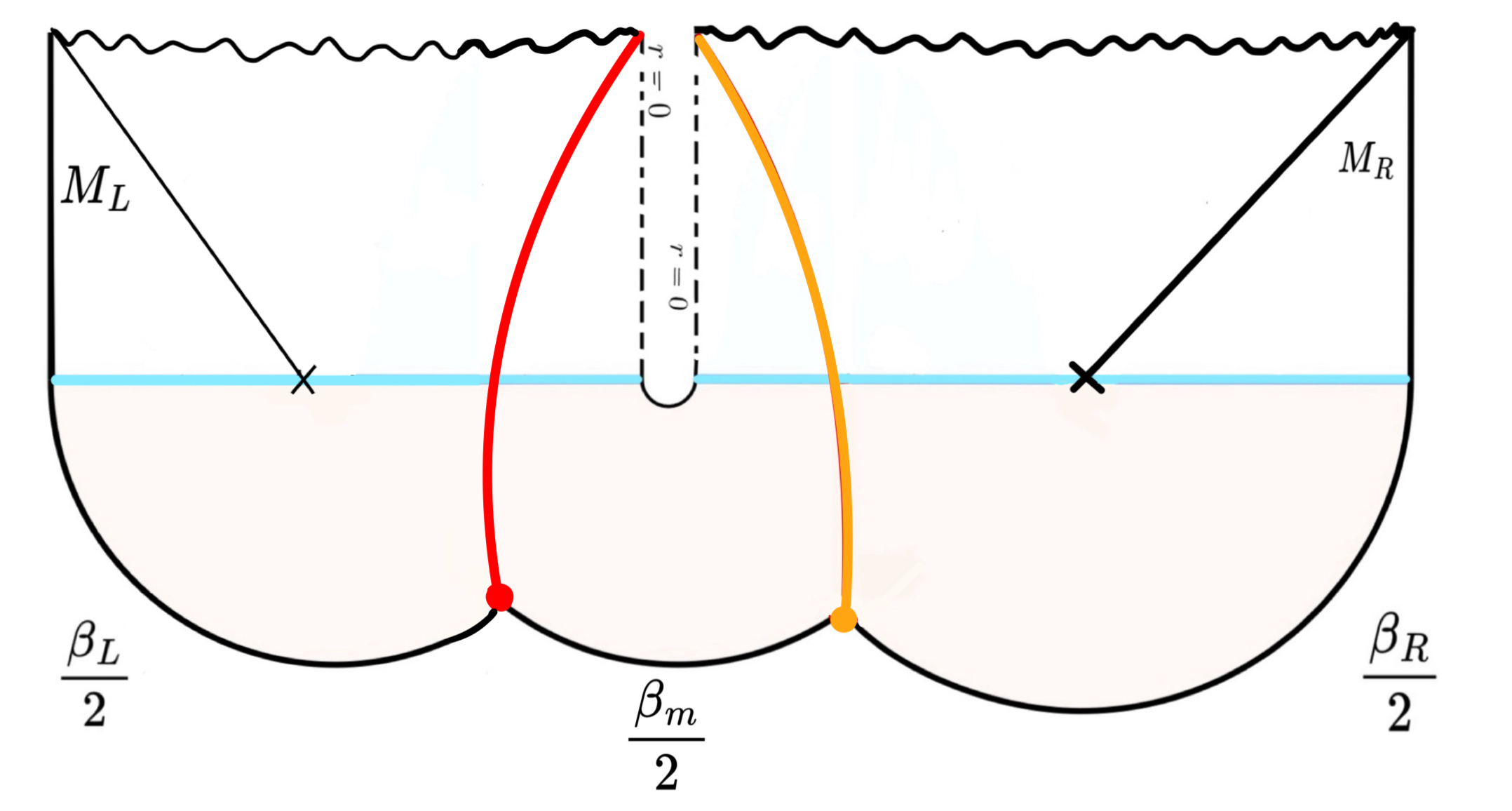}
    \caption{}
    \label{fig:disc_BH}
    \end{subfigure}
    \caption{({\bf a}) Lorentzian geometry for $\beta_L,\beta_R < \beta_{HP}$ and $\beta_m < \beta_{HP}$ corresponding to a two-sided AdS black hole with a long interior supported by the shell states. ({\bf b}) Lorentzian geometry for $\beta_L,\beta_R < \beta_{HP}$ and $\beta_m > \beta_{HP}$ corresponding to two disconnected AdS black holes.}
\end{figure}

Similarly, the saddle points computing the norms of the two-shell states are constructed analogously to the single-shell two-sided case. The single-shell saddles are constructed by gluing together portions of $L,R$ disks whereas the two shell saddles are constructed by gluing the both the $L,R$ disks into central disk via the Israel junction conditions (Fig.~\ref{fig:s2-s2-saddle}). In the large shell mass limit the shell homology regions pinch off ($\Delta T_{i,L} \,, \Delta T_{i,m} \, , \Delta T_{j,m} \, , \Delta T_{j,R}\to 0$) and the shells contribute universally to the action, as they did in earlier sections. Hence $\overline{\langle \mathbf{i,j}| \mathbf{i,j} \rangle_{2s}}=\overline{Z}(\beta_L)\overline{Z}(\beta_R)\overline{Z}(\beta_m)Z_{m_i}Z_{m_j}$. The Lorentzian geometry for the two-shell states is determined by the leading saddles for each of the three Z-partition functions, which depends on whether the preparation temperatures $\beta_L,\beta_m,\beta_R$ are above/below the Hawking-Page transition.  This produces $2\times2\times2=8$ different classes of two-shell states. These Lorentzian geometries can be thought of as taking any of the 4 classes of single-shell two-sided states discussed in \cite{Toolkit}, cutting the geometry along the shell, and gluing in between the two sides either a portion of the AdS black hole ($\beta_m < \beta_{HP}$) or thermal AdS ($\beta_m > \beta_{HP}$), again giving 8 possibilities in all.  For example, choosing $\tilde{\beta}_L,\tilde{\beta}_R < \beta_{HP}$ and $\beta_m > \beta_{HP}$ results in a Lorentzian geometry consisting of two disconnected AdS black holes (Fig.~\ref{fig:disc_BH}), while $\beta_m < \beta_{HP}$ corresponds to a two-sided AdS Black hole with a very long interior (Fig.~\ref{fig:long_WH_2s_2s_BH}). 

The $\beta_m > \beta_{HP}$ regime is particularly interesting as the Lorentzian geometry associated to $| \mathbf{i,j} \rangle_{2s}$ consists of disconnected $L$ and $R$ geometries identical to the geometry of the tensor product state $|i\rangle_L \otimes|j\rangle_R$ if we choose $\beta_L=\tilde{\beta}_L$ and  $\beta_R=\tilde{\beta}_R$. Hence each state in this  basis for $\mathcal{H}_{LR}$ has a geometry that is in one-to-one correspondence with the geometries in the basis for $\mathcal{H}_{\mathcal{B}_L}\otimes \mathcal{H}_{\mathcal{B}_R}$ discussed above. Any other state in $\mathcal{H}_{LR}$ can therefore be expanded into a linear superposition of states corresponding to factorised Lorentzian geometries.

\begin{figure}
    \centering
    \includegraphics[width=0.7\linewidth]{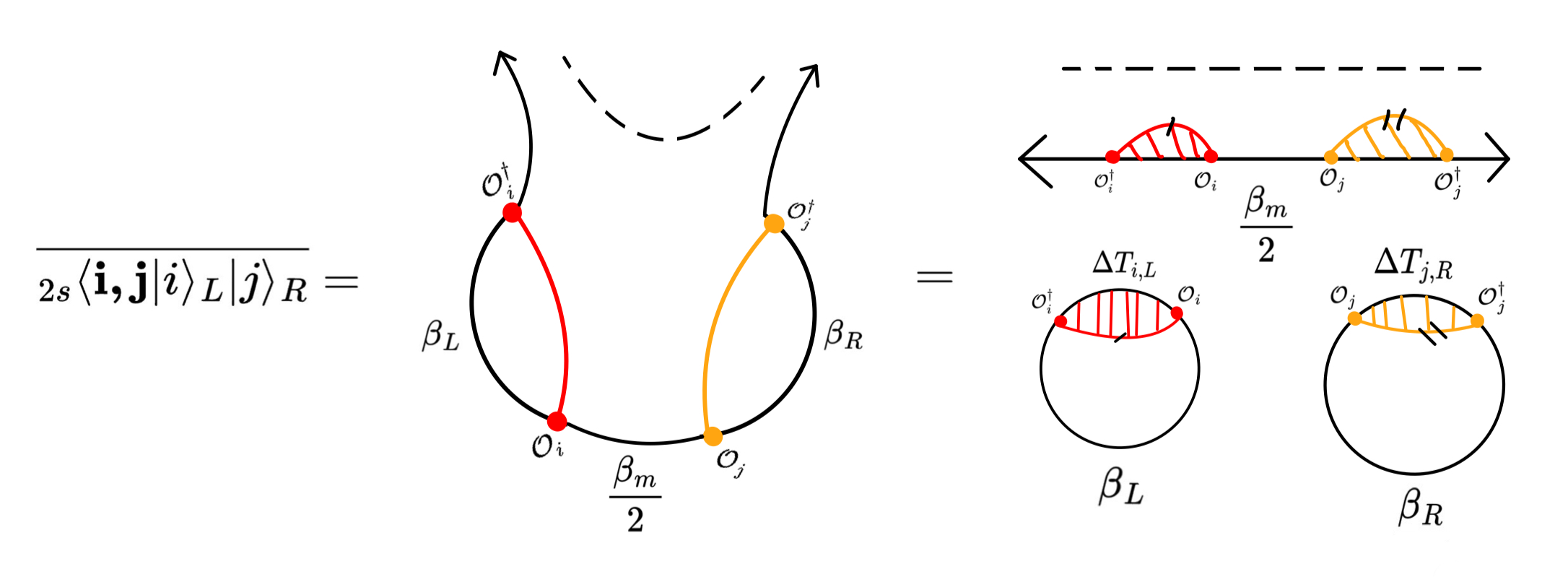}
    \caption{Saddle for the direct overlap (\ref{eq:facoverlap}) is constructed by gluing two disks into a strip of length $\beta_m/2$. }
    \label{fig:direct_overlap2}
\end{figure}

We have recovered a factorised basis by tuning down the entanglement between the two factors. However, for any finite $\beta_m$ there should still be \textit{some} entanglement, unlike in a tensor product state.  The fact that $| \mathbf{i,j} \rangle_{2s}$ and $|i\rangle_L \otimes|j\rangle_R$ cannot be the same state but still their semiclassical geometry is identical might seem puzzling. The resolution is that the semiclassical approximation of considering only the leading saddle does not capture the finer correlations that distinguish between these two states.  To see this we consider the saturation of the Cauchy-Schwarz inequality for the inner product $_{2s}\langle \mathbf{i,j}|i\rangle_{L}|j\rangle_{R}$ when $\beta_L=\tilde{\beta}_L$ and  $\beta_R=\tilde{\beta}_R$. The saddles computing $\overline{_{2s}\langle \mathbf{i,j}|i\rangle_{L}|j\rangle_{R}}$ are constructed by gluing two disks into a strip of length $\beta_m/2$ along the two shells  (Fig.~\ref{fig:direct_overlap2}).  In the large shell mass limit we obtain:
\beq \label{eq:facoverlap}
\frac{\overline{_{2s}\langle \mathbf{i,j}|i\rangle_{L}|j\rangle_{R}}}{\sqrt{\overline{\langle \mathbf{i,j}|\mathbf{i,j}\rangle_{2s} \langle i|i\rangle_{L} \langle j|j\rangle_{R}}}}=\frac{\overline{S(0)}\overline{Z}(\beta)_L\overline{Z}(\beta_R)\overline{S}(\beta_m/2)}
{\sqrt{\overline{S(0)}\overline{Z}(\beta_L)}\sqrt{\overline{S(0)}\overline{Z}(\beta_R)}\sqrt{\overline{Z}(\beta_L)\overline{Z}(\beta_R)\overline{Z}(\beta_m)}}=\frac{\overline{S}(\beta_m/2)}{\sqrt{\overline{Z}(\beta_m)}}.
\eeq
 
In the regime $\beta_m > \beta_{HP}$ where these states have identical semiclassical geometries, the dominant saddle for $\overline{Z}(\beta_m)$ is thermal AdS, for which  $\sqrt{\overline{Z}(\beta_m)}=\overline{Z}(\beta_m/2)$ and $\overline{S}(\beta_m/2)=\overline{Z}(\beta_m/2)$\footnote{Thermal AdS is just a section of the strip of length $\beta$ with the endpoints identified.} and thus $\frac{\overline{_{2s}\langle \mathbf{i,j}|i\rangle_{L}|j\rangle_{R}}}{\sqrt{\overline{\langle \mathbf{i,j}|\mathbf{i,j}\rangle_{2s} \langle i|i\rangle_{L} \langle j|j\rangle_{R}}}}=1$. As the Cauchy-Schwarz inequality is saturated,  $| \mathbf{i,j} \rangle$ and $|i\rangle_L \otimes|j\rangle_R$ must be linearly dependent when $\beta_m > \beta_{HP}$. However, if you keep the sub-dominant black hole saddle for $\overline{Z}(\beta_m)$ this is not longer the case; the states are distinguishable. Also note that in the $\beta_m < \beta_{HP}$ regime the leading disk saddle for $\overline{Z}(\beta_m)$ is the Euclidean AdS black hole, while the strip saddle is still given by Euclidean AdS, and therefore (\ref{eq:facoverlap}) will be smaller than unity, as the semiclassical geometry of the two-shell and tensor product state are not equal when $\beta_m < \beta_{HP}$.

\section{Discussion} \label{sec:discussion}
In this paper we have shown that the Hilbert space of quantum gravity with two asymptotic boundaries factorises into two copies of the single-boundary  Hilbert space. To show this we used a toolkit introduced in \cite{Toolkit} that drastically simplifies non-perturbative gravity path integral calculations, and allows extraction of  fine-grained equalities from the coarse-grained path integral. In \cite{Toolkit} these tools were used to show that the two-sided shell states described in
\cite{Sasieta:2022ksu,Balasubramanian:2022gmo,Balasubramanian:2022lnw,Antonini:2023hdh,Balasubramanian:2024rek} and Appendix~\ref{sec:2Sss} span the entire intrinsically two-boundary, fine-grained Hilbert space. Similarly, in this paper we used these methods to show that a single-sided version of the shell states discussed in \cite{Chandra:2022fwi} span the entire single-sided, fine-grained Hilbert space. The semiclassical geometry of the single-sided shell states is either thermal AdS entangled with 
a compact baby universe, or a single-sided black hole, depending on the preparation temperature of the state. As the single-sided shell states form a basis for any preparation temperature, our results imply that a black hole can be written as a superposition of horizonless geometries, and vice versa. The authors of \cite{Toolkit} reached the same conclusion for the two-sided case, and used it to argue against the existence of a ``geometry" observable in the quantum theory. The same logic applies to the single boundary gravity theory.

As an application of these results, we showed that the  two-boundary, fine-grained gravity Hilbert space factorises into two copies of the single-boundary, fine-grained gravity Hilbert space. This  result followed from novel non-perturbative wormhole contributions to the path integral that connect the asymptotic boundaries preparing  single- and two-sided states. The toolkit  from \cite{Toolkit} rendered the sum over topologies tractable, and allowed us to extend coarse-grained path integral calculations to a fine-grained result.  Since we have shown that the two-boundary gravity Hilbert space factorises, and  have also explicitly constructed bases with semiclassical interpretations as connected or disconnected geometries, we conclude that  connected geometries can be written as superpositions of disconnected ones and vice versa. Therefore,  there cannot be a linear operator that measures ``connectedness'' of a state in quantum gravity.

Interestingly, after normalizing the shell states every wormhole geometry encountered in this work, whether connecting two-sided  states (as in \cite{Toolkit}), single-sided states, or a mix, evaluates to products of the $Z$-partition function (see Sec.~\ref{sec:review}) which is thought to compute the thermal partition function in theories of gravity \cite{Gibbons:1976ue}. This  echos recent proposals \cite{deBoer:2024mqg,deBoer:2023vsm} that the gravity path integral captures the maximal ignorance matrix model constrained by a set of input parameters, one of which is the expectation value for thermal partition function as computed by the gravity path integral. Wormhole contributions to the gravity path integral should be captured by moments of the matrix model. It would be helpful to  explicitly show from the proposed matrix model why the quantities receiving the wormhole contributions considered in this work evaluate to products of the thermal partition function.

Recently \cite{Colafranceschi:2023moh} showed that the operator algebra on the Hilbert space obtained by cutting open the gravity path integral is of Type I, if the theory satisfies a certain general set of axioms. Type I algebras admit a trace, which \cite{Colafranceschi:2023moh} used to give a new understanding of the HRT entropy formula \cite{Ryu:2006ef,Ryu:2006ef,Hubeny:2007xt,Lewkowycz:2013nqa} from a purely gravitational point of view.  However, their results were not sufficient to establish that the two-boundary Hilbert space is  a product of single boundary factors. It would be interesting if our results could help to establish which axioms should be added to those in \cite{Colafranceschi:2023moh} to imply factorization.

We have shown factorization of the two-sided Hilbert space using just  the gravitational path integral, and  our results apply equally to asymptotically flat or AdS gravity, without any assumption about holographic duality. Interestingly, the factorization  we found is {\it required} if there is a dual quantum mechanics living on the asymptotic boundary of  spacetime, as there is for AdS gravity. We do not know if such duls exist for asymptotically flat gravity but the factorisation we showed is consistent with their existence.

\acknowledgments 
TY thanks the Peter Davies Scholarship for continued support. VB was supported in part by the DOE through DE-SC0013528 and QuantISED grant DE-SC0020360, and in part by the Eastman
Professorship at Balliol College, University of Oxford.

\begin{appendix}
\section{Two-sided shell states} \label{sec:2Sss}
\begin{figure}[h]

        \begin{subfigure}[b]{.43\linewidth}
            \centering
            \includegraphics[width=0.9\linewidth]{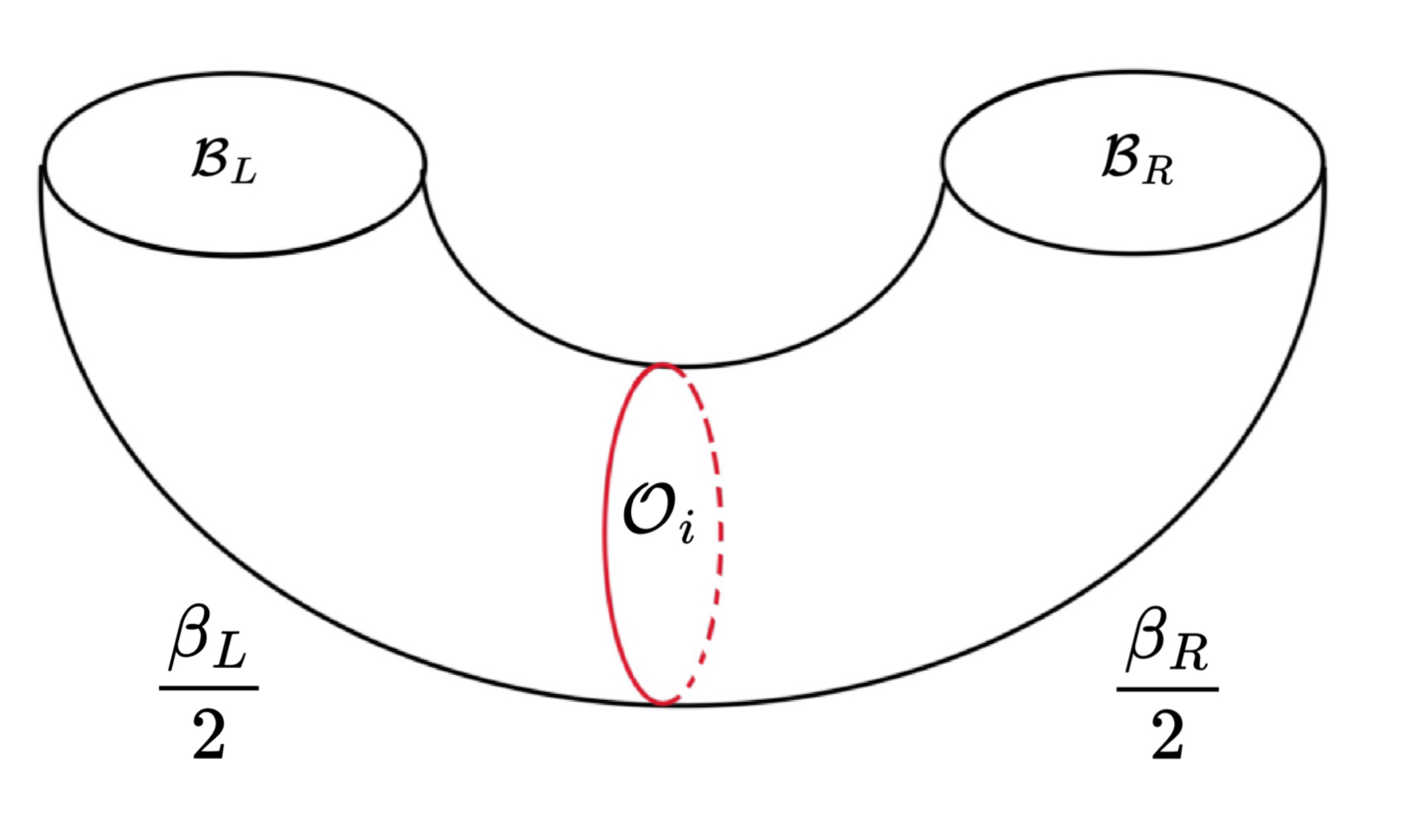}
            \caption{}
       
        \end{subfigure}
 \hfill
        \begin{subfigure}[b]{.5\linewidth}
            \centering  
            \includegraphics[width=0.9\linewidth]{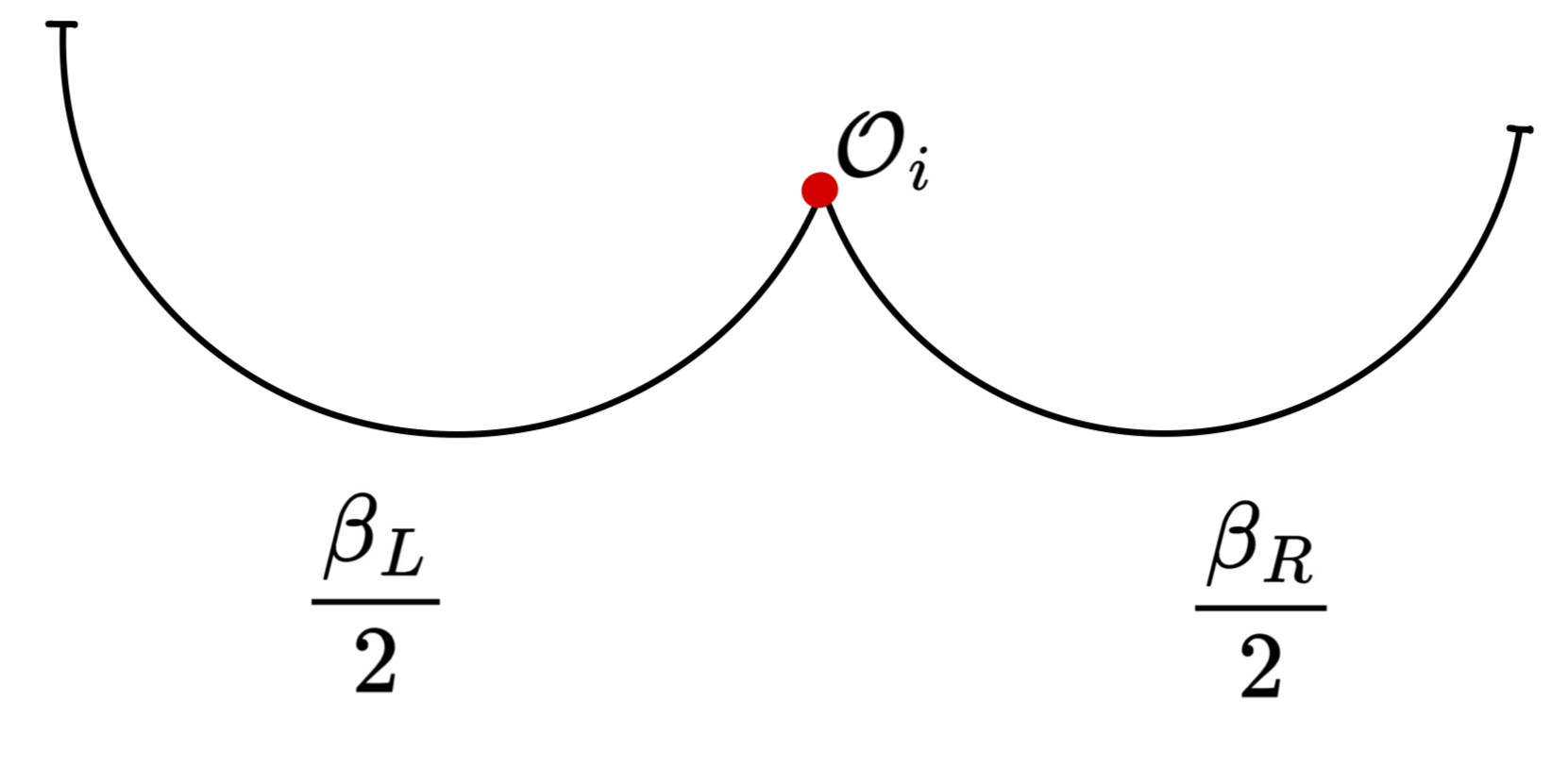}
            \caption{}
       
        \end{subfigure}
    \caption{Asymptotic boundary condition for the gravity path integral defining the shell state.  ({\bf a}) Cut-open Euclidean boundary with topology $\mathbb{I}_{\frac{\beta_L+\beta_R}{2}}\times\mathbb{S}^{d-1}$ for preparation of the shell states. The shell operator $\mathcal{O}_{i}$  is pictured in red. In AdS/CFT we can also perform the path integral in the boundary CFT with insertion of a $\mathbb{S}^{d-1}$ symmetric operator dual to the shell.  ({\bf b}) Euclidean boundary with the $\mathbb{S}^{d-1}$ suppressed. We adopt this convention for the rest of the paper. Here $\beta_{L,R}/2$ are  Euclidean ``preparation times''. Figure adapted from \cite{Toolkit}.}
    \label{fig:shell_b.c}
\end{figure}

\begin{figure}[h]
    \centering
    \includegraphics[width=0.3\linewidth]{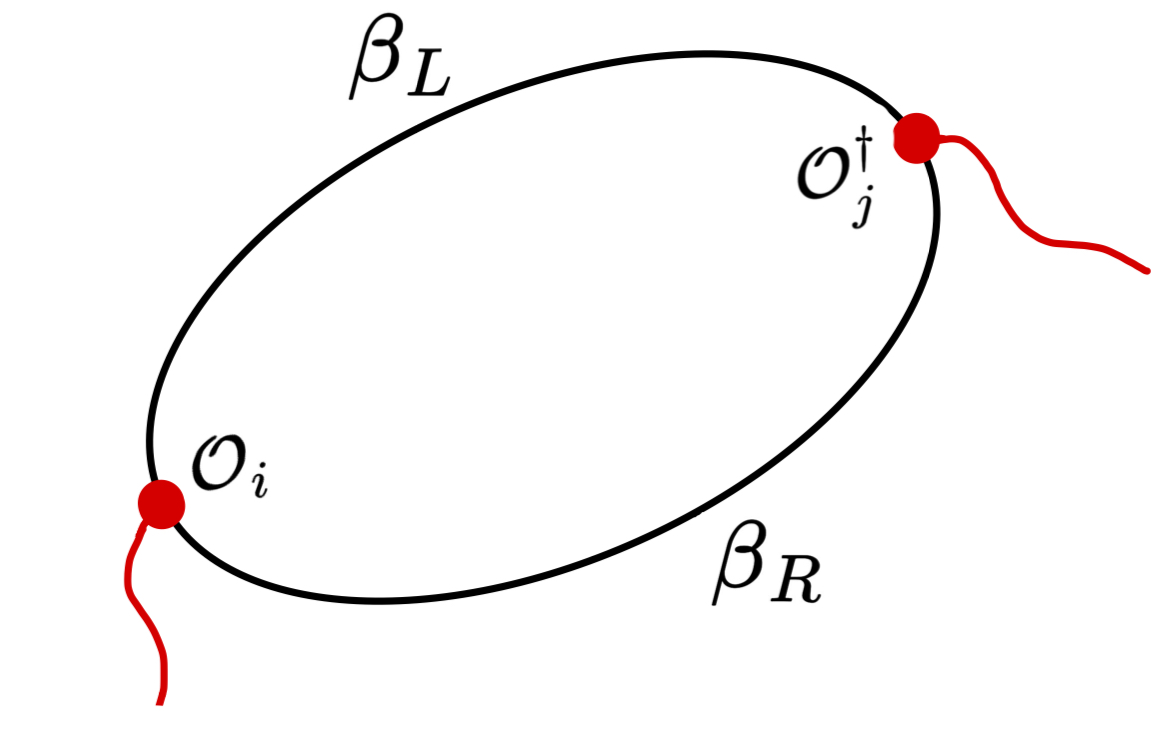}
    \caption{Shell asymptotic boundary condition for $\overline{\langle j|i\rangle} $, consisting of the operator insertions $\mathcal{O}_{i}$ and  $\mathcal{O}^{\dagger}_{j}$ separated by asymptotic time extent $\beta_{L}$ and $\beta_{R}$ respectively. The red lines represent the shells propagating into the bulk. Figure adapted from \cite{Toolkit}.}
    \label{fig:shell_bdry}
\end{figure}

\begin{figure}[h]
            \centering
            \includegraphics[width=1\linewidth]{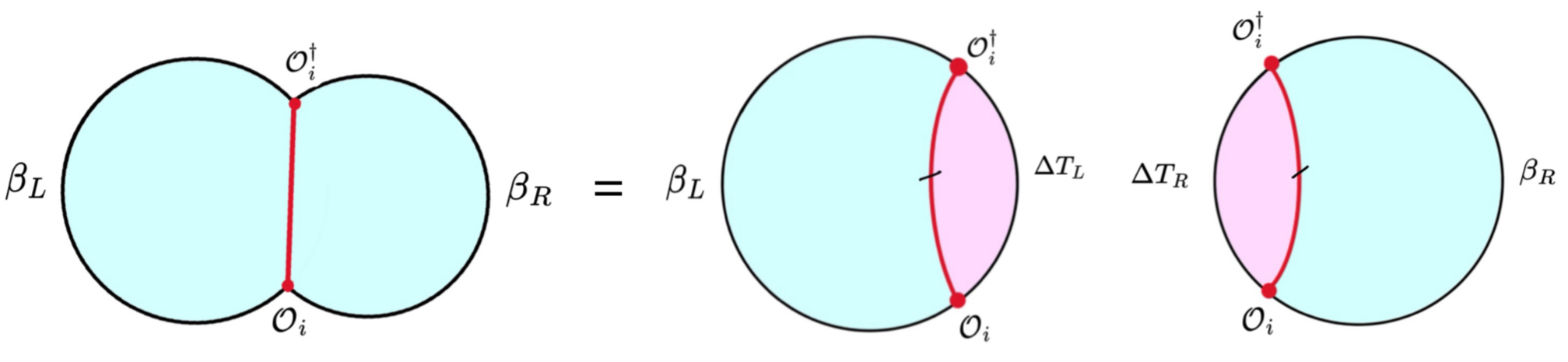}
            \caption{The saddlepoints for the shell norm path integral $\overline{\langle i|i\rangle}$ are constructed by gluing together two disks. In particular the shell homology regions (purple) on each disk are discarded and the resulting geometries glued together along the shell worldvolume. Figure adapted from \cite{Toolkit}.}
            \label{fig:shell_norm}
\end{figure}

\begin{figure}
    \centering
    \includegraphics[width=1\linewidth]{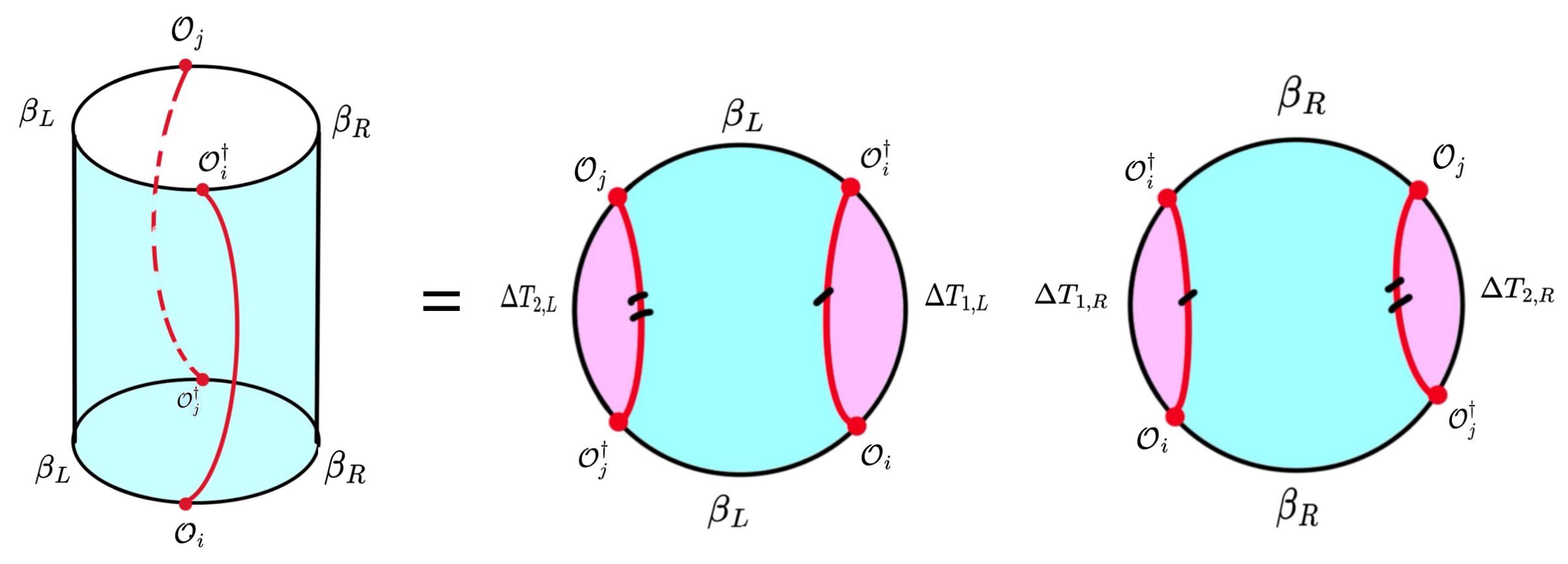}
    \caption{Construction of shell wormhole saddles $Z_2$. We glue two disks together by discarding the regions indicated in purple and  attaching the remainder along the shell worldvolumes (red, black dashes indicate the glued shells) while obeying the Israel junction conditions.  Saddlepoints, i.e., solutions to the equation of motion, exist with this topology \cite{Balasubramanian:2022gmo}. Figure adapted from \cite{Toolkit}.}
    \label{fig:shell_wormhole}
\end{figure}

We use the shell state basis for the two-sided Hilbert space  developed in  \cite{Sasieta:2022ksu,Balasubramanian:2022gmo,Balasubramanian:2022lnw,Antonini:2023hdh,Balasubramanian:2024rek,Toolkit}.\footnote{In the AdS/CFT context, these are related to Partially Entangled Thermal States (PETS) \cite{Goel:2018ubv}.}  These states are defined by the Euclidean gravity path integral with boundary $\mathbb{I}\times \mathbb{S}^{d-1}$. The cut surface of this boundary has two $\mathbb{S}^{d-1}$ components which we call $\mathcal{B}_{L,R}$ (Fig.~\ref{fig:shell_b.c}).  To create states we insert $\mathbb{S}^{d-1}$ symmetric dust shell operators $\mathcal{O}_{S}$ of mass $m_i \sim \mathit{O}(1/G_{N})$ on the boundary, at distances $\frac{\beta_L}{2}$ and $\frac{\beta_R}{2}$ from the $\mathcal{B}_{L,R}$ cuts. By varying $m_i$ we  obtain an infinite family of shell states $\ket{i}$.
To calculate entries of  the Gram matrix $G_{ij}\equiv \langle i|j\rangle$ of overlaps between the shell states,  we must sew together the $\ket{i}$ and $\ket{j}$ state preparation path integrals at their cuts. We will  say that the overlap path integral has a  \textit{shell boundary} because it includes boundary operator insertions $\mathcal{O}_{j}$ and  $\mathcal{O}^{\dagger}_{i}$ separated by asymptotic times  $\beta_{L}$ and $\beta_{R}$ on the closed boundary manifold (Figs.~\ref{fig:shell_bdry}, \ref{fig:shell_norm}).
To leading order 
\begin{equation}
\overline{\langle i|j \rangle} =\delta_{ij} Z_{1} \,.
\label{eq:defZ1}
\end{equation} 
if the shell inertial masses $m_{i,j}$ are sufficiently large and different, because it takes $|m_{i}-m_{j}|$ bulk interactions in Planck units to match such shells  \cite{Balasubramanian:2022gmo}.  However,  higher topology  contributions  (Fig.~\ref{fig:shell_wormhole}) stabilized by the shell matter  modify the overlap to
\begin{equation}
\overline{|\langle i|j \rangle|^2}=\overline{\langle i|j \rangle\langle j|i \rangle} = Z_{2} + \delta_{ij}Z_{1}^2 \, ,
\end{equation}
where $Z_{2}$ is a wormhole saddlepoint contribution \cite{Balasubramanian:2022gmo,Toolkit}.  The wormhole is constructed by identifying shell worldvolumes across two disks subject to the Israel junction conditions  (Fig.~\ref{fig:shell_wormhole}). The saddlepoint geometry on either side of the shell can be anything that locally solves the equations of motion; hence by symmetry these are portions of any of the saddlepoints contributing to the Z-partition function in Sec.~\ref{sec:review}.

\begin{figure}[h]
    \centering
    \includegraphics[width=0.4\linewidth]{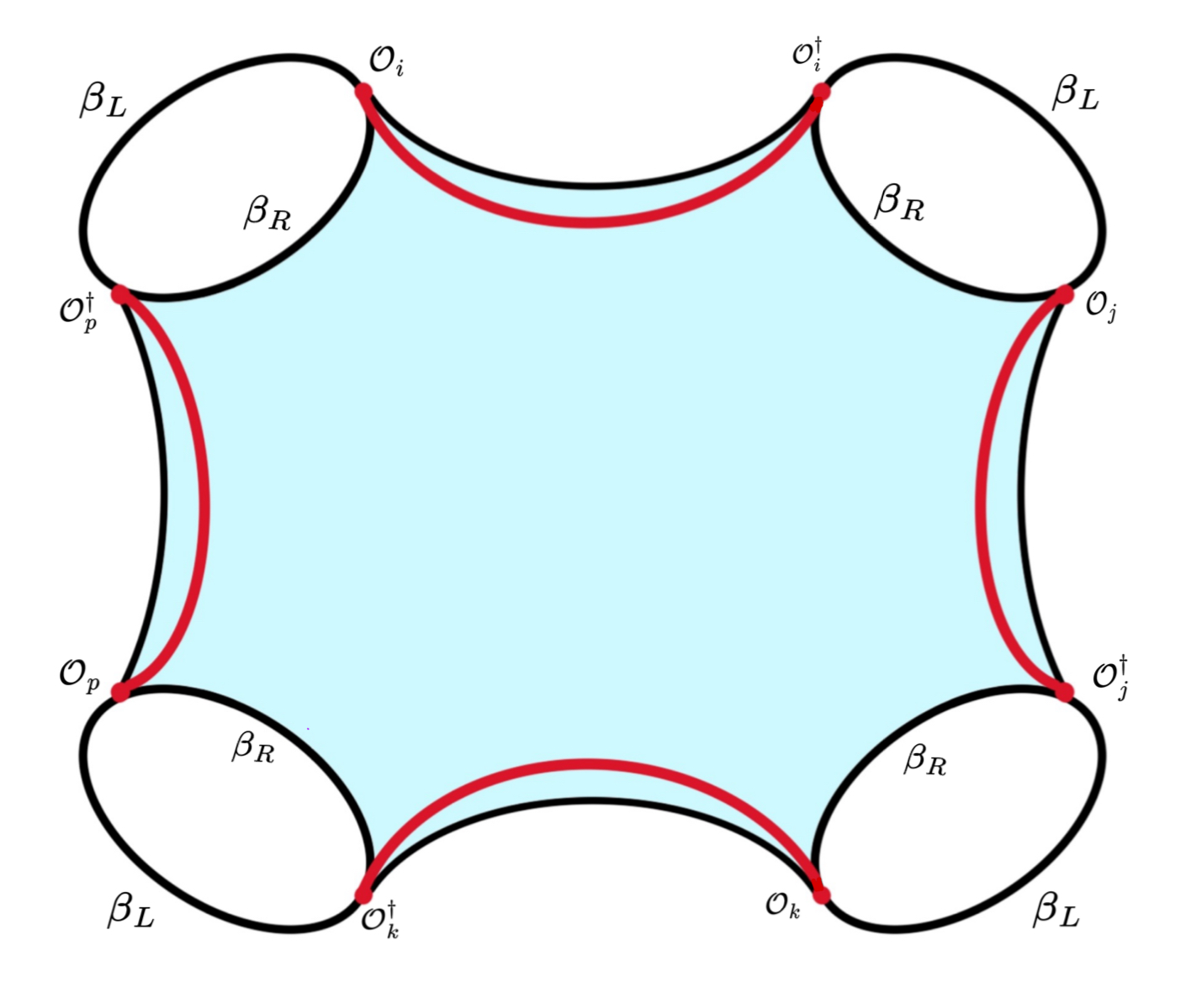}
    \caption{Fully connected wormhole geometry for $\overline{(G^n)_{ii}}$ consisting of $n$ shell boundaries connected by a single bulk, pictured here for $n=4$. Figure adapted from \cite{Toolkit}.}
    \label{fig:pinwheel}
\end{figure}

Similarly, we can construct  saddlepoints contributing to   $\overline{(G^n)_{ii}}=\overline{\braket{i|j_1}\braket{j_1|j_2} \cdots \braket{j_n|i}}$  (no sum on $i,j_k$) by connecting $n$ asymptotic boundaries (arising from each factor in $(G^n)_{ii}$) with a single multiboundary wormhole. The $k$-th shell can only propagate through the wormhole from $\mathcal{O}_{k}$ to $\mathcal{O}^{\dagger}_{k}$ on different asymptotic boundaries, and so the index structure of $\overline{(G^n)_{ii}}$ is reflected in the geometry of the wormhole (Fig.~\ref{fig:pinwheel}).  We will call this wormhole geometry the \textit{pinwheel}. There are also disconnected saddlepoints connecting different subsets of the boundaries together which look like collections of smaller pinwheels.

\begin{figure}[t]
\begin{subfigure}{\linewidth}
    \centering
    \includegraphics[width=0.5\linewidth]{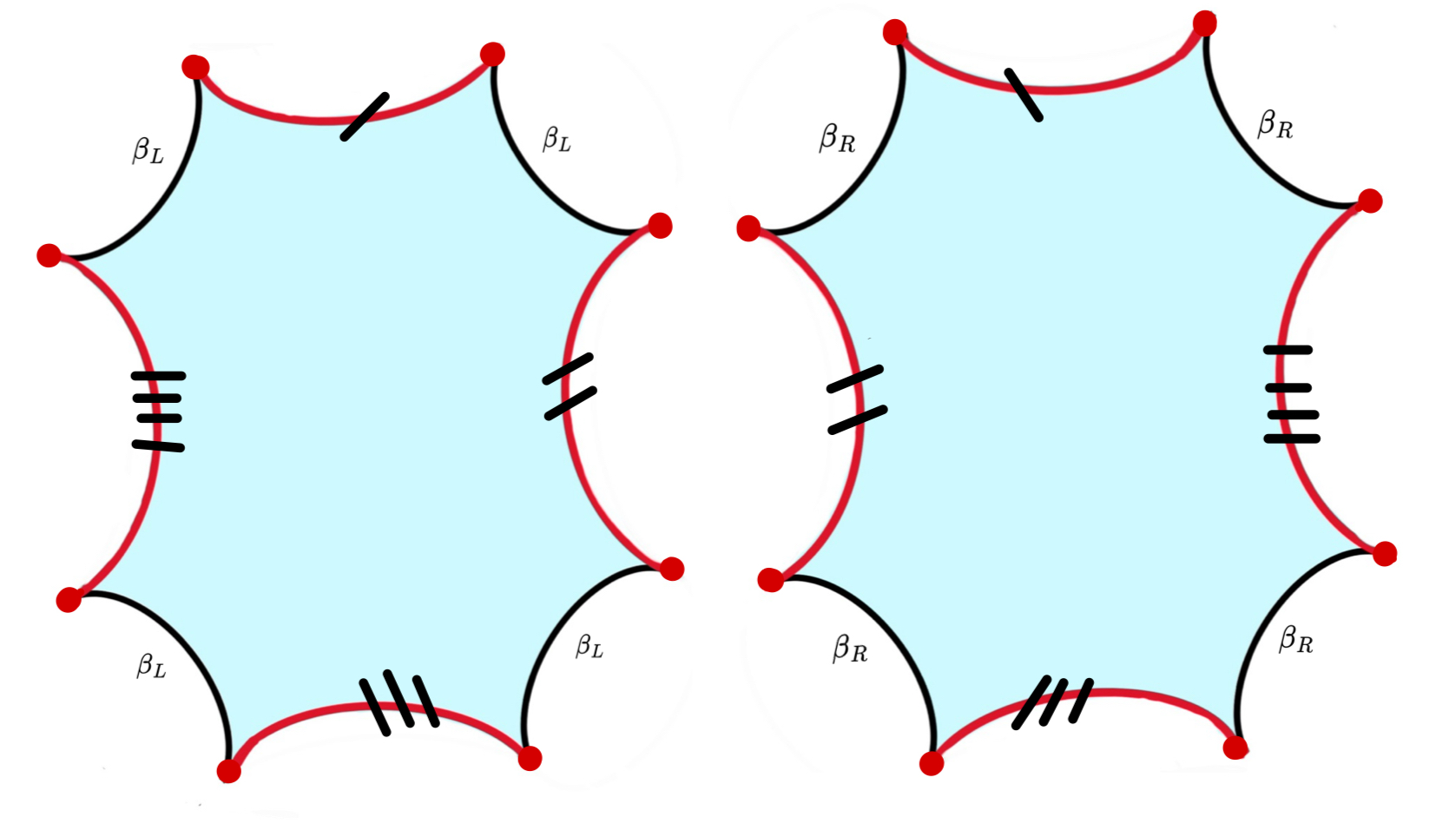}
    \caption{}
    \label{fig:pinwheelsheet}
\end{subfigure}
\begin{subfigure}{\linewidth}
\centering
    \includegraphics[width=0.6\linewidth]{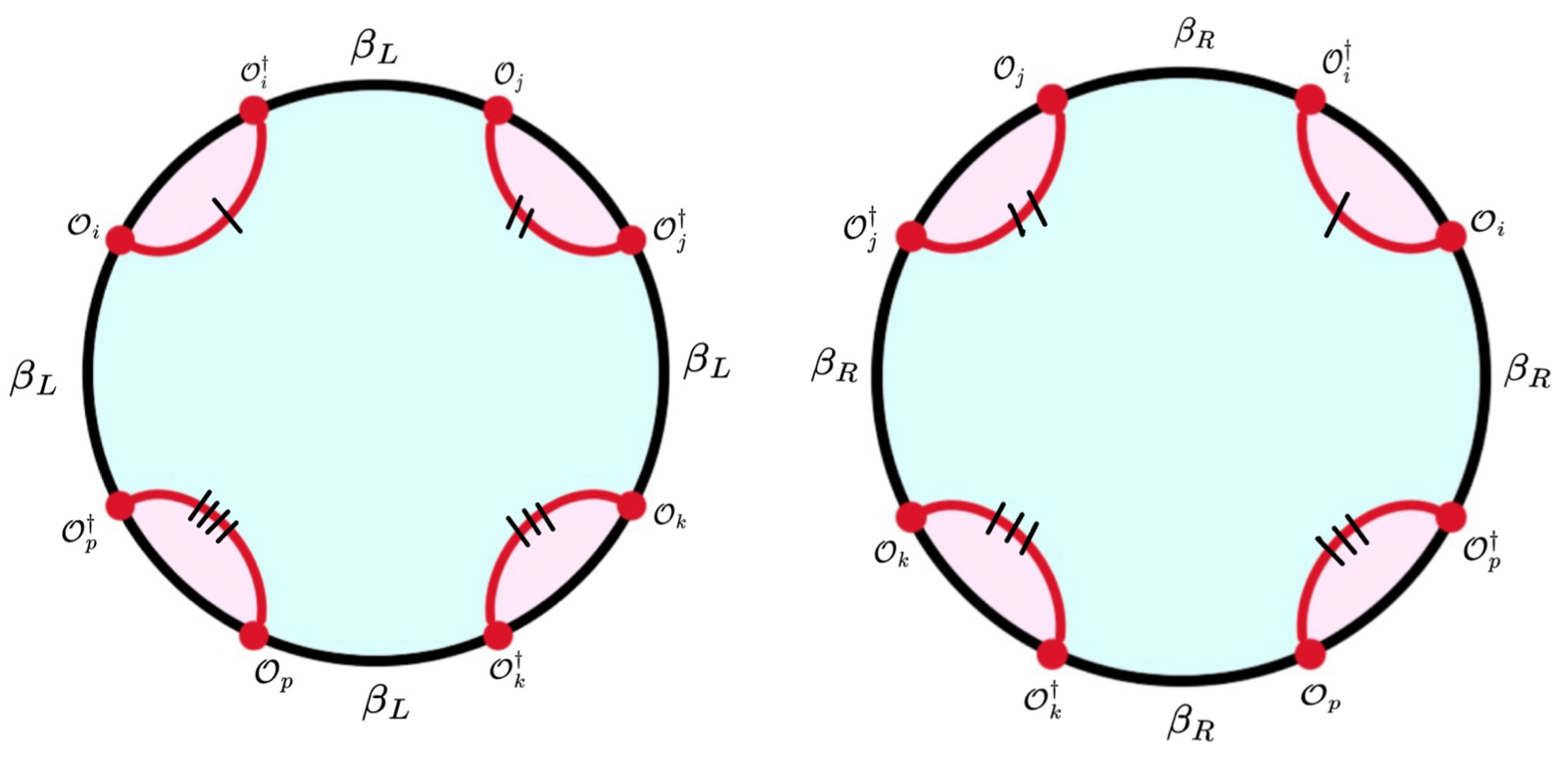}
    \caption{}
    \label{fig:pinwheeldisks}
    \end{subfigure}
    \caption{Schematic of the construction of the pinwheel wormhole. ({\bf a}) Sheet diagrams for the pinwheel obtained by cutting Fig.~\ref{fig:pinwheel} along the shell worldvolumes.  ({\bf b}) Each of the pinwheel sheets can be regarded a portion of a disk. The pinwheel saddle constructed by gluing to disks together along the corresponding shell worldvolumes by the junction conditions. Figure adapted from \cite{Toolkit}.}
\end{figure}

To construct a pinwheel saddlepoint  
we slice Fig.~\ref{fig:pinwheel} open along the shell worldvolumes producing  $L$- and  $R$- \textit{sheet diagrams} in Fig.~\ref{fig:pinwheelsheet}. We can also produce  these sheets by removing the pink shell homology regions from the $L$ and $R$ disks in Fig.~\ref{fig:pinwheeldisks}.    
The $L$ ($R$) disk boundary contains $n$ shell operator insertions  $\mathcal{O}_{1,2, \cdots n}, \mathcal{O}^{\dagger}_{1,2, \cdots n}$ with  $\mathcal{O}_{i}$ and $\mathcal{O}^{\dagger}_{i}$ separated by a time $\Delta T_{L,i}$ ($\Delta T_{R,i}$) and $\mathcal{O}^{\dagger}_{i},\mathcal{O}_{i+1} $  separated by a time $\beta_L$ ($\beta_R$). So to construct the wormhole we can  start with the disks in  Fig.~\ref{fig:pinwheeldisks} with  shell homology regions removed, and then glue along the indicated shell worldvolumes by using the Israel junction conditions \cite{Israel:1966rt}. The shell propagation times $\Delta T_i$ are then determined dynamically by the shell equations of motion.  We can apply this procedure started with any saddle geometry to the Z-partition function in Sec.~\ref{sec:review}.

As shown in \cite{Balasubramanian:2022gmo,Balasubramanian:2022lnw,Antonini:2023hdh}  the  turning points of the shell trajectories approach the asymptotic boundary  of the disk as the shells masses increase, so that the shell homology regions pinch off. In this limit, $\Delta T_i \to 0$, and each shell contributes simply contributes a factor $Z_{m_i}\sim e^{-2(d-1)\log(\mathrm{G_N} m_i)}$.  This behavior is independent of the Z-partition saddle geometries that are  glued along the shells \cite{Balasubramanian:2022gmo,Balasubramanian:2022lnw,Antonini:2023hdh}, so that the fully connected saddlepoint contribution to the n-boundary path integral decomposes into two Z-partition functions and a shell contribution: 
\beq \label{eq:2s_WH_nonormal}
\overline{(G^n)_{ii}}|_{connected} = \overline{Z}(n\beta_L)\overline{Z}(n\beta_R )\prod_{i=1}^n Z_{m_i} \,.
\eeq

As $m_i \to \infty$  the diagram in Fig.~\ref{fig:shell_norm} computing  $\braket{i|i}$ also simplifies -- the propagation time of the shell vanishes, so that the diagram decomposes into two disks of circumference $\beta_{L,R}$ attached via a pointlike shell trajectory. Thus (\ref{eq:defZ1}) becomes $\braket{i|j} = \delta_{ij} Z_1 = \delta_{ij} \overline{Z}(\beta_L) \overline{Z}(\beta_R)Z_{m_i}$ in the sum over saddlepoints approximation.  We can also see this by setting $n=1$ in the above expression for $\overline{(G^n)_{ii}}|_{connected}$. Normalizing the states as $\ket{k}/\sqrt{Z_1}$, we get
\beq \label{eq:2s_shell_WH}
\overline{(G^n)_{ii}}|_{connected} = \frac{\overline{Z}(n\beta_L)\overline{Z}(n\beta_R)}{\overline{Z}(\beta_L)^n \overline{Z}(\beta_R)^n} \, ,
\eeq
where again the overbars indicate a coarse-grained averages path integral computation.  Each Z partition sum on the right side is a sum over all saddlepoints, accounting for all the saddle contributions to $\overline{(G^n)_{ii}}|_{connected}$.

\section{All order extensions} \label{appendixb}
The main text showed various results in the sum over saddlepoints approximation to the gravity path integral. Here we  extend those results to the full path integral by using the ``surgery'' method of \cite{Toolkit}.  Our arguments are  adapted from Sec.~5 of \cite{Toolkit}, to which we refer  the reader for details.

\subsection{Single-sided shell completeness} \label{sec:allOspan}
The proof in the main text that the single-sided shell states form a basis was restricted to sum over saddle points approximation to the gravity path integral. In this section we argue that the full path integrals on the LHS and RHS of (\ref{eq:idinsert}) and (\ref{eq:idsqr}) are identical, and hence (\ref{Hshellspan}) is exact in the fine-grained theory. Throughout this section we adopt the notation of Sec.~\ref{sec:complete}. All our considerations will be in the strict $\kappa \to \infty$ limit in which only wormhole geometries that do not break an index loop contribute. The argument made in Sec.~\ref{sec:complete} is easily seen to include perturbative corrections around the $\hat{Z}_{n+2}(\Psi)$  saddle points by considering the limit in which the shells are parametrically heavy. In this limit the shells propagate in the asymptotic region, which is protected from quantum corrections, and contribute universally to the action. The shells therefore only serve to glue the the parts of the different strips together. In this limit corrections to the $\hat{Z}_{n+2}(\Psi)$  saddle points correspond to the corrections to the time evolved overlap $\overline{\langle \Psi|\Psi\rangle}(\frac{\tilde{\beta}_R+{(n+1)\beta_R}}{2})$. Hence the corrections contribute equally to both sides of (\ref{eq:idinsert}) and similarly for (\ref{eq:idsqr}). 

\begin{figure}[h]
    \begin{subfigure}[c]{0.45\linewidth}
    \centering
    \includegraphics[width=\linewidth]{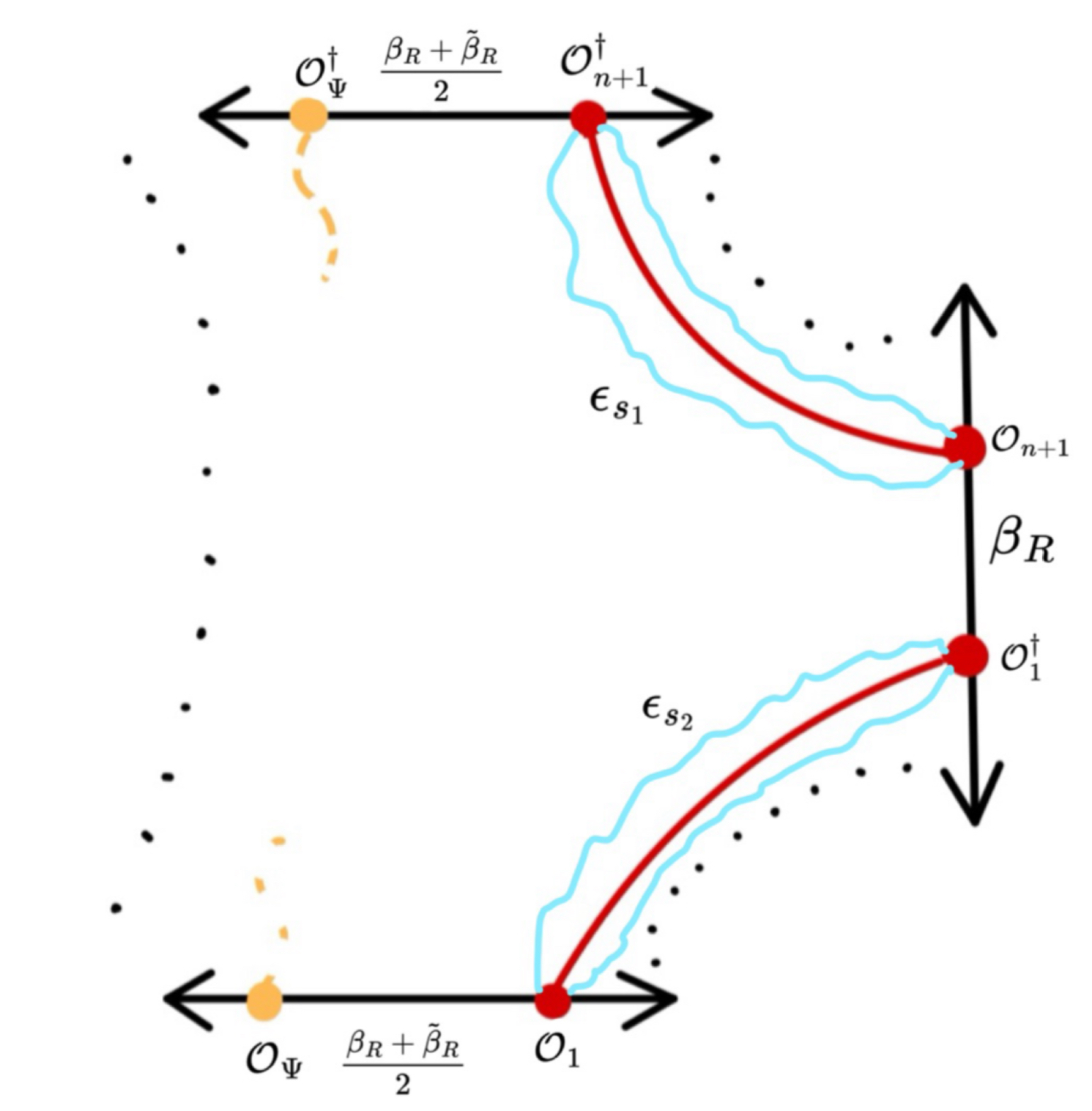}
    \caption{}
    \label{fig:cutout}
    \end{subfigure}
    \begin{subfigure}[c]{0.45\linewidth}
    \centering
    \includegraphics[width=\linewidth]{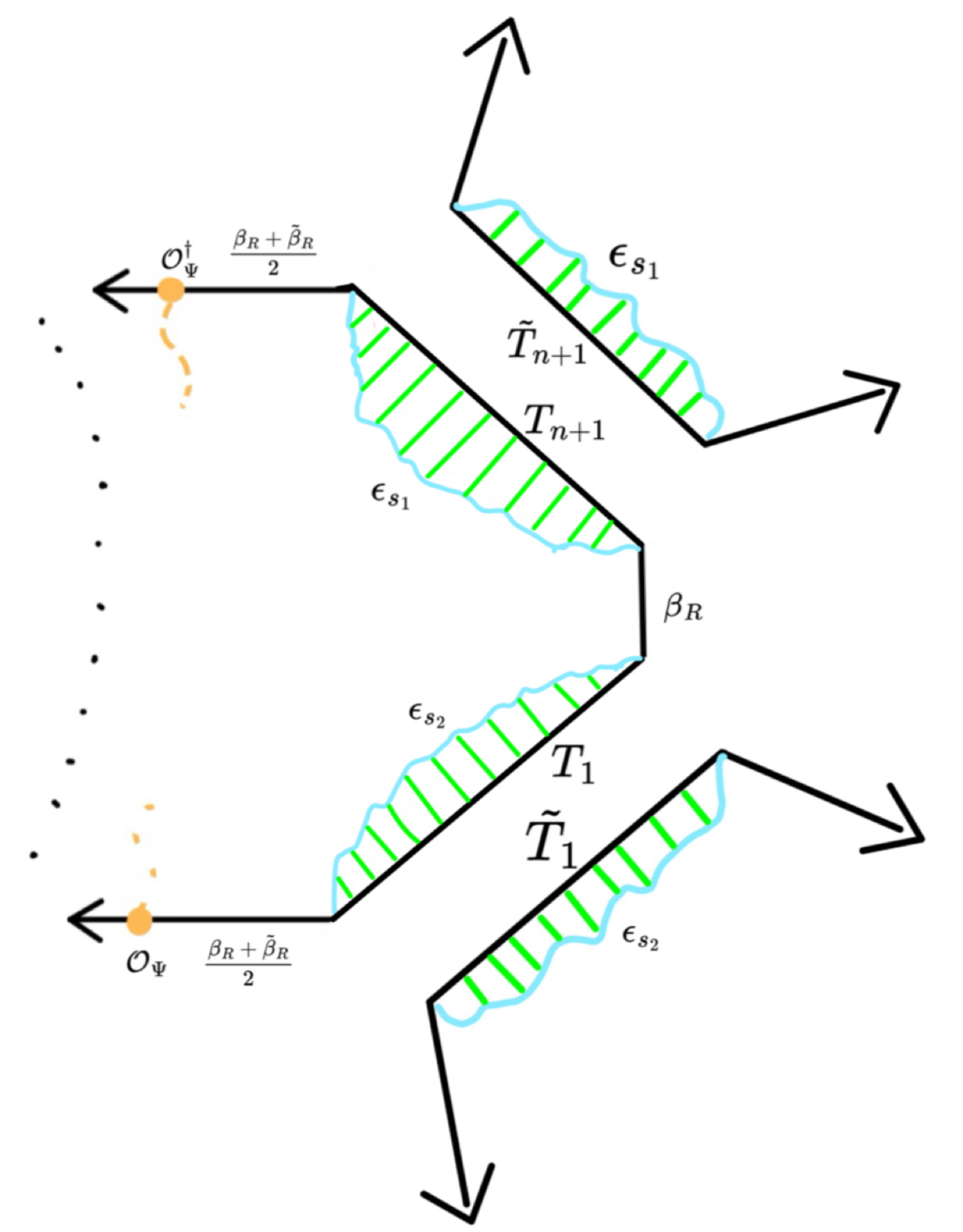}
    \caption{}
    \label{fig:sheet_embed}
    \end{subfigure}
    \hfill
   
    \caption{ ({\bf a}) Cartoon of the $\epsilon_{s_i}$ cut-outs around the shell worldvolumes for some contribution to (\ref{eq:spanpathint}), depicted for $n=2$. ({\bf b}) The homology regions are filled in (green) to complete the bulk regions into two vacuum strips and one central strip.}
\end{figure}

We might even extend the equality (\ref{Hshellspan}) to topologies that do not support a saddlepoint using the surgery tool from \cite{Toolkit}. This argument is  adapted from that of Sec.~5.1 in \cite{Toolkit} and we invite the reader to look at that section for details. Mainly, we argue that the structure of the full gravity path integral 
\beq \label{eq:spanpathint}
\overline{G^{n}_{ij}\langle \Psi|i\rangle\langle j|\Psi\rangle} = \int_{g, \phi \, \to \, G^{n}_{ij}\langle \Psi|i\rangle\langle j|\Psi\rangle} \mathcal{D}g \mathcal{D}\phi\mathcal \, \,  e^{-I_{tot}[g]}
\eeq
at a fixed positive integer $n$ is such that in the $n \to -1$ limit this path integral reduces to the full path integral for $\overline{\langle \Psi|\Psi\rangle}( \tilde{\beta}_R)$, where $g, \phi \to G^{n}_{ij}\langle \Psi|i\rangle\langle j|\Psi\rangle$ denotes that we path integrate over metrics and quantum fields satisfying the asymptotic boundary conditions. As we are in the $\kappa\to \infty$ limit we only have to consider the fully connected contributions. First, we organize the full sum over connected geometries by topology and consider a geometry $g$. If the shells are parametrically heavy enough the shell matter path integral $\int \,\mathcal{D}\phi$ subject to the asymptotic shell insertion boundary condition localizes to geodesics on the background $g$. Recall that the saddle points for (\ref{eq:spanpathint}) were constructed by gluing $n+1$ shell strips into a so-called central strip. We will use the same terminology here to denote the analogous regions on $g$. We imagine dividing up $g$ by cutting out a small region around the $i$-th shell worldvolume of size $\epsilon_{s_i}$ (of which there are $n+1$ in total), see Fig.~\ref{fig:cutout}. The complement of these regions is denoted by $g_{bulk}$ and looks like a cut-out of a central strip and cut-out shells strips. The cut-outs $\epsilon_{s_i}$ are chosen such that all handles of $g$ are contained within in $g_{bulk}$. 

 As the Einstein-Hilbert action consists of local and boundary terms we may write the total gravitational action of $g$ as the sum of the action of these individual patches: 
\beq
I_{tot}[g]= \sum_{i=1}^{n+1} I_{\epsilon_{s_i}}+ I_{g_{bulk}} \, .
\eeq
By shrinking the cut-out regions $\epsilon_{s_i}$ to include just the shell world-volumes the contribution to the action from the shells is given by the co-dimension 1 integral $ \int_{\mathcal{W}_i[g]} \sigma_i$ of the shell matter density ($\sigma_i$) over the world-volume ($\mathcal{W}_i[g]$)
 \beq
 \sum_{i=1}^{(n+1)} I_{\epsilon_{s_i}}=\sum_{i=1}^{(n+1)} \int_{\mathcal{W}_i} \sigma_i
 \eeq
and hence we write
\beq \label{eq:5}
\zeta[G^{n}_{ij}\langle \Psi|i\rangle\langle j|\Psi\rangle] = \int_{g \to \, G^{n}_{ij}\langle \Psi|i\rangle\langle j|\Psi\rangle} \mathcal{D}g\mathcal \, \, e^{-I_{bulk}} \prod_{i=1}^{n+1} e^{-\int_{\mathcal{W}_i[g]} \sigma_i}.
\eeq

Now consider the $g_{bulk}$ region, which consists cut-outs of both the central and shell strip regions (and possibly handles connecting them) with incomplete asymptotic boundaries, see Fig.~\ref{fig:cutout}. To make contact with the desired result, we imagine completing the boundary of each of these sections into a full strip boundary. Take the central strip cut-out for example, for every shell that was cut out we add in ``homology" regions $g_{\mathcal{C}_i}$ between an added-in piece of asymptotic boundary of length $T_i$ and the shell cut-out contour $\epsilon_{s_i}$. There $n+1$ such pieces needed to complete the central strip boundary resulting an asymptotic boundary of length $(n+1) \beta_R + \tilde{\beta}_R +\sum_{i=1}^{n+1}T_i$, see Fig.~\ref{fig:sheet_embed}. We can sum over all possible completions of the central strip cut-out $\subset g_{bulk}$ into a completed strip  by summing over all possible homology region geometries $g_{\mathcal{C}_i}$ 
ending on the worldvolume $\mathcal{W}_i$: $ \int\mathcal{D}g_{\mathcal{C}_i} \, e^{-
I{g_{\mathcal{C}_i}}}\equiv Z_{\mathcal{C}_i}[g] $. Similarly the $i$-th cut-out shell strip  can be completed into a full strip by adding in a homology region $g_{\mathcal{S}_i}$ corresponding to a boundary section of length $\tilde{T}_i$ and  we define $ \int\mathcal{D}g_{\mathcal{S}_i} \, e^{-
I_{g_{\mathcal{S}_i}}}\equiv Z_{\mathcal{S}_i}[g] $. While these terms were not present 
in the original path integral we can rewrite (\ref{eq:5}) in a helpful way by inserting $1=\frac{\prod_{i=1}^{n+1}Z_{\mathcal{C}_i}[g]}{\prod_{i=1}^{n+1}Z_{\mathcal{C}_i}[g]}$ and  $1=\frac{\prod_{i=1}^{n+1}Z_{\mathcal{S}_i}[g]}{\prod_{i=1}^{n+1}Z_{\mathcal{S}_i}[g]}$ :

\begin{equation}
\begin{aligned}
&\int_{g\to \, G^{n}_{ij}\langle \Psi|i\rangle\langle j|\Psi\rangle} \mathcal{D}g\, e^{-I_{\text{bulk}}} \prod_{i=1}^{n+1} e^{-\int_{\mathcal{W}_i[g]} \sigma_i} 
= \int_{g\to \, G^{n}_{ij}\langle \Psi|i\rangle\langle j|\Psi\rangle} \mathcal{D}g\, \frac{\prod_{i=1}^{n+1} Z_{\mathcal{C}_i}[g]}{\prod_{i=1}^{n+1} Z_{\mathcal{C}_i}[g]} 
\cdot \frac{\prod_{i=1}^{n+1} Z_{\mathcal{S}_i}[g]}{\prod_{i=1}^{n+1} Z_{\mathcal{S}_i}[g]} 
e^{-I_{\text{bulk}}} \prod_{i=1}^{n+1} e^{-\int_{\mathcal{W}_i[g]} \sigma_i} \\
&= \int_{g\to \, G^{n}_{ij}\langle \Psi|i\rangle\langle j|\Psi\rangle} \, \mathcal{D}g\,e^{-I_{\text{bulk}}} \prod_{i=1}^{n+1} \int \mathcal{D}g_{\mathcal{Y},i} \mathcal{D}g_{\mathcal{C},i} \, 
e^{-I_{g_{\mathcal{C},i}}} e^{-I_{g_{\mathcal{S},i}}} 
\left( \frac{e^{-\int_{\mathcal{W}_i[g]} \sigma_i}}{Z_{\mathcal{C}_i}[g] Z_{\mathcal{S}_i}[g]} \right)
\end{aligned}
\end{equation}

The path integral $\int \mathcal{D}g\, \mathcal{D}g_{\mathcal{C},i}\, \mathcal{D}g_{\mathcal{S},i} $ therefore sums over all geometries satisfying the boundary condition of a completed central strip of length $(n+1) \beta_R + \tilde{\beta}_R +\sum_{i=1}^{n+1}T_i$ and $n+1$ ``vacuum" strips with no operator insertion where the $i$-th strip has length $\tilde{T}_i$. Note that due to the cut-out these boundary conditions do not contain any shell insertions, and that the sum over geometries includes those with handles connecting, say, the central strip bulk to the $k$-th vacuum strip. In terms of the notation used through this work, this is nothing more than the boundary condition $\braket{\Psi|\Psi}(n+1) \beta_R + \tilde{\beta}_R +\sum_{i=1}^{n+1}T_i)\prod_{i=1}^{n+1}S(\tilde{T}_i)$, where $\braket{\Psi|\Psi}(\cdots)$ was defined to denote the norm of a time evolved version of the state $\ket{\Psi}$. Although there is the possibility of handles in the $g_{bulk}$ region connecting the central and shells strips, the cut-outs $\epsilon_{s_i}$ where shrunken to just include the shell worldvolumes. The shell worldvolume contributions to the action are therefore insensitive to these handles. The above path integral can therefore alternatively be written as an integral over metrics $\hat{g}$ satisfying the $\hat{g} \to \braket{\Psi|\Psi}(n+1) \beta_R + \tilde{\beta}_R +\sum_{i=1}^{n+1}T_i)\prod_{i=1}^{n+1}S(\tilde{T}_i)$ asymptotic boundary condition:
\beq
\int_{\hat{g}\to \braket{\Psi|\Psi}(n+1) \beta_R + \tilde{\beta}_R +\sum_{i=1}^{n+1}T_i)\prod_{i=1}^{n+1}S(\tilde{T}_i)} \, \mathcal{D}\hat{g}\,e^{-I_{\hat{g}}} \prod_{i=1}^{n+1}  
\left( \frac{e^{-\int_{\mathcal{W}_i[\hat{g}]} \sigma_i}}{Z_{\mathcal{C}_i}[\hat{g}] Z_{\mathcal{S}_i}[\hat{g}]} \right)\, ,
\eeq

and we can now see that the $n \to -1$ limit recovers 
\beq
\int_{\hat{g}\to \braket{\Psi|\Psi}( \tilde{\beta}_R )} \, \mathcal{D}\hat{g}\,e^{-I_{\hat{g}}} = \overline{\langle \Psi | \Psi \rangle} \, .
\eeq
This shows full equality between the path integrals $\overline{\langle \Psi|\prod_{\mathcal{H}_{R}}|\Psi\rangle}=\overline{\langle \Psi|\Psi\rangle}$. The argument can be extended to show $\overline{(\langle \Psi|\prod_{\mathcal{H}_{R}}|\Psi\rangle-\langle \Psi|\Psi\rangle)^2}=0$ by including all handles connecting $\langle \Psi|\prod_{R}|\Psi\rangle$ with $\langle \Psi|\prod_{R}|\Psi\rangle$ (or $\langle \Psi|\Psi\rangle$ in the cross term) into the $I_{bulk}$ region and repeating the above argument. This concludes our argument that the  $\kappa \to \infty$ set of shell states at any preparation temperature spans the fine-grained Hilbert space $\mathcal{H}_{LR}$ in the full gravitational path integral.

\subsection{All order argument for $\mathcal{H}_{2s} \subseteq  \mathcal{H}_L\otimes \mathcal{H}_R$} \label{allOtensorspan2s}
In this section we adopt the notation of Sec.~\ref{sec:1s_span_2s}. In the limit where the shells are parametrically heavy the extension to include perturbations around the saddlepoints follows immediately by the same logic as in Sec.~\ref{sec:allOspan}. We can again extend further to topologies that do not allow saddlepoints. The argument in Sec.~\ref{sec:allOspan} was made by cutting up the connected contributions to (\ref{eq:spanpathint}) along the shell world volumes, resulting in a central strip and $n+1$ vaccum strips. Consider now the connected contributions to 
\beq\overline{ G^{n}_{L,ac}G^{m}_{R,il} \langle \mathbf{p}|_{2s} a\rangle_{L} |i\rangle_{R}   \langle c|_{L}\langle l|_{R} |\mathbf{p}\rangle_{2s}}\eeq
 as in Fig.~\ref{fig:tensorspanbcloop}. If we cut a given geometry along both the $L$ and $R$ shell worldvolumes we end up with a set of cut-out $L$ and $R$ strips and a central region that looks like a cut-out circle containing the two-sided shell $\mathbf{p}$ insertions. Whereas before in the $n\to -1 $ limit the strips disappeared and the cut-out of the central strip was glued together into the operator strip boundary condition defining $\braket{\Psi|\Psi}$, here we find that the cut-out circle gets glued into the boundary circle defining $\langle \mathbf{p}| \mathbf{p}\rangle_{2s}$ as $n,m \to -1$.  In particular, the asymptotic boundary of this circle and of the $L,R$ strips can again be completed by adding in homology regions of time extent $T_k,\tilde{T}_{i,L},\tilde{T}_{j,R}$ respectively in a completely analogous manner as above. For the cut-out circle this results in an asymptotic boundary condition that is simply the norm $\langle \mathbf{p}| \mathbf{p}\rangle_{2s}$ of a time-evolved version of the $| \mathbf{p}\rangle_{2s}$ state. Hence we end up rewriting the path integral (\ref{eq:1s_span_2s})  as one that sums over geometries and topologies with a boundary consisting of several disconnected components: (a)  a circular asymptotic boundary condition consisting of the $\mathcal{O}_{\mathbf{p}}$ and $\mathcal{O}^{\dagger}_{\mathbf{p}}$ operators separated by an $L,R$ time of $\sum_{k=1}^{n+1}T_k+\tilde{\beta}_L,\sum_{l=1}^{m+1}T_l+\tilde{\beta}_R$ respectively, (b)  $n$ copies of a vacuum $L$ strip, and (c)  $m$ copies of a $R$ vacuum strip.

 In the limit that $n,m \to -1$ the shell dependent terms vanish and we are left with the path integral computing $\overline{ \langle \mathbf{p}| \mathbf{p}\rangle_{2s}}$. We therefore have 
$ \overline{\langle \mathbf{p}| \prod_{\mathcal{H}_L\otimes \mathcal{H}_R}| \mathbf{p}\rangle_{2s} 
}=\overline{\langle \mathbf{p}| \mathbf{p}\rangle_{2s}}$ to as a full gravity path integral equality. The argument can again be extended to show 
$ \overline{\left(\langle \mathbf{p}| \prod_{\mathcal{H}_L\otimes 
\mathcal{H}_R}| \mathbf{p}\rangle_{2s} -\langle \mathbf{p}| \mathbf{p}\rangle_{2s}\right)^2}$ by including any handles connecting the various boundary insertions into the $g_{bulk}$ region.

\subsection{All order argument for $\mathcal{H}_L\otimes \mathcal{H}_R \subseteq  \mathcal{H}_{2s} $} \label{allO2sspan1s}
In this section we adopt the notation of Sec.~\ref{sec:2s_span_1s}. 
The argument extending (\ref{eq:abc}) and (\ref{eq:2sintensorsqr}) now proceeds analogously to Secs.~\ref{sec:allOspan} and \ref{allOtensorspan2s}. Again, in the limit where the shells are parametrically heavy the extension to include perturbations around the saddlepoints follows from the same logic as in Sec.~\ref{sec:allOspan}.
As the relevant connected contributions to (\ref{eq:abc}) and (\ref{eq:2sintensorsqr}) are wormholes of mixed single- and two-sided boundary conditions, the extension to include topologies that do not allow saddlepoints is done by combining elements of the argument in Sec.~\ref{sec:allOspan} above and that of Sec.~5.1 of \cite{Toolkit}.  In particular, for any geometry that contributes we can produce a so-called ``sheet diagram" by cutting the geometry along the two-sided shell worldvolumes, resulting in one $L$ and $R$ sheet, potentially connected by handles. We now again imagine cut-outs around each two-sided shell, and complete the two sheets into two strips by filling in homology regions. The rest of the argument now carries over straightforwardly from  above: in the $n \to -1$ limit we again find that the shell contributions vanish and the boundary condition reduces to the strip boundary conditions defining $\overline{\langle i|i\rangle_{L} \langle j| j\rangle_{R}}$. The argument for $\overline{\left(\langle i|_{L} \langle j|_{R} \prod_{\mathcal{H}_{2s}} |i\rangle_{L}|j\rangle_{R}-\langle i|i\rangle_{L} \langle j| j\rangle_{R}\right)^2}=0$ is again done by including any handles connecting the various boundary insertions into the $g_{bulk}$ region.

\end{appendix}

\bibliographystyle{jhep}
\bibliography{references}

\end{document}